\documentclass[twocolumn, 10pt]{article}
\usepackage{inputenc}
\usepackage{cite}
\usepackage{graphicx}
\usepackage{caption}
\usepackage{amsmath,amsfonts}
\usepackage{amssymb}
\usepackage{float}
\usepackage[squaren]{SIunits}
\usepackage[T1]{fontenc}
\usepackage{epstopdf}
\usepackage[english]{babel}
\usepackage[export]{adjustbox}
\title{Ultimate frequency resolution}
\author{Eddy Collin}
\date{} 
\begin{document}

\begin{titlepage}

\begin{center}
\vspace*{2.5cm}
\textbf{\LARGE Non-linear Frequency Transduction of Nano-mechanical Brownian Motion \\}
\vspace*{0.5cm} 
Olivier Maillet$^1$, Xin Zhou$^1$, Rasul Gazizulin$^1$, Ana Maldonado Cid$^1$, Martial Defoort$^{1,2}$, Olivier Bourgeois$^1$ \& Eddy Collin$^{1,*}$
\vspace*{0.5cm}

\textit{\small{$^1$Universit\'e Grenoble Alpes, Institut N\'eel - CNRS,  F-38042 Grenoble, France\\ $^2$Now at: CEA, LETI, MINATEC Campus, 17 rue des Martyrs, 38054 Grenoble Cedex 9, France \\ *: Corresponding Author} 
}

\vspace*{0.25cm}
\end{center}
\textbf{ 
We report on experiments addressing the non-linear interaction between a nano-mechanical mode and position fluctuations.
The {\it Duffing} non-linearity transduces the Brownian motion of the mode, and of other non-linearly coupled ones, into frequency noise.
This mechanism, {\it ubiquitous to all weakly-nonlinear resonators thermalized to a bath}, results in a phase diffusion process altering the motion: two limit behaviors appear, analogous to {\it motional narrowing} and {\it inhomogeneous broadening} in NMR. 
Their crossover is found to depend non-trivially on the ratio of the frequency noise correlation time to its magnitude. Our measurements obtained over an unprecedented range covering the two limits match the theory of Y. Zhang and M. I. Dykman, Phys. Rev. B {\bf 92}, 165419 (2015), with no free parameters.
We finally discuss the fundamental bound on frequency resolution set by this mechanism, which is {\it not marginal} for bottom-up nanostructures.
}

\end{titlepage}

\newpage

\section{Introduction}
\label{intro}

Emerging from the tremendous development of micro/nano technologies, nano-electro-mechanical systems (NEMS) have opened unique capabilities to both engineers and physicists. In the first place, they serve as ultra-sensitive probes for force sensing \cite{mosernanotube} with applications e.g. to mass, charge, and even single electronic spin detection \cite{mass,charge,spin}. 
In the second place, these objects are extremely fruitful (weakly) non-linear devices that are able to implement useful functions like e.g. mechanical frequency mixing \cite{blick}, amplification \cite{buks} and bit storage \cite{yamaguchi}. On the fundamental level, high-quality NEMS structures can be thought of as {\it model systems} in which basic phenomena can be advantageously reproduced; one example being the ubiquitous bifurcation mechanism \cite{BifurcCleland,BifurcChan,bifurc}.   

Ultimately, when coupled to a quantum-limited detection scheme such as a microwave cavity or a Single-Electron Transistor, their sensitivity can be brought to the quantum limit \cite{SQLimit,SQLimit2}. This leads to a unique platform realizing the ``ultimate force detector'' foreseen by C. Caves in the 80's \cite{caves80}.
Such moving structures that are macroscopic relative to the atomic scale but follow the laws of quantum mechanics are currently under development for tests of quantum foundations \cite{QMgrounds,QMGrounds2,QMGrounds3}. Furthermore, they are thought to be a unique new quantum electronics component enabling e.g. coherent photon conversion from the microwave to the optical domain \cite{clelandQPh,Painter}. 
		
Essentially all applications require in the first place the resonance frequency of the mechanical mode in use to be {\it as stable as possible}. As such, the understanding of the sources of frequency fluctuations in nano-mechanical devices becomes {\it an essential technical topic} \cite{NoiseClelandJAP,mosernanotube,bachtold-highQ,Kippen,novotny,hentz}. 
But in the first place, it is also {\it a fundamental research goal}: the measured frequency noise in actual devices is much larger than all expectations \cite{hentz,tang,dykbachtold,phasediff}, demonstrating even non-linear features for carbon-based systems \cite{venstra,nonlinDephas}. Thus, attempts have been made to model noise sources \cite{nanotubeTheory,graphenefluctu}, or to create model experiments experimentally demonstrating the underlying mechanisms \cite{novotny,usdecoh,MotNarrowChan,vinante}. 
		
Clever driving schemes taking advantage of non-linearities have been devised to significantly suppress frequency noise \cite{RoukesPRLnoise,RoukesPRLnoiseII}.
But what shall be an ``ideally frequency-noise minimizing'' nano-mechanical system in the first place? 
We know that at lowest order, the dynamics of a mechanical structure can be described by a family of {\it normal modes} which are nothing but independent harmonic oscillators. Pushing to the next order, these modes are weakly non-linear (so-called {\it Duffing} resonators) and are dispersively coupled one to the other 
\cite{VenstraNlin, KunalNlin, RoukesNlin}. 
Since all of the modes are unavoidably coupled to a thermal reservoir (ideally the same one), Brownian motion of each of the modes will transduce into a frequency noise on all the others \cite{novotny,vinante,hentz,dykmanfluctu,nanotubeTheory}, and also on itself. Even in a system realized with ideal materials having no internal sources of noise, this built-in mechanism shall fix {\it an ultimate limit} to the mechanical resonance frequencies stability at $T \neq 0$.
Only in the limit of $T \rightarrow 0$, when all the modes are in their quantum ground state, do the dispersive couplings lead to a simple frequency renormalization of the resonances through the zero-point-fluctuations of each of them: a sort of {\it mechanical Lamb shift} that dresses all the modes \cite{lamb}.
				
In the present article, we report on a {\it model experiment} in which we use very high quality silicon-nitride NEMS cooled down to Kelvin temperatures. 
A {\it single} mode is driven by a stochastic force, leading to effective temperatures as high as $10^9~$K for this mode only. We extract the effect of this ``artificial out-of-equilibrium heating'' on the {\it mode itself}, both by measuring the spectrum of the motion and by measuring the simultaneous response of the same mode to a sine-wave excitation. 
The effect on a {\it nearby mode} is measured with the sine-wave excitation scheme. 
The setup is carefully calibrated \cite{RSICollin}, while the devices' characteristics are obtained by both measurements and calculations; the agreement with theory is obtained {\it with no free parameters}. 
Besides, the experiment is performed on different devices proving the reproducibility of the results. 

We demonstrate experimentally the two regimes of the Brownian motion transduction, named after analog phenomena present in Nuclear Magnetic Resonance (NMR): ``motional narrowing'' and ``inhomogeneous broadening'' \cite{MotNarrowChan}. 
Based on Ref. \cite{dykmanfluctu} and simple expansions of Euler-Bernoulli theory (including non-linear coefficients \cite{KunalNlin, RoukesNlin, CrossLifshitz}) we give the analytic tools enabling the calculation of the ``ultimate frequency stability'' reached by any doubly-clamped device, depending on stress, dimensions and temperature $T$ \cite{Suppl}.
For bottom-up structures like e.g. carbon nanotubes with high aspect ratio, this limit {\it is not marginal} \cite{nanotubeTheory}.  
\vspace*{-5mm}
	
\section{Results}
\label{Results}
	\subsection{The nano-electro-mechanical systems}	
 	
		\begin{figure*}[t!]
		\center
	\includegraphics[width=8.cm]{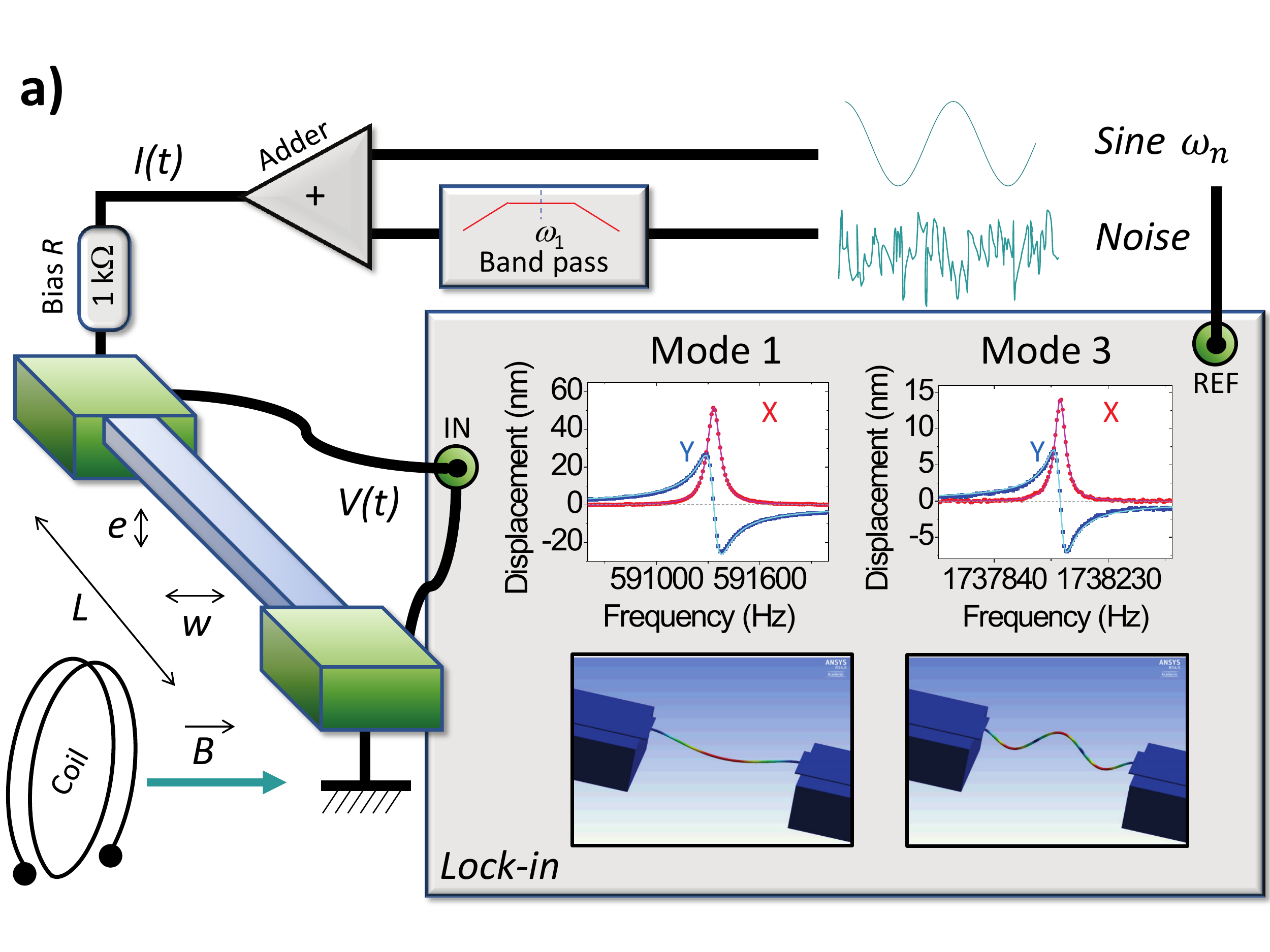}     \includegraphics[width=8.7cm]{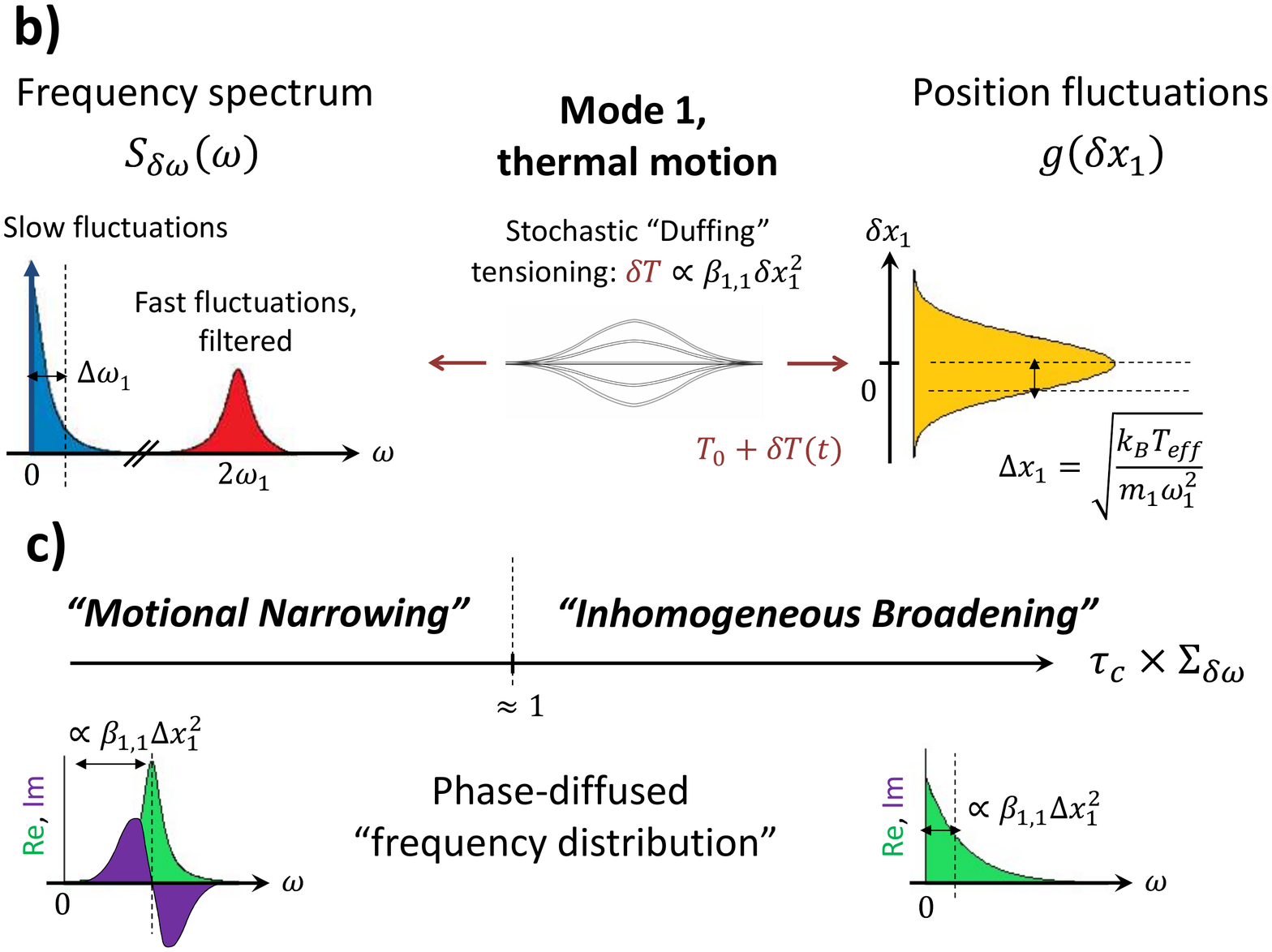}
			\caption{\small{
			(Color online). {\bf a)} The nano-mechanical beams (left) are driven by means of a d.c. magnetic field $B$ and an a.c. current consisting of the sum of two components (top): a sine-wave which frequency is swept around a chosen mode ($n=1$ or $n=3$ here) and a Gaussian white noise filtered around mode $n=1$. The motion is detected with a lock-in amplifier through the induced voltage $V$, leading to the two quadratures $X$ and $Y$ for each mode $n=1$ or $n=3$ (right; lines are Lorentz fits and images Ansys\textregistered $~$numerical simulations). Data corresponding to beam 300$~\mu$m-n$^\circ$1 in the linear regime.
		{\bf	b)} The Gaussian noise force applied onto the mode (here mode 1, center) is equivalent to an effective temperature $T_{ef\!f}$ (right). The motion transduces into a frequency noise (spectrum on the left) because of the Duffing non-linearity $\beta_{1,1}$ due to tensioning. Only the low frequency part of these fluctuations is relevant (in blue, with the d.c. average marked by an arrow), the high frequency term (red) is filtered-out by the dynamics of the mode (adiabatic picture in the rotating frame of the motion). 
		{\bf	c)} Depending on the amplitude of frequency fluctuations $\Sigma_{\delta \omega}$ (their standard deviation $\propto [\int\! S_{\delta \omega} d\omega]^{1/2}$) with respect to their correlation time $\tau_c$ (here $1/\Delta \omega_1$, with $\Delta \omega_1$ the linewidth of the noisy mode) two regimes are distinguished: ``motional narrowing'' and ``inhomogeneous broadening''. This is due to the underlying dynamics of phase diffusion experienced by the mechanical mode, leading to the averaged ``frequency distribution'' depicted  below the horizontal arrow (green and violet). }}
			\label{fig_setup}
			\label{fig_scheme}
		\end{figure*}

The devices under study are doubly-clamped silicon-nitride nano-beams having width $w=$300$~$nm and thickness $e_{SiN}=$100$~$nm. Two high-stress (1$~$GPa) beams of $L=$300$~\mu$m length have been measured plus a 250$~\mu$m one (samples 300$~\mu$m-n$^\circ$1, 300$~\mu$m-n$^\circ$2 and 250$~\mu$m-n$^\circ$1), together with one low-stress (100$~$MPa) $L=$15$~\mu$m shorter beam (15$~\mu$m-n$^\circ$1). 
A thin layer of aluminum ($e_M$ about 30$~$nm to 90$~$nm thick) has been added on top to create electrical contacts. The experiments are performed at 4.2$~$K in cryogenic vacuum (pressure $<10^{-6}~\milli\bbar$).
 
Fig. \ref{fig_setup} a) shows a schematic of the setup. For each device and each mode $n$ (or $m$) studied, we perform a careful calibration based on the technique developed in Ref. \cite{RSICollin}. We can thus infer forces $F_n$ and displacements $x_n$  in S.I. units, and compute the devices characteristics (namely mass $m_n$, spring constant $k_n$, non-linear coefficients $\beta_{n,m}$). These match the expected calculated values; note that a particular care has been taken in the calibration of the noise source. The only fit parameter is indeed an overall correction of the force noise not exceeding 15$~$\% in amplitude (same order as in Ref. \cite{bifurc}).
Actuation and detection are performed with the magnetomotive scheme \cite{RSICollin,SAACleland}. A drive current (composed of both the Gaussian noise component centered around resonance frequency $\omega_1$ and a sine-wave of frequency $\omega$ close to $\omega_n$, with $n=1$ or $n=3$) is injected in the NEMS metallic layer through a home-made adder and a $1~\kilo\ohm$ bias resistor. In an in-plane d.c. magnetic field orthogonal to the beams, this generates an out-of-plane driving force $F_n (t)$ with harmonic component $F_n^0 \cos(\omega t)$.
 The motion is detected through the induced voltage by means of a standard lock-in detection. We obtain the two quadratures, in-phase (X) and out-of-phase (Y) with respect to the local oscillator.

In order to preserve our calibration capabilities, the lock-in has also been used for the spectral measurements $S^n_X(\omega)$ of the Brownian motion of mode $n=1$. 
Moreover, this enables to measure fluctuations on each of the two quadratures $X$, $Y$ independently (plus their cross-correlations). When the sinusoidal excitation is weak (or nonexistent), the spectra on X and Y are equivalent and no correlations are detected; this is the range of validity of the work presented here. 
However, signatures of squeezed statistics of motion \cite{fluctu_Buks} can be observed on measured spectra when the sinusoidal excitation is too large. Details on the measurement technique, calibrations and calculated parameters can be found in S.M. \cite{Suppl}.  
	
\subsection{Dispersive coupling driven by stochastic motion}	

Linear motion of thin nano-beams is very well described by the Euler-Bernoulli equation \cite{clelandbook}. The basic ingredients involved are the inertia (through the density $\rho_{beam}$), the Young's modulus $E_{beam}$ and the tension $T_0$ generated by the in-built stress. 	
For doubly-clamped beams, the non-linear behaviour is well understood: it arises from the stretching of the device under transverse motion $x$ \cite{RoukesNlin,KunalNlin,CrossLifshitz}. This {\it geometric non-linearity} results in a tensioning $T_0+\delta T$ of the beam with $\delta T \propto x^2$, which can be incorporated into the beam equation \cite{CrossLifshitz}. This leads to a frequency shift of the modes that is proportional to the square of the displacement. When only two modes $n,m$ are under study, it writes: 
\begin{eqnarray}
\omega_n  & = &  \omega_n^0 + \beta_{n,n} x_n^2 + \beta_{n,m} x_m^2  \label{freqshift} , \\
\omega_m  & = &  \omega_m^0 + \beta_{m,m} x_m^2 + \beta_{m,n} x_n^2   \label{freqshift2},
\end{eqnarray}
where we introduced  $\omega_n^0$, $\omega_m^0$ the linear resonance frequencies, and the Duffing non-linear coefficients $\beta_{i,j}$ \cite{RoukesNlin}. We remind for the interested reader the mathematical derivation of these expressions in S.M. \cite{Suppl}.

Eqs. (\ref{freqshift}-\ref{freqshift2})
can be adapted when one of the motions, say $x_n$, is a stochastic variable: $x_n=x_n^0+ \delta x_n$, with $x_n^0$ the certain component and $\delta x_n$ the Gaussian and centered random component. 
In order to introduce the phenomenon, let us first consider the case depicted in Fig. \ref{fig_scheme} b), where only one mode $n$ is addressed. 
We apply on $n=1$ a Gaussian random force $\delta F_n(t)$ of spectrum $S^n_F(\omega)$ whose strength can be converted into an effective temperature $T_{ef\!f}$ through the Fluctuation-Dissipation Theorem $S^n_F=2 k_B T_{ef\!f} m_n \Delta \omega_n$. 
$\Delta \omega_n$ is the linewidth of the resonance of mode $n$ (with $Q_n=\omega_n^0/\Delta \omega_n$ the quality factor), and $S^n_F(\omega)$ is white around the mode studied only (and negligible elsewhere). The mechanical mode thus experiences position fluctuations (Brownian motion) linked to $S^n_F$ through the mechanical susceptibility, whose spectrum $S_X^n(\omega)$ is peaked around $\omega_n^0$. Since $T_{ef\!f} \gg 4.2~$K the experimental temperature, we safely neglect all other sources of fluctuations while enabling a thorough tuning of the Brownian motion amplitude of mode $n$ only.

	
		\begin{figure}[h!]		 
			 \includegraphics[width=9cm]{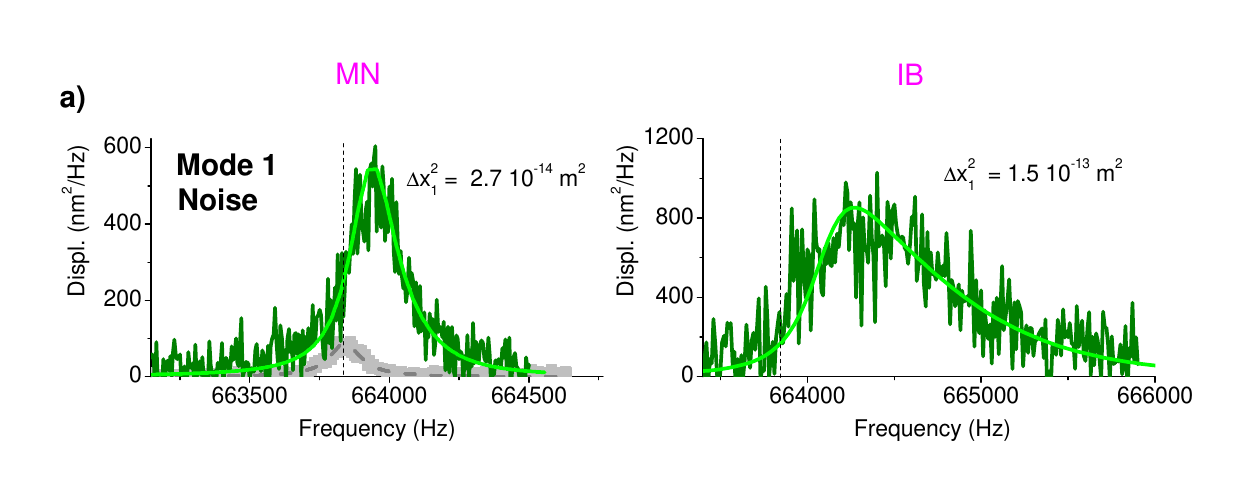}			 			 \includegraphics[width=9cm]{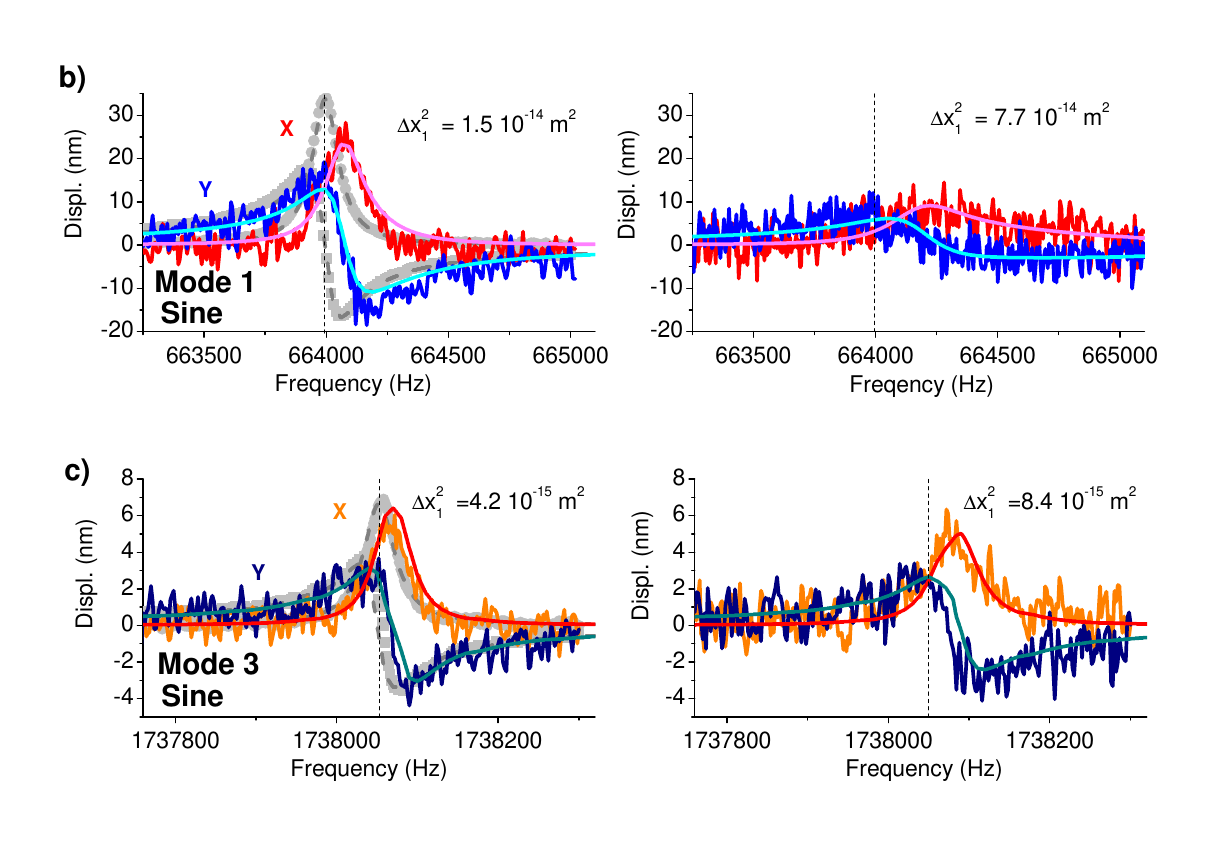}
			\caption{\small{
			(Color online). \textbf{(a)} Brownian motion spectra measured on mode $n=1$ of sample 300$~\mu$m-n$^\circ$2 (Duffing spectra). \textbf{(b)} In-phase (X) and quadrature (Y) components measured for mode $n=1$ while driving noise on the same $n=1$ mode for sample 300$~\mu$m-n$^\circ$2 (intra-mode). \textbf{(c)} Same measurement performed on mode  $n=3$ while driving fluctuations on $n=1$ for sample 300$~\mu$m-n$^\circ$1 (inter-mode coupling).  The standard deviation $\Delta x_1^2$ (i.e Brownian motion level) is increased from left to right (essentially from MN to IB regime, see Fig. \ref{fig_sum1}), and sinusoidal drives are kept in the linear regime. The grey data are the references obtained for very weak noise levels. The verticals are resonance position without Brownian transduction, and lines theoretical calculations (see text). }}
			\label{fig_reslines}
		\end{figure}

\subsection{Transduction mechanism}

From 
the Duffing equations, the random motion $\delta x_n$ is transduced into a frequency noise $S_{\delta \omega}(\omega)$. Since this dependence is quadratic, the frequency noise is neither Gaussian nor centered. Its spectrum depicted in Fig. \ref{fig_scheme} b) consists in a low frequency part and a high frequency component peaked around $2 \omega_n$. The high-frequency fluctuations are essentially filtered out by the mode dynamics, as can be seen in a Rotating Wave Approximation (RWA).
Thus, driving the mode with a sine wave force $F_n^0 \cos(\omega t)$ weak enough to remain in the linear response limit, the motion $x_n^0$ will adiabatically follow the slow frequency fluctuations experiencing both a frequency shift and a spectral broadening \cite{dykmanfluctu}.
The measurement scheme itself is always slow enough to ensure that all fluctuations are spanned while acquiring data. Note that the Brownian fluctuations do not need to be small for the theory to apply.

The phenomenon is non-trivial, and depends strongly on the correlation time of the fluctuations $\tau_c = 1/\Delta \omega_n$. 
Defining $\Sigma_{\delta \omega}=4  \beta_{n,n} \Delta x_n^2$ a frequency noise amplitude parameter (essentially their standard deviation $\propto [\int\! S_{\delta \omega} d\omega]^{1/2}$), two regimes should be distinguished depending on the magnitude of the product $\tau_c \times \Sigma_{\delta \omega}$, see Fig. \ref{fig_scheme} c). 
The process can be understood in terms of phase-diffusion for the mode studied, the dynamics being averaged over all realizations of the fluctuating resonance frequency $\delta\omega$, namely $x_n^0(t) \propto \left\langle e^{\displaystyle i\int_0^t\delta\omega(t'')\mathrm{d}t''}\right\rangle$ \cite{dykmanfluctu}. 
The frequency-domain data can thus be described by a convolution of the linear response by a complex valued distribution of frequencies, as seen from the NEMS (bottom of Fig. \ref{fig_scheme} c):
\begin{equation}
\mathrm{FT} \left[\frac{\exp(+ \Gamma_n t)}{\cosh( a_n t ) + \frac{\Gamma_n}{a_n} \left( 1+2 i \alpha_n  \right) \sinh( a_n t )  }\right]\left(\omega \right) , \label{olives}
\end{equation}
FT meaning Fourier Transform, with $\Gamma_n =\Delta \omega_n /2$ the mode's relaxation rate, $a_n  =  \Gamma_n \sqrt{1+4 i \alpha_n}$ and $\alpha_n   =   \frac{\Sigma_{\delta \omega}}{2 \Gamma_n} = \tau_c \times \Sigma_{\delta \omega}$ the {\it motional narrowing parameter}.

By analogy with Nuclear Magnetic Resonance, when $\tau_c \times \Sigma_{\delta \omega} \ll 1$ the certain component's dynamics is said to be in the ``motional narrowing'' limit (MN), while for $\tau_c \times \Sigma_{\delta \omega} \gg 1$ it lies in the ``inhomogeneous broadening'' limit (IB). 
In the former case, the fluctuations are too fast to enable the resolution of the small frequency changes $\Sigma_{\delta \omega}$ \cite{MotNarrowChan, motnarrowQubit}: the random variable's dynamics looses memory too quickly, and only a fraction of the frequency fluctuations impacts the driven motion. 
This leads to a {\it certain} frequency shift which is nothing but the average of the frequency fluctuations proportional to $\Delta x_n^2$, together with a (weaker, second order) symmetric broadening {\it quadratic} in $\Delta x_n^2$ (bottom-left ``distribution'' in Fig. \ref{fig_scheme} c). 
In the latter case, the fluctuations are slow enough so that the full range of frequency fluctuations can be explored by the  $x_n^0$ sine-wave response \cite{dyk_adiab,usdecoh}: there is a large asymmetric broadening, which reflects the actual distribution of frequency fluctuations (bottom-right in Fig. \ref{fig_scheme} c).
When mode $m=3$ is sine-wave driven and detected while force noise is still applied onto mode $n=1$, the treatment is identical with the replacement $\Sigma_{\delta \omega}=  2 \beta_{m,n} \Delta x_n^2$ \cite{dykmanfluctu}. Besides, an equation similar to Eq. (\ref{olives}) holds for the direct calculation of non-linear Brownian spectra \cite{dykmanfluctu}.
A brief description of the theoretical tools developed in Ref. \cite{dykmanfluctu} is given in S.M. \cite{Suppl}.

In the next Section, we present the experimental data and the theoretical calculations corresponding to these two situations. 
The displacement noise spectrum of mode $n=1$ is also directly measured. We reach the limit where this spectrum itself is {\it imprinted by the  Duffing non-linearity} \cite{novotny}, and match it to the theory \cite{dykmanfluctu}. 
Since Brownian motions of two $m \neq n$ distinct modes are not correlated, from these elementary measurements one can then deduce the {\it generic} situation where $N$ thermalized modes of the same structure are coupled together.

		\begin{figure}[h!]
	\includegraphics[width=8cm]{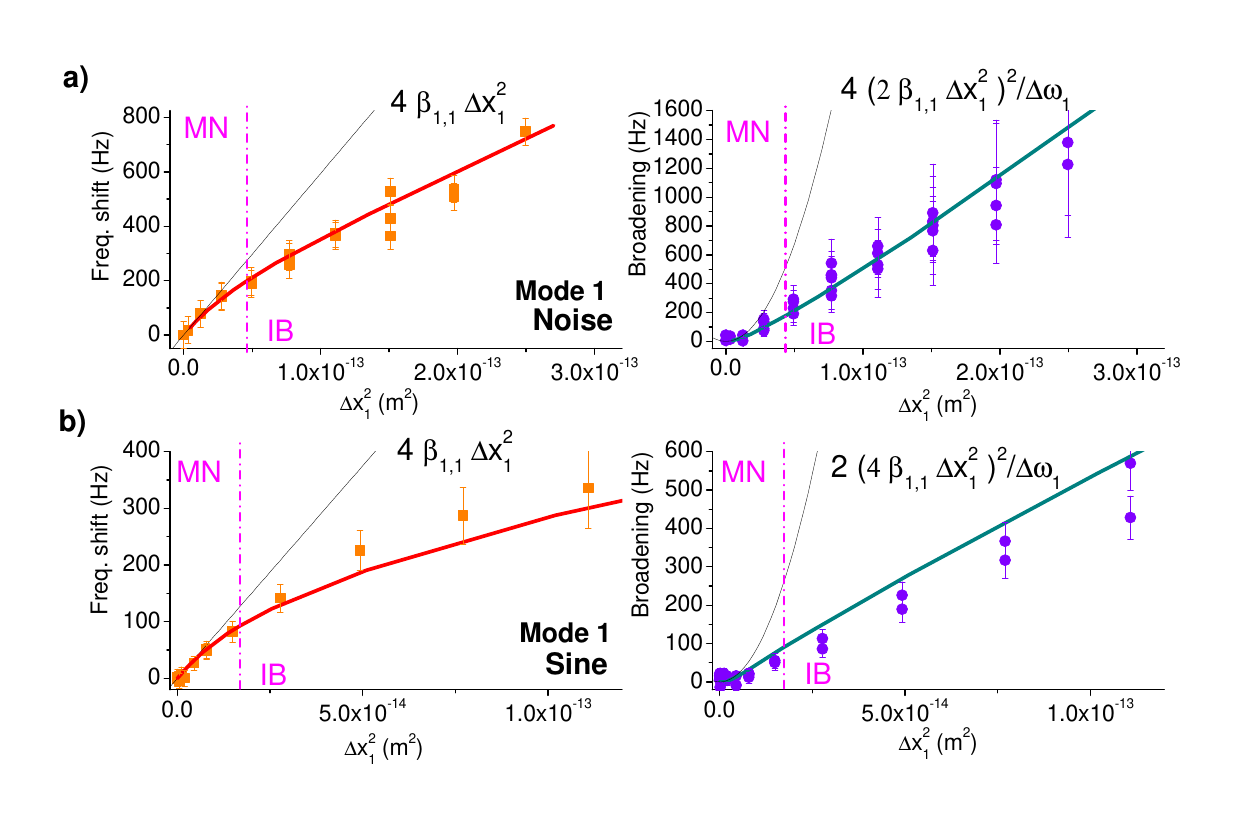} \includegraphics[width=8cm]{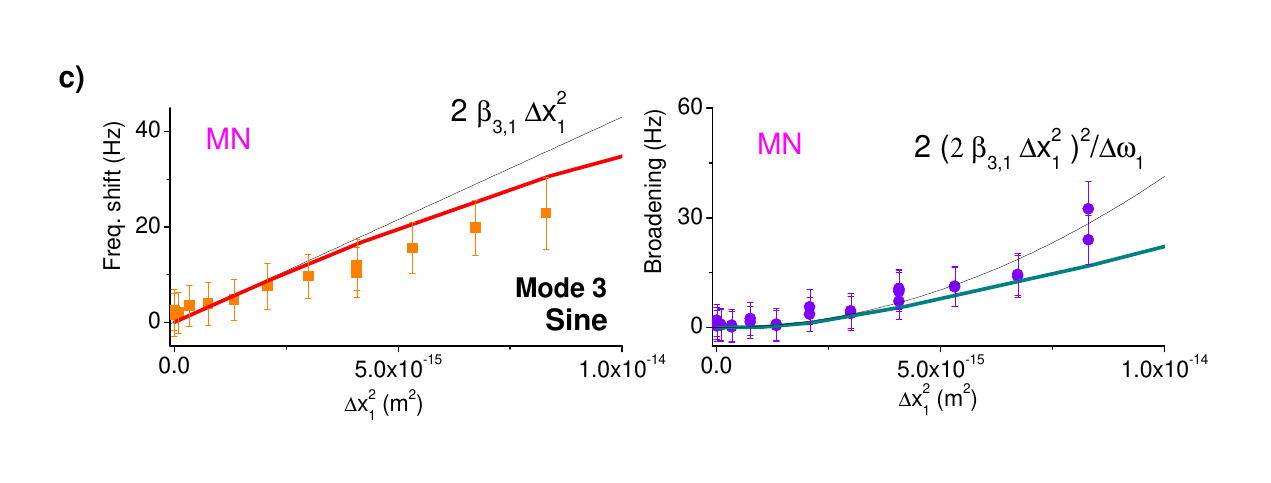}
			\caption{\small{
			(Color online). Frequency shift (left) and broadening (from FWHH, right) for \textbf{(a)} the (Duffing) spectra measured on mode $n=1$ for sample 300$~\mu$m-n$^\circ$2, \textbf{(b)} the sine-wave excitation of mode $n=1$, with Brownian motion on the same mode $n=1$ for sample 300$~\mu$m-n$^\circ$2 (intra-mode case), and \textbf{(c)} similar result for sine-wave excitation of mode $m=3$, with Brownian motion of mode $n=1$ for sample 300$~\mu$m-n$^\circ$1 (inter-mode). 
			The thin lines are the ``motional narrowing'' (MN) analytic expansions, with the dashed verticals corresponding to the cross-over towards ``inhomogeneous broadening'' (IB) when $\tau_c \times \Sigma_{\delta \omega} =1$. The full lines are from the complete theoretical model (see text). }}
			\label{fig_sum1}
  		\end{figure}

\subsection{Measured resonance properties}	
				
In Fig. \ref{fig_reslines} a) we present the direct measurement of the Brownian noise spectra $S_X^n(\omega)$ on mode $n=1$ for sample 300$~\mu$m-n$^\circ$2. 
No sine-wave excitation is applied, neither on $n=1$ nor on $m=3$ modes. 
The noise level is quoted in terms of standard deviation $\Delta x_1^2$ instead of $T_{ef\!f}$ (or force noise intensity) since this is the physical parameter of importance. 
For small Brownian excitations, the peak remains Lorentzian. 
However, when the amplitude of motion becomes large, the non-linear term $\beta_{1,1}$ starts to impact the lineshape: the peak broadens and becomes asymmetric \cite{novotny,vinante}. 
As expected, the resonance peak globally shifts towards higher frequencies (see Fig. \ref{fig_sum1} a) for a summary of the spectrum characteristics). 
The lines are the exact theory from Ref. \cite{dykmanfluctu}, computed with no free parameters: we call them {\it ``Duffing spectra''} \cite{Suppl}.
Note that no deviations from standard Gaussian statistics are measured in these conditions, as it should be for high-$Q$ devices 
\cite{nonGaussFP}: spectra on the X quadratures are equivalent to the ones measured on Y, and no cross-correlations are detected \cite{Suppl}. 

We turn next to the case of the intra-mode coupling. We still drive mode $n=1$ with white noise, but we also measure and drive it with a sine-wave signal. Mode $m=3$ is left unexcited. Data and theory from Ref. \cite{dykmanfluctu} are compared in Fig. \ref{fig_reslines} b) with no free parameters.
The X lineshapes look like the peaks obtained in the ``Duffing spectrum'' case, Fig. \ref{fig_reslines}. The effect of the added force noise on the mode is again twofold: first, the resonance peak slightly shifts towards higher frequencies, and second it broadens (consequently flattens) and acquires an asymmetric shape.
In Fig. \ref{fig_sum1} b) we summarize the characteristics of the measured resonance lines on device 300$~\mu$m-n$^\circ$2 (obtained from the X quadrature).

Measured resonance lines and calculations in the inter-mode case (sine-wave driving and measuring mode $m=3$ while adding force noise on mode $n=1$) are shown in Fig. \ref{fig_reslines} c). They resemble very much the intra-mode results of Fig. \ref{fig_reslines}, even though the quality of the data did not enable to reach as high fluctuation levels (see Fig. \ref{fig_sum1} c). More data can be found in S.M. \cite{Suppl}. 

The three basic situations are compared in Fig. \ref{fig_sum1}: we show the characteristics of the measured spectra and resonance lines on 300$~\mu$m devices in terms of frequency shift (position of the maximum of the resonance peak) and broadening (measured from the Full Width at Half Height, FWHH).
The same characteristics for 250$~\mu$m and 15$~\mu$m devices are also shown in S.M. \cite{Suppl}: since the non-linear coefficients depend strongly on the length $L$ of the structures, this demonstrates the robustness of the effect.

		\begin{figure}[t!]
		\centering
	\includegraphics[width=8cm]{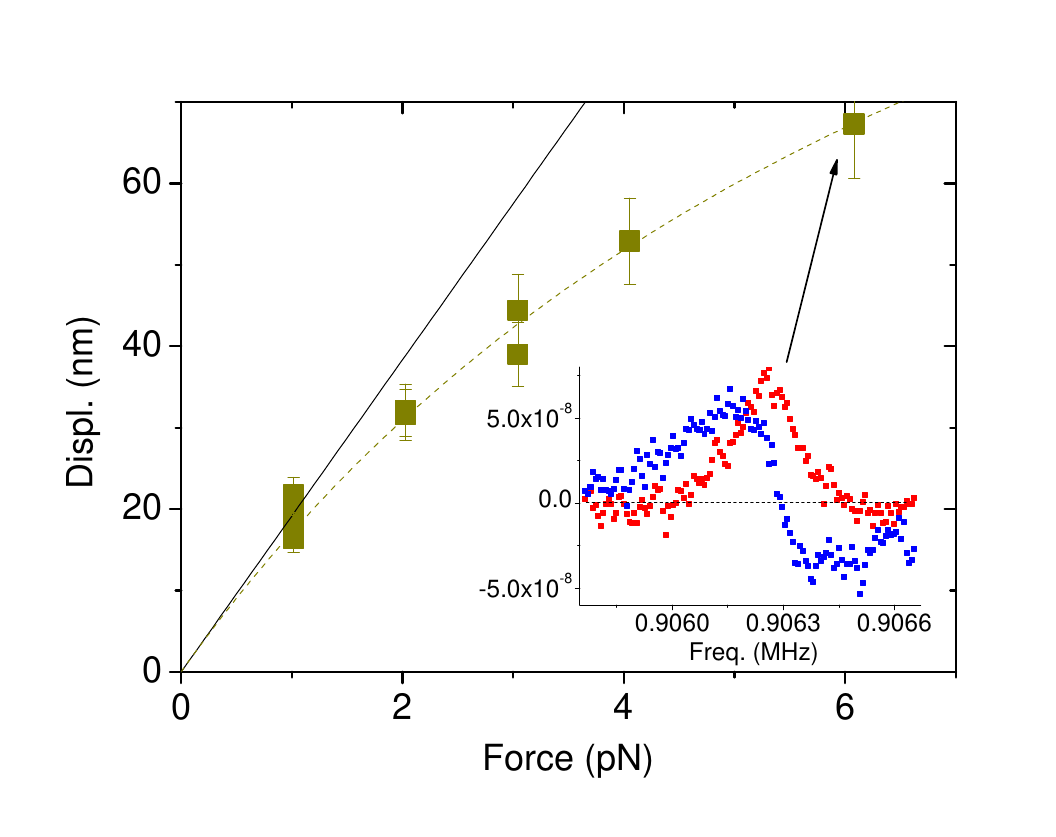}
			\caption{\small{\textbf{Large sine-wave excitation forces} 
			(Color online).	Amplitude of the sine-wave response peak as a function of excitation force for mode $n=1$ of device	250$~\mu$m-n$^\circ$1. Force noise is applied on the same mode $n=1$ such that $\Delta x_1^2 = 2.2\, 10^{-15}~$m$^2$. The full line is the linear theory, which clearly does not fit the data (squares and dashed guide). The inset corresponds to the largest drive measured peak. }}
			\label{fig_large}
		\end{figure}

The global agreement between data and theory is remarkable. 
Essentially, ``Duffing spectra'', intra-mode and inter-mode Brownian frequency transduction display the same characteristic features.
This highlights that the main ingredient is the {\it dynamics of the noisy mode}, not the one of the chosen probe.
From Fig. \ref{fig_sum1}, we see that we span the whole range of the phenomenon from ``motional narrowing'' to ``inhomogeneous broadening''. 
In the ``motional narrowing'' limit, indeed the first order effect is a global frequency shift proportional to $\Delta x_n^2$. On the other hand, in the ``inhomogeneous broadening'' range the main feature is the asymmetric broadening which is nothing but the image of the frequency distribution (inhomogeneity in time-domain, as opposed to position-domain for NMR \cite{usdecoh}).
Further technical discussions of these two limits can be found in S.M. \cite{Suppl}.  

However, the theory of Ref. \cite{dykmanfluctu} applies for sinusoidal excitation strengths lying within the linear response range. 
When the motion amplitude is increased beyond this limit, new phenomena are expected to take place like e.g. the parametric squeezing of the Brownian motion \cite{fluctu_Buks}. One signature obtained experimentally that fails to be reproduced by the theory is shown in Fig. \ref{fig_large}: for large sine-wave excitations, 
the amplitude of the detected mechanical peak {\it lies below} the calculation, as if the impact of frequency noise was stronger than expected. 
In S.M. \cite{Suppl}, we show that the noise spectra measured on mode $n$ are indeed altered by the back-action of the sine-wave response $x_n^0$; the $X$ and $Y$ quadratures are not equivalent anymore, and cross-correlations are non-zero at some peculiar frequencies. 
Further work both theoretical and experimental is required to explore this new dynamical range.

\subsection{Application to a thermalized family of modes}

For a physical thermal bath, the device is always in the ``motional narrowing'' limit. In this case, the linear response of mode $n$ to a weak sinusoidal drive remains Lorentzian, with a resonance frequency ``dressed'' by the Brownian motion of all modes (global frequency shift proportional to $T$). This is essentially analogous to a mechanical Lamb shift \cite{lamb}, in the classical domain. Furthermore the linewidth of the resonance is impacted by a $T^2$ term, a ``thermal decoherence'' effect.

Reproducing results from Ref. \cite{dykmanfluctu}, these can be written at lowest order in terms of simple expansions, respectively for mode $n$:
\begin{eqnarray}
\omega_n &=& \omega_n^0 +  4 \beta_{n,n} \Delta x_n^2 \nonumber \\
&+& \sum_{m \neq n} 2 \beta_{n,m}  \Delta x_m^2 + \sum_{m'} 2 \bar{\beta}_{n,m'}  \Delta y_{m'}^2 \label{lamb}, \\
\Delta \omega_n & = & \Delta \omega_n^0 + 2 \frac{\left(4\beta_{n,n} \Delta x_n^2\right)^2}{\Delta \omega_n^0} \nonumber \\
&\!\!\!\!\!\!\!\!\!\!\!\!\!\!\!\!\!\!\! +&\!\!\!\!\!\!\!\!\!\!\!\!\!\! \sum_{m \neq n} 2\frac{\left(2\beta_{n,m} \Delta x_m^2\right)^2}{\Delta \omega_m^0} + \sum_{m'} 2\frac{\left(2\bar{\beta}_{n,m'} \Delta y_{m'}^2\right)^2}{\Delta \bar{\omega}_{m'}^0} . \label{decohT}
\end{eqnarray}
The validity of these expansions has been experimentally verified in the present work for two modes only, Fig. \ref{fig_sum1}.
They can be extended in this simple way to many modes since the Brownian motion between $n\neq m$ is uncorrelated.
For the sake of completeness, we also added the sum over the {\it other family of transverse modes} (in $\vec{y}$ direction), which coefficients are designed with a bar, and the index with a prime (the position standard deviation simply writes $\Delta y_{m'}^2$). 
Indeed, the nonlinear coupling between flexural modes of different families has been studied recently \cite{other_modes}. 
We shall not discuss the coupling to longitudinal and torsional modes, which is outside of the scope of beam mechanics; 
these depend directly on the Poisson's ratio, and shall be very weak.

Eqs. (\ref{lamb}-\ref{decohT}) can be easily evaluated for doubly-clamped beams by means of mode parameters calculated using the non-linear extension of Euler-Bernoulli beam theory \cite{RoukesNlin,KunalNlin,CrossLifshitz}.
With the simple equipartition result $\Delta x_n^2 = k_B T/k_n$, $\Delta y_{m'}^2 = k_B T/\bar{k}_{m'}$ 
we can rewrite these expressions such that:
\begin{eqnarray}
\frac{\omega_n - \omega_n^0}{\omega_n^0}  &\propto &    \left(\frac{E_{beam} A}{2 L^3}\right) \frac{(k_B T)}{(2 k_n^2)} , \label{dressing} \\
\frac{\Delta \omega_n - \Delta \omega_n^0}{\Delta \omega_n^0}  & \propto &  \left(\frac{E_{beam} A}{2 L^3}\right)^2 \frac{(k_B T)^2}{(2 k_n^4)} Q_n^2 , \label{thermdecoh}
\end{eqnarray}
with $A=w \, e$ the cross-section. In S.M. we summarize the mode parameters obtained in the two extreme limits of Euler-Bernoulli: low-stress (beam) and high-stress (string) \cite{Suppl}. Two key facts have to be highlighted: first, the prefactor in Eqs. (\ref{dressing}-\ref{thermdecoh}) that gives the strength of the effect depends on materials properties and strongly on {\it geometry}. Second, increasing the stress in the structure {\it does reduce} the sensitivity to Brownian transduction. 

\section{Conclusion}		

By artificially heating a single mode of a NEMS structure, we have demonstrated experimentally the non-linear frequency transduction of the Brownian motion of this mode onto itself and onto a nearby one.
Beyond harmonic mode-coupling \cite{VenstraNlin,KunalNlin,RoukesNlin}, the correlation time $\tau_c$ of fluctuations impacts the dynamics. Two regimes are observed depending on the strength of the stochastic force applied: ``motional narrowing'' when the frequency fluctuations are small with respect to $1/\tau_c$, and ``inhomogeneous broadening'' when they are large. The data are compared to the theory from Ref. \cite{dykmanfluctu} that spans the whole range, and we demonstrate excellent agreement without free parameters. To our knowledge, the present work is the first one presenting a complete experimental analysis of this fundamental (classical) phenomenon, analogous to Nuclear Magnetic Resonance (quantum); effective temperatures up to $10^9~$K for the mechanical mode under study have been required to reach the ``inhomogeneous broadening'' limit. 

When extending these results to the case of a family of modes thermalized to a bath at temperature $T$, we find that for typical high-stress top-down structures like the ones used here, the Brownian transduction phenomenon is clearly negligible.
However, for much smaller low-stress structures with high aspect-ratio, the effect is foreseen to be limiting in sensing applications \cite{Suppl}. Besides, the certain frequency shift arising from the thermal dressing can also be a source of frequency instability if temperature $T$ is fluctuating. Finally 
when addressing the issue of {\it intrinsic} sources of decoherence, it is mandatory to control this phenomenon. 
Note that other Authors reached the same conclusions with a different approach specific to nanotubes \cite{nanotubeTheory}.

\section*{Acknowledgements}

\small{We thank J. Minet and C. Guttin for help in setting up the experiment, as well as J.-F. Motte, S. Dufresnes and T. Crozes from facility Nanofab for help in the device fabrication. We also warmly thank Mark Dykman for very useful discussions. We acknowledge support from the ANR grant MajoranaPRO No. ANR-13-BS04-0009-01 and the ERC CoG grant ULT-NEMS No. 647917. This work has been performed in the framework of the European Microkelvin Platform (EMP) collaboration.}

%

\bibliographystyle{ieeetran}

\begin{center}
\onecolumn
{\huge Supplementary Information for \\}
{\huge Non-linear Frequency Transduction of Nano-mechanical Brownian Motion}
\end{center}

\section{Device, setup and magnetomotive scheme}
\label{expesetup}

\begin{figure}[h!]
				\centering
			 \includegraphics[width=8cm]{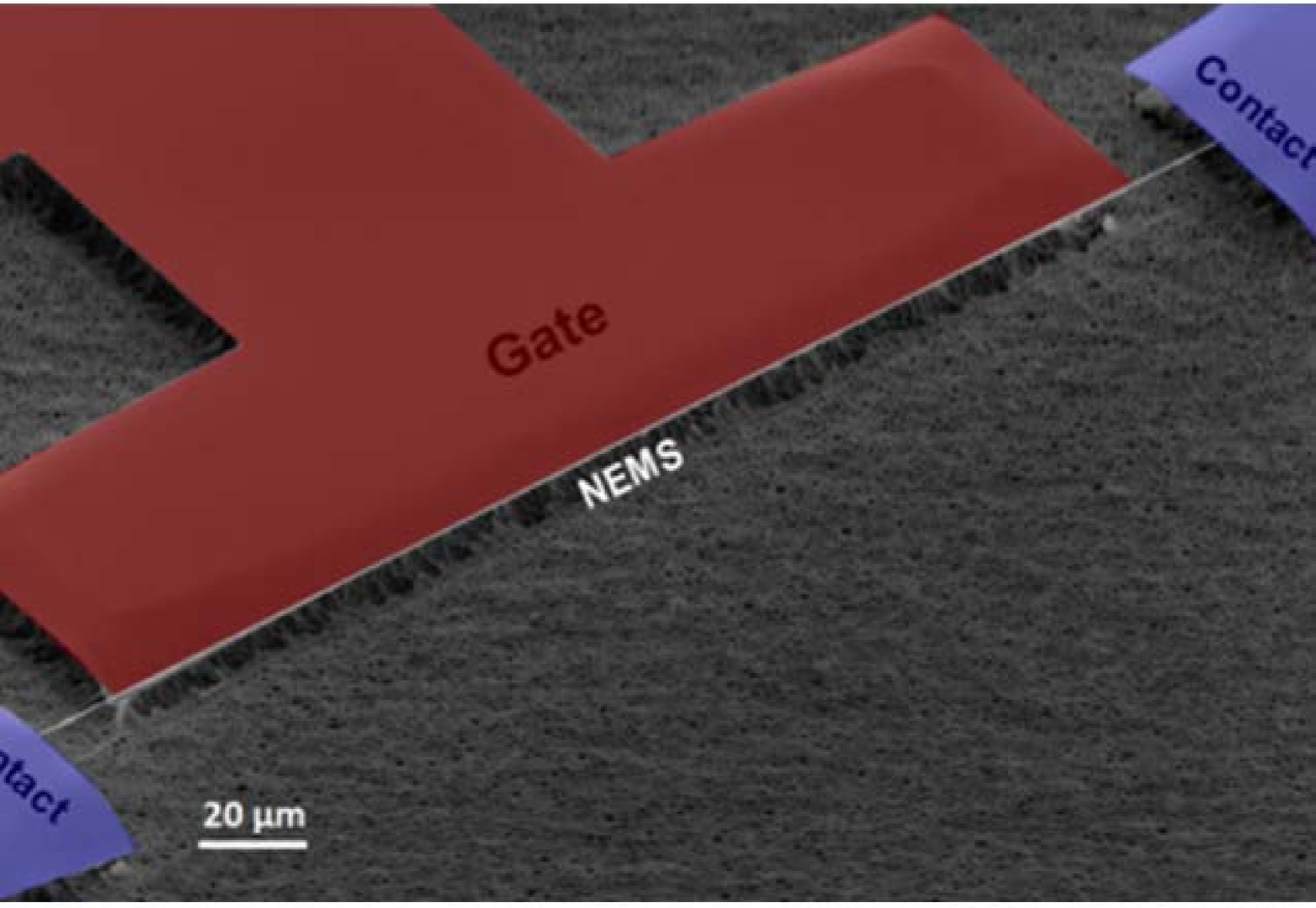}			 			 \includegraphics[width=7.5cm]{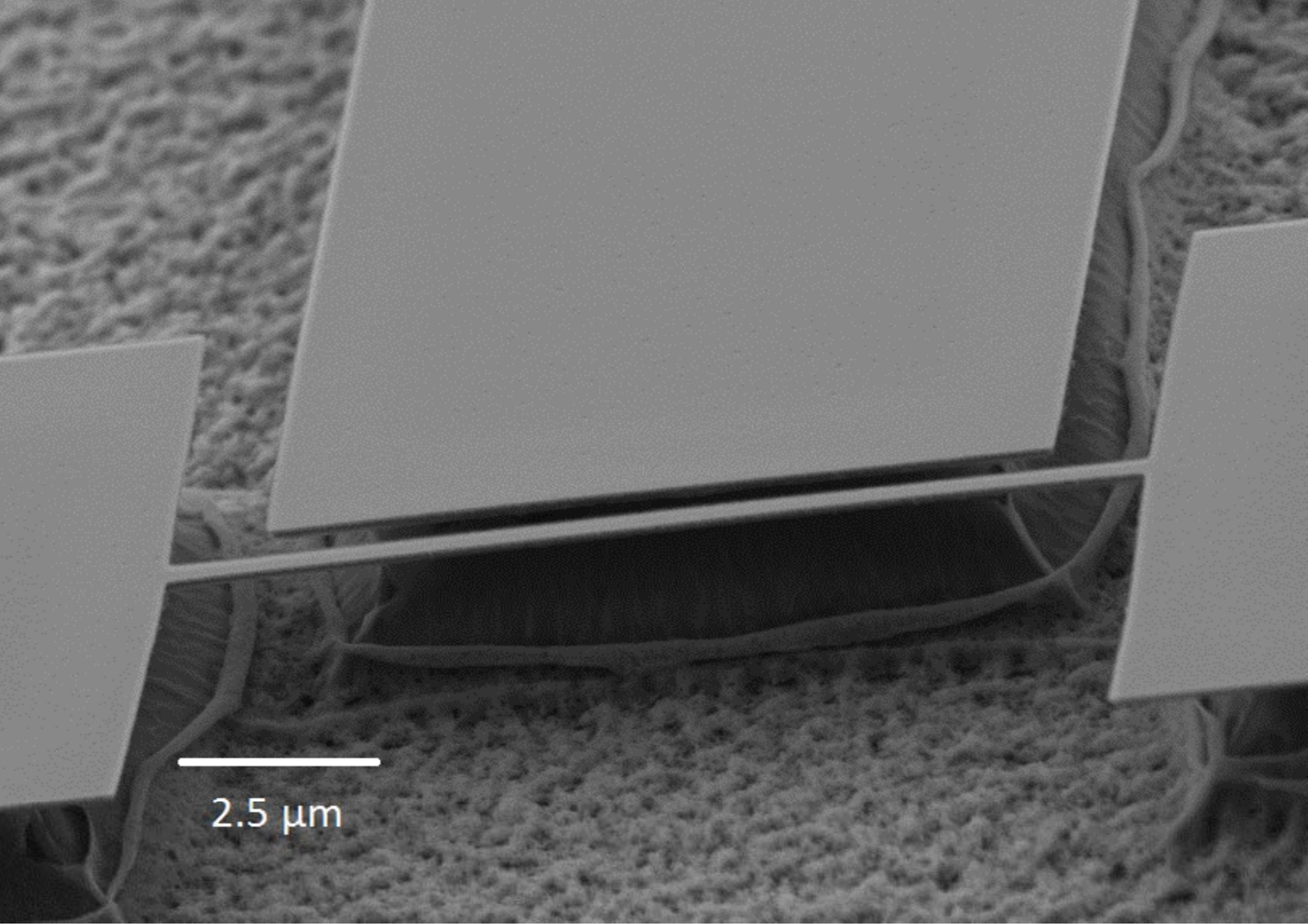}
			\caption{\small{\textbf{Devices.}
			SEM pictures of the devices. Left: 300$~\mu$m beam (in false colour, the different elements of the device are indicated). Right: 15$~\mu$m beam. The gates are not used in the present work.}}
			\label{fig_nems}
		\end{figure}

In Fig. \ref{fig_nems} we show two Scanning Electron Micrograph (SEM) pictures of typical 300$~\mu$m and 15$~\mu$m devices.
The gates, used in other works (e.g. \cite{usdecoh}) are not connected in the present work. Keeping them floating or grounded is found to be equivalent.
 
Actuation and detection are performed with the magnetomotive scheme \cite{RSICollin,SAACleland}. The force amplitude acting on mode $n$ in the sine-wave scheme writes $F^0_n=\zeta_n L B I_0$ with $I(t)=I_0 \cos(\omega t)$ the drive current applied, $L$ the length of the beam and $B$ the magnetic field. $\zeta_n$ is a mode-dependent parameter which can be readily computed from the mode shapes (see Section \ref{sectionBernie}). The detected voltage $V(t)$ is simply the image of the velocity $v_n(t)$ of the mode: $V(t)=-\zeta_n L B \, v_n(t)$. We detect it in a standard lock-in scheme, synchronized to the drive frequency $\omega$. We choose for commodity our definitions of the $X$ and $Y$ quadratures of the {\it position} to match the phase of the {\it velocity}, and thus the one of the detected signal. 
Note that velocity $v_n=i \omega \, x_n$ and displacement $x_n$ require an additional definition: we choose the position of maximum displacement of the beam (in the first mode, it is simply the motion at the center of the beam). The actual values of the $\zeta_n$ coefficients obviously depend on this definition. \\
		
		\begin{figure}[h!]
		\centering
	\includegraphics[width=6.5cm]{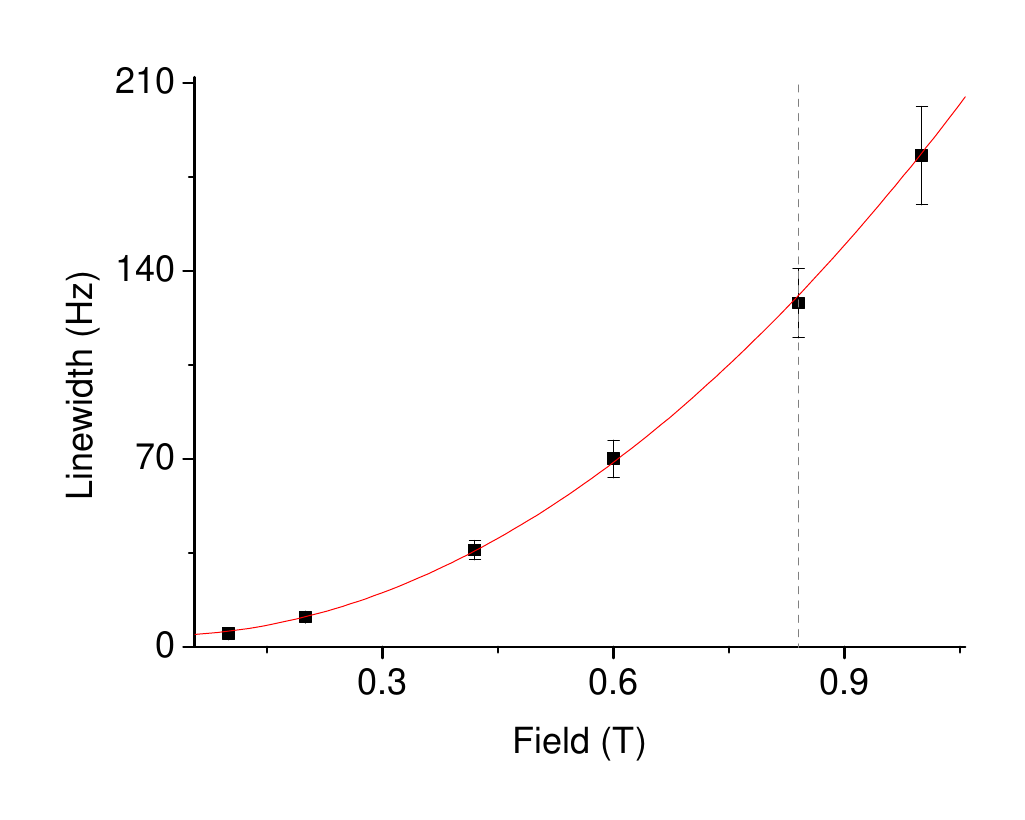}
			\caption{\small{\textbf{Loading from electric circuit.} 
			Linewidth as a function of magnetic field for device 300$~\mu$m-n$^\circ$2. The full line is a quadratic fit (see text). The dashed vertical shows were most data have been taken.}}
			\label{fig_load}
		\end{figure}
		
In the magnetomotive scheme, the resonance is loaded by the impedance of the external circuit \cite{SAACleland}. As a result, the linewidth measured presents a quadratic field dependence. We show this result for device 300$~\mu$m-n$^\circ$2 in Fig. \ref{fig_load}. As such, the linewidths quoted in the paper are mostly due to the electric circuit: for instance device 300$~\mu$m-n$^\circ$2 has been measured mainly around 1 Tesla (at 0.84$~$T, vertical in Fig. \ref{fig_load}). But data has also been acquired at 0.6$~$T and 0.42$~$T to verify that all features scale properly with the resonance linewidth. From the parabola in Fig. \ref{fig_load} one can deduce the circuit impedance as seen from the NEMS; we obtain about 2$~$k$\Omega$ which corresponds to the bias resistor plus the NEMS metallic layer resistance.  \\

The calibration scheme we use is described in Ref. \cite{RSICollin}. The concept on which it is based is very simple: sending a non-resonant current at frequency $\omega'$ to the device we heat it and track the heating by means of the frequency shift of the resonance (which is $T$-dependent). Doing this procedure for various $\omega'$, and taking the DC heating curve as reference and scaling all the others onto it, we can easily extract the transmission coefficient $\left|G(\omega')\right|$ of the injection line. This technique is very reliable: it is based on a {\it genuinely local property}, directly linked to the {\it power injected in the nanomechanical beam}. 
Since the aluminum on the device is non-superconducting at these temperatures, by measuring its Ohmic contribution we can then infer the transmission of the detection line.

Measuring the height $x_{max}$ of the X peak we can easily obtain the spring constant $k_n$ of the mode through $x_{max}= F^0_n Q_n/k_n$ with $Q_n=\omega^0_n/\Delta \omega_n$ the quality factor of mode $n$. From $(\omega^0_n)^2=k_n/m_n$ we deduce the mass associated to the mode. \\
		
		\begin{figure}[h!]
		\centering
	\includegraphics[width=15cm]{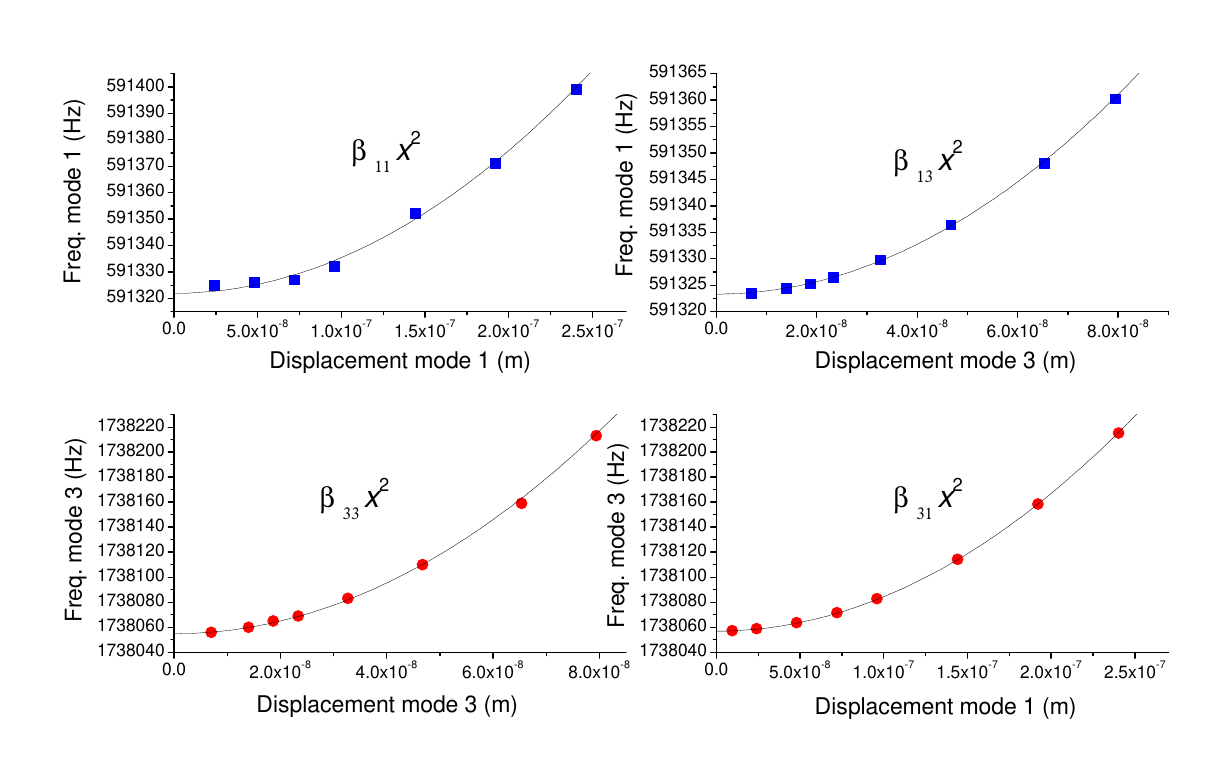}
			\caption{\small{\textbf{Nonlinear $\beta_{i,j}$ coefficients.} 
			Measurements of the $\beta_{i,j}$ for device 300$~\mu$m-n$^\circ$1. The full lines correspond to quadratic fits enabling to extract the numerical values quoted in the paper.}}
			\label{fig_betas}
		\end{figure}

The experimental definition of the $\beta_{i,j}$ coefficients needed is shown in Fig. \ref{fig_betas}, for device 300$~\mu$m-n$^\circ$1. 
$\beta_{n,n}$ is obtained by measuring the position of the resonance maximum as a function of its amplitude squared, for mode $n$. 
Driving mode $n$ with a strong sine-wave signal, we measure the quadratic frequency shift induced on mode $m$ detected with a weak sine-wave drive (such that $\beta_{m,m}$ can be neglected) \cite{RoukesNlin,KunalNlin,CrossLifshitz}. The fit leads to coefficient $\beta_{m,n}$. Reversing the procedure, one obtains $\beta_{n,m}$.\\
		
The parameters $k_n$, $m_n$ and $\beta_{m,n}$ can be compared to the theoretical expectations given in Sec. \ref{formulas}.
Typically, $k_n$ and $m_n$ match within at most $\pm 10~$\% while the $\beta_{i,j}$s match within $\pm 20~$\% (which accounts for a total error on calibrations of order 10$~$\%). In some (rare) cases, discrepancies of about $\pm 50~$\% for the nonlinear coefficients have been seen between experiment and theory. Note also that device 300$~\mu$m-n$^\circ$1 has a frequency noticeably lower than 300$~\mu$m-n$^\circ$2: its parameters are consistent with a slightly larger mass, certainly due to an imperfection of the beam (e.g. etching residue of silicon, inhomogeneity in shape).
Typical parameters taken for the theoretical evaluation are $\rho_{SiN}=3~$g/cm$^3$, $E_{SiN}=200~$GPa, and in-built stress $0.9~$GPa (high-stress sample), $120~$MPa (low-stress). Note that mathematically, both types of samples are in the so-called high-stress (or string) limit. We take for the metal $\rho_{Al}=2.7~$g/cm$^3$, $E_{Al}=50~$GPa.
 The stess term includes thermal expansion mismatch.
 Computed average parameters for the composite beams are obtained from textbook results, e.g. Timoshenko's. We summarize the experimental modal parameters in Tab. \ref{ValuesModes}. \\

\begin{table}[h!]
\begin{center}
\begin{tabular}{|c|c|c|c|c|}    \hline
                       & sping $k_n$ & freq. $\omega_n/2 \pi$ & FWHH $\Delta \omega_n/2 \pi $ at $0.84~$T  & non-lin. $\beta_{n,m}/2 \pi $   \\   \hline \hline
$n=1$ device 300$~\mu$m-n$^\circ$1           & $0.40~$N/m       & $0.59~$MHz         & $100~$Hz    &  $\beta_{1,1} =  1.35\,10^{15}~$Hz/m$^2$   \\    \hline
$n=1$ device 300$~\mu$m-n$^\circ$2           & $0.45~$N/m       & $0.66~$MHz         & $140~$Hz    &  $\beta_{1,1} =  1.70\,10^{15}~$Hz/m$^2$   \\    \hline
$n=1$ device 15$~\mu$m-n$^\circ$1            & $1.2~$N/m       & $6.9~$MHz         & $550~$Hz    &  $\beta_{1,1} =  5.0\,10^{19}~$Hz/m$^2$   \\    \hline
$n=3$ device 300$~\mu$m-n$^\circ$1           & $3.4~$N/m       & $1.74~$MHz         & $40~$Hz    &  $\beta_{3,1} =  2.75\,10^{15}~$Hz/m$^2$   \\    \hline
$n=1$ device 250$~\mu$m-n$^\circ$1           & $0.45~$N/m       & $0.9~$MHz         & $190~$Hz    &  $\beta_{1,1} =  8.5\,10^{15}~$Hz/m$^2$   \\    \hline
$n=3$ device 250$~\mu$m-n$^\circ$1           & $4.0~$N/m       & $2.7~$MHz         & $25~$Hz    &  $\beta_{3,1} = 1.6\,10^{16}~$Hz/m$^2$   \\    \hline
\end{tabular}
\caption{\label{ValuesModes} Measured mode parameters relevant to the data presented here. All agree fairly well with analytic computation (see text).}
\end{center}
\end{table}

		\begin{figure}[h!]
		\centering
	\includegraphics[width=8.5cm]{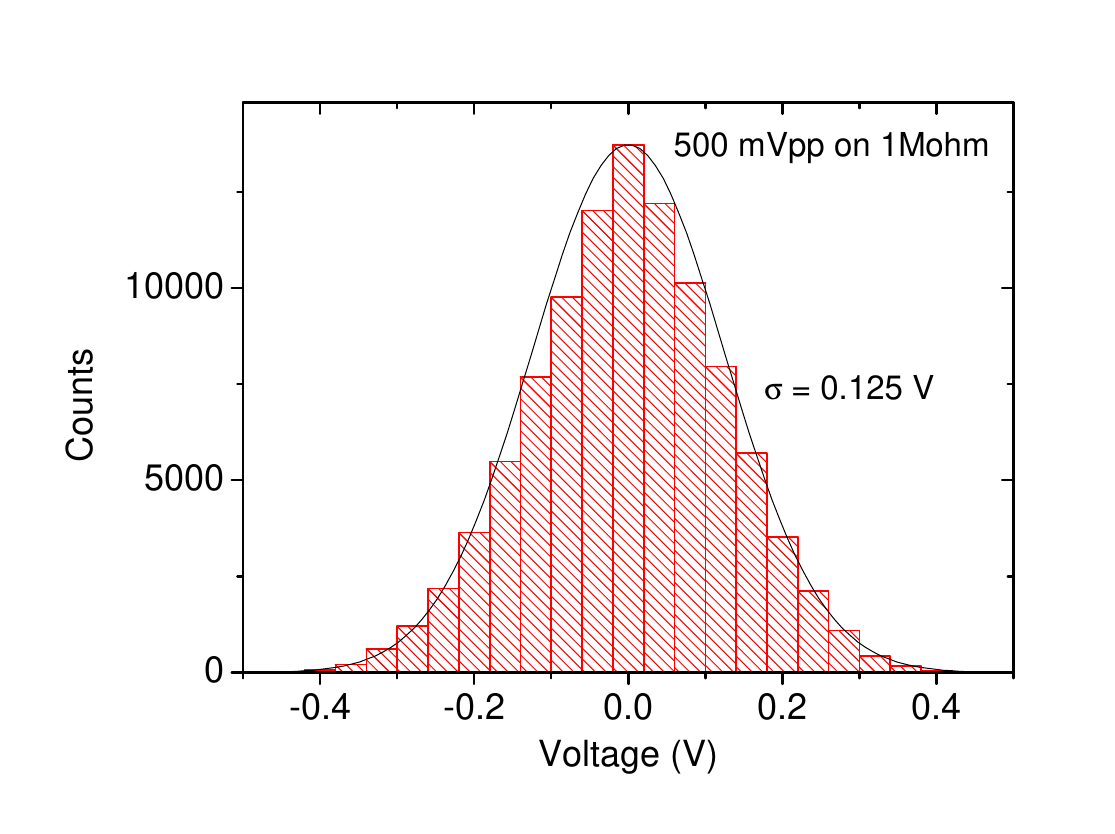}
			\caption{\small{\textbf{Gaussian noise.} 
			Characterization of Gaussian voltage noise. Histogram obtained with a large bandwidth scope; with 1$~$M$\Omega$ input. The apparatus was the HP34401A, with 0.5$~$Vpp/50$~$$\Omega$ settings (delivering thus 1$~$Vpp on 1$~$M$\Omega$). The line is a Gaussian fit with $\sigma=0.125~$V. }}
			\label{fig_gaussian}
		\end{figure}
		
\begin{figure}[h!]
		\centering
	\includegraphics[width=9.5cm]{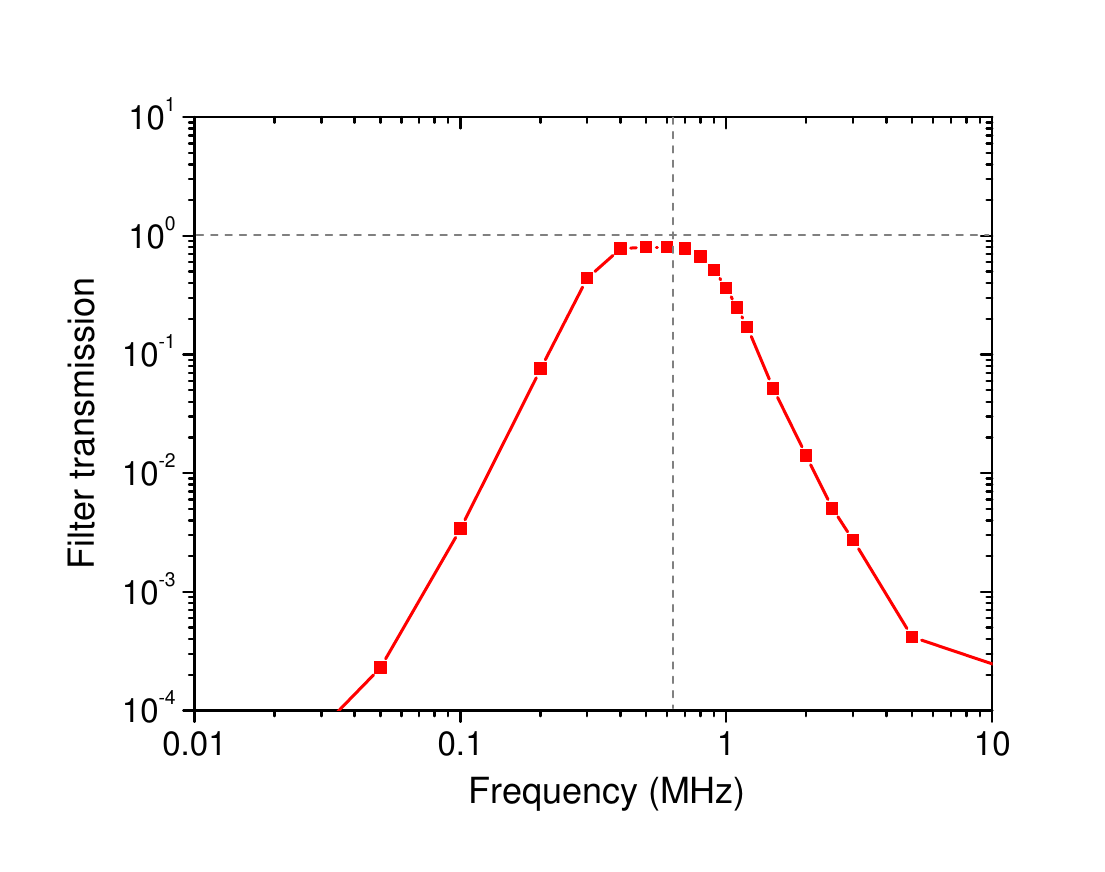}
			\caption{\small{\textbf{Home made filter.} 
			Transmission through our home-made filter. The curve is obtained from the ratio of input sine-wave drive amplitude to output sine-wave signal amplitude, in high-impedance mode. The dashed vertical represents the frequency of mode $n=1$ of our 300$~\mu$m beams. Note that the mode linewidth $\Delta \omega_1$ is much smaller than the bandwidth of the filter.}}
			\label{fig_filter}
		\end{figure}

A key in our work is the careful calibration of the noise level. We used two voltage AWG (Agilent HP34401A and Tektronix AFG3252), in noise mode, which were both calibrated. The case of the HP34401A is presented in Fig. \ref{fig_gaussian}. We first check that the noise is indeed Gaussian, and centered, with a large bandwidth scope. We find the conversion factor from applied voltage (in Vpp) to standard deviation $\sigma$. This is performed in high-impedance mode, since the setup itself is high impedance (as opposed to 50$~$$\Omega$). We then verify with a spectrum analyser (and our lock-in technique, see below) that the spectrum is flat up to the generator cutoff frequency $\omega_c$ (about $2 \pi \times$ 10$~$MHz for the Agilent and 150$~$MHz for the Tektronix). By definition, this cutoff is chosen such that the noise spectrum writes $S_V(\omega)=\pi \times \sigma^2/\omega_c$ in V$^2$/(Rad/s) (the $\pi$ comes from standard Fourier Transform (FT) definitions). However, typically from DC to MegaHertz there is a tiny slope which accounts for about 5$~$\% losses in amplitude. The home made filter is also measured experimentally using a sine-wave signal and both the scope and the lock-in (Fig. \ref{fig_filter}). It is essentially flat from 0.4$~$MHz to 0.75$~$MHz, with an insertion loss of -10$~$\% in amplitude (for our 0.6$~$MHz devices; similar properties apply to the shorter beam resonating around 7$~$MHz using another filter \cite{bifurc}). \\

We can easily convert the noise voltage drive into a noise current $\delta I(t)$, and then in turn into a noise force $\delta F_n(t) = \zeta_n L B \delta I(t)$ onto mode $n$. The force noise correlator is then $C_F^n(\tau)=\left\langle \delta F_n(t) \delta F_n(t+\tau)\right\rangle$ and the spectrum $S_F^n(\omega)=\mathrm{FT}[C_F^n](\omega)$, thus $S_F^n(\omega)=(\zeta_n L B )^2\, S_I(\omega)$. By construction, we only need to evaluate $S_I(\omega)$ within the bandwidth of the filter, where mode $n=1$ is excited. Elsewhere, the current is almost zero thus $S_F^m(\omega) \approx 0$ for $m \neq n$. \\

In order to measure the Brownian motion, we designed a straightforward technique converting our lock-in into a (phase-resolved) spectrum analyser. The aim of this technique is to be able to keep all our calibrations methods the same from sine-wave measurements to noise measurements, enabling us to be as quantitative as possible. Furthermore, this noise measurement being phase-resolved, it should enable to detect any deviation from perfectly Gaussian statistics.
The signal $\delta U(t)$ that arrives at the level of the lock-in is in fact composed of four terms: (1) the actual $\delta V(t) = -\zeta_n L B \, \delta v_n(t)$ to be studied generated by $\delta F_n(t)$, 
(2) an Ohmic component $R_{eq} \delta I(t) $ plus (3) a capacitive component $\int \delta I(t) /C_{eq} \, dt$, and finally (4) all other sources of noise $\delta U_{noise}(t)$ at the level of the detection which are not correlated to the current $\delta I(t)$. The three last components of the signal are called backgrounds, and arise from the wiring of the experiment (the resistive metallic layer, the transmission line itself from 4.2$~$K to room temperature and the amplifier noise).
$R_{eq}$ and $C_{eq}$ are the series transform of the actual setup as seen from the lock-in, and can be estimated from known circuit parameters. 
Making a RWT on the noise current, we write $\delta I(t) = \delta I^c(t) \cos (\omega t) - \delta I^s(t) \sin (\omega t) $ and similarly for other variables. $\delta v_n(t) $ is readily calculated from the mechanical susceptibility relation $\delta x_n(t) = \chi_n (t) \ast \delta F_n (t)$. The total signal $\delta U(t)$ is thus:
\begin{eqnarray}
\delta U(t) & = & \left( -(\zeta_n L B)^2 \, \left[ + \omega  Im [\chi_n (\omega)] \delta I^c(t) -\omega Re [\chi_n (\omega)] \delta I^s(t)  \right] +R_{eq} \delta I^c(t) + \frac{\delta I^s(t)}{C_{eq} \omega } + \delta U_{noise}^c \right) \cos(\omega t) \nonumber \\
&\!\!\!\!\!\!\!\!\!\!\!\!\!\!\!\!\!\!\!\!\! - &\!\!\!\!\!\!\!\!\!\!\!\!\!\! \left( -(\zeta_n L B)^2 \, \left[ + \omega  Re [\chi_n (\omega)] \delta I^c(t) + \omega Im [\chi_n (\omega)] \delta I^s(t) \right] +R_{eq} \delta I^s(t) - \frac{\delta I^c(t)}{C_{eq} \omega } + \delta U_{noise}^s \right) \sin (\omega t)  , \label{signalnoise}
\end{eqnarray}
neglecting the derivatives of slow terms $\delta \dot{I}^c(t)$, $\delta \dot{I}^s(t)$ with respect to $\omega \delta I^c(t)$, $\omega \delta I^s(t)$. By definition the susceptibility terms write $Re [\chi_n (\omega)]= \frac{1}{m_n} (\omega_n^2-\omega^2)/\left[(\omega_n^2-\omega^2)^2+ (\Delta \omega_n \, \omega )^2 \right]$ and $Im [\chi_n (\omega)]= \frac{1}{m_n} (-\Delta \omega_n \, \omega )/\left[(\omega_n^2-\omega^2)^2+ (\Delta \omega_n \, \omega )^2 \right]$ in the linear response limit, when the Duffing term can be neglected. Calling $\delta U^c(t)$ and $\delta U^s(t)$ the two components in parenthesis in Eq. (\ref{signalnoise}), it is straightforward to compute the correlators $\left\langle \delta U^c(t) \delta U^c(t+\tau)\right\rangle $, $\left\langle \delta U^s(t) \delta U^s(t+\tau)\right\rangle $ and $\left\langle \delta U^c(t) \delta U^s(t+\tau)\right\rangle $ and take their Fourier Transform. One readily obtains:
\begin{eqnarray}
S^{c,c} (\omega) & = &  +(\zeta_n L B)^4 \omega^2 \left[ Re [\chi_n (\omega)]^2 + Im [\chi_n (\omega)]^2 \right] \, S_I (\omega) \nonumber \\
&+& (\zeta_n L B)^2 \omega \left[ -2 R_{eq} Im [\chi_n (\omega)]  + 2 \frac{1}{C_{eq} \omega} Re [\chi_n (\omega)]  \right] \, S_I (\omega) \nonumber \\
& +&  \left[  \left( R_{eq} \right)^2 + \left(\frac{1}{C_{eq} \omega}\right)^2 \right] \, S_I (\omega) + S_{\delta U_{noise}} (\omega), \label{spectrummeasure}\\
S^{s,s} (\omega) & = & S^{c,c} (\omega) , \\
S^{c,s} (\omega) & = & 0 ,
\end{eqnarray}
having used the properties of the injected noise current, of spectrum $ S_I (\omega) $, white and uncorrelated between the two identical quadratures (the amplifier noise is also assumed regular, of spectrum $S_{\delta U_{noise}} [\omega]$). The sought information is in Eq. (\ref{spectrummeasure}), first term, with $\left| \chi_n (\omega) \right|^2  = Re [\chi_n (\omega)]^2 + Im [\chi_n (\omega)]^2$. But the measurement contains also a cross-term between mechanics and background (second term), and a pure background component (last term). Eq. (\ref{spectrummeasure}) can be rewritten:
\begin{eqnarray}
S^{c,c} (\omega) & = & \left\{ +(\zeta_n L B)^4 \omega^2 \left| \chi_n (\omega)\right|^2   -2 R_{eq} (\zeta_n L B)^2 \omega \, Im [\chi_n (\omega)] \right\}  \, S_I (\omega) \nonumber \\
&+&  \left\{ 2 \frac{1}{C_{eq} \omega}(\zeta_n L B)^2 \omega \, Re [\chi_n (\omega)]  \right\} \, S_I (\omega) \nonumber \\
& +&  \left[  \left( R_{eq} \right)^2 + \left(\frac{1}{C_{eq} \omega}\right)^2 \right] \, S_I (\omega) + S_{\delta U_{noise}} (\omega) .
\end{eqnarray}
The first term is a peak, maximum at resonance, that we call the main term. The second is an imprint of the quadrature signal which is zero on resonance (we call it cross term).
Finally the last term is a true background, non-mechanical, independent of field. Note that because of the loading effect which generates a peak width $\propto B^2$, the two first components tend towards a constant for large fields, with the main contribution having a stronger $B^4$ growth.

		\begin{figure}[h!]
		\centering
	\includegraphics[width=7.5cm]{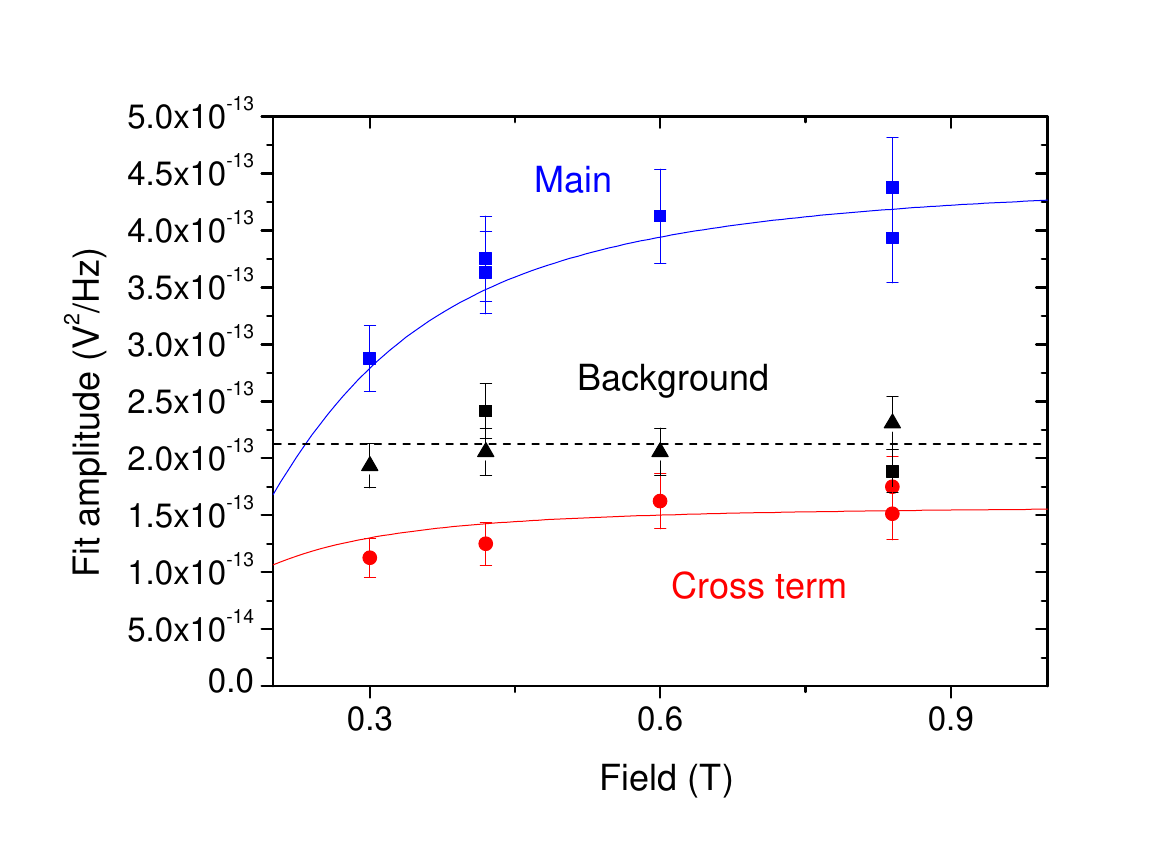}
			\caption{\small{\textbf{Fit of spectral components vs field.} 
			Data taken on sample 300$~\mu$m-n$^\circ$2, for a noise drive current of $2.8\,10^{-19}~$A$^2$/Hz. Fit parameters are discussed below, and can be related to circuit parameters. }}
			\label{fig_amplfit}
		\end{figure}

These components are fit on the data in Fig. \ref{fig_amplfit}. The fit is consistent with a model circuit of a 500$~\Omega$ NEMS series resistance in parallel with 500$~$pF of coaxial line capacitance. These values are within a factor of $2$ from actual circuit parameters, as is the overall measured background magnitude. This discrepancy is expected in the present wiring of the setup, because of the direct cross-talk (distributed along the lines) from injection signal to detection signal that affects the background detection level. 
At large fields, the main contribution is thus essentially $\propto \left| \chi_n (\omega)\right|^2$, which defines the position spectra $S(\omega)$ presented in this work.\\

		\begin{figure}[h!]
		\centering
	\includegraphics[width=7.5cm]{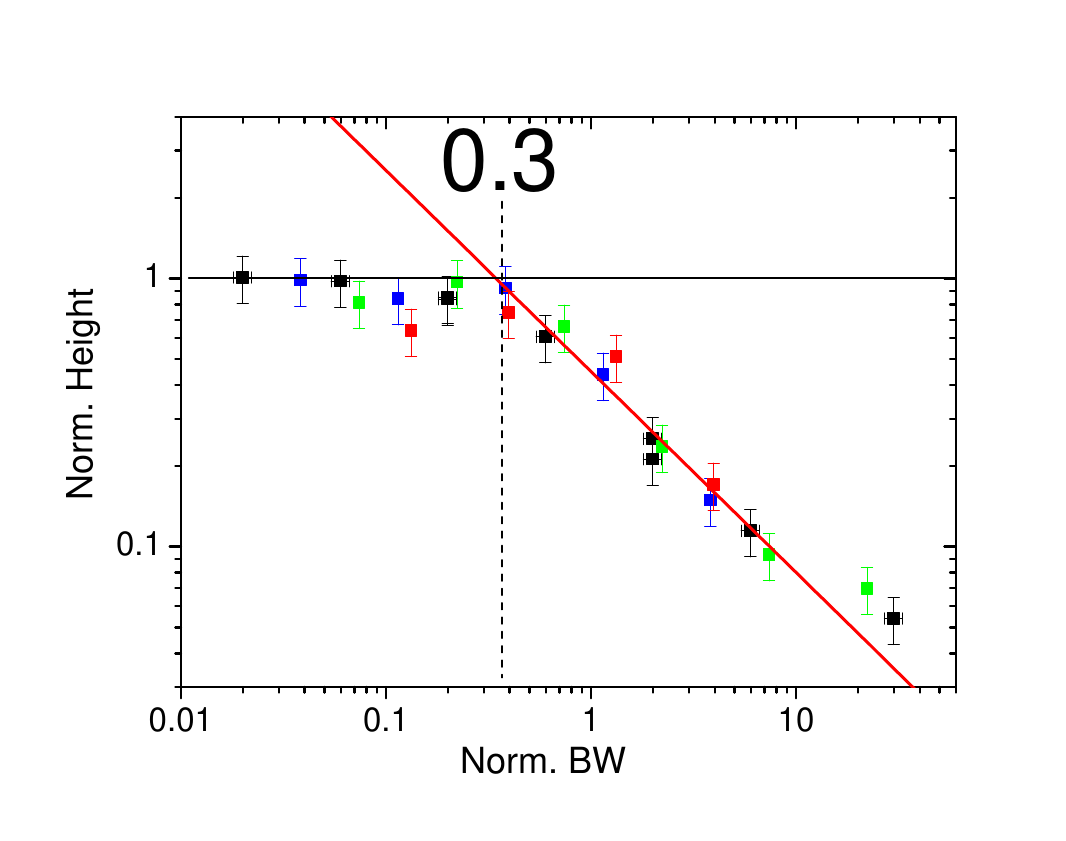}
			\caption{\small{\textbf{Measured peak height versus lock-in bandwidth.} 
			Height of the motion spectra normalized to its maximum value, with respect to the bandwidth of the lock-in detection normalized to the mechanical linewidth (Sample 300$~\mu$m-n$^\circ$2 same current noise as Fig. \ref{fig_amplfit}). The different colors correspond to different field loadings: black 0.84$~$T, blue 0.6$~$T, green 0.42$~$T and red 0.3$~$T. Over this range the resonance FWHH changes from 140$~$Hz to 20$~$Hz. The lines are guides, with a crossover around 0.3.}}
			\label{fig_amplBW}
		\end{figure}

In practice, the lock-in measurement is simply equivalent to a phase-resolved spectrum analyser. As such, the bandwidth $BW$ with which the data is acquired is a key parameter. We proceed by digitizing for each given settings (drive, frequency) both X and Y signals demodulated at $\omega$ on a Ni-DAQ card which is much faster than the lock-in filter itself (typ. 100$~$kHz). We take rather long traces (typ. few thousand points $N$) from which we can calculate the auto-correlators $C_X(j)=<X_i \, X_{i+j}>$, $C_Y(j)=<Y_i \, Y_{i+j}>$, and the cross-correlator $C_{XY}(j)=<X_i \, Y_{i+j}>$ (with $< \cdots > = \sum_i \cdots /N$ using cyclic index notations when $i+j > N$). The Stanford SR 844 filter is used with the $24~$dB/oct. option, and from the chosen time-constant we compute the effective bandwidth $BW$ of the detector. By definition $S_X(\omega) =\pi\, C_X(0)/BW$, $S_Y(\omega) = \pi\, C_Y(0)/BW$ and $S_{XY}(\omega) = \pi\, C_{XY}(0)/BW$. When the statistics is {\it regular}, we find $S_X(\omega)=S_Y(\omega)$ and $S_{XY}(\omega)=0$ as we should (see Section \ref{spectraSC}).

		\begin{figure}[h!]
		\centering
	\includegraphics[width=7.5cm]{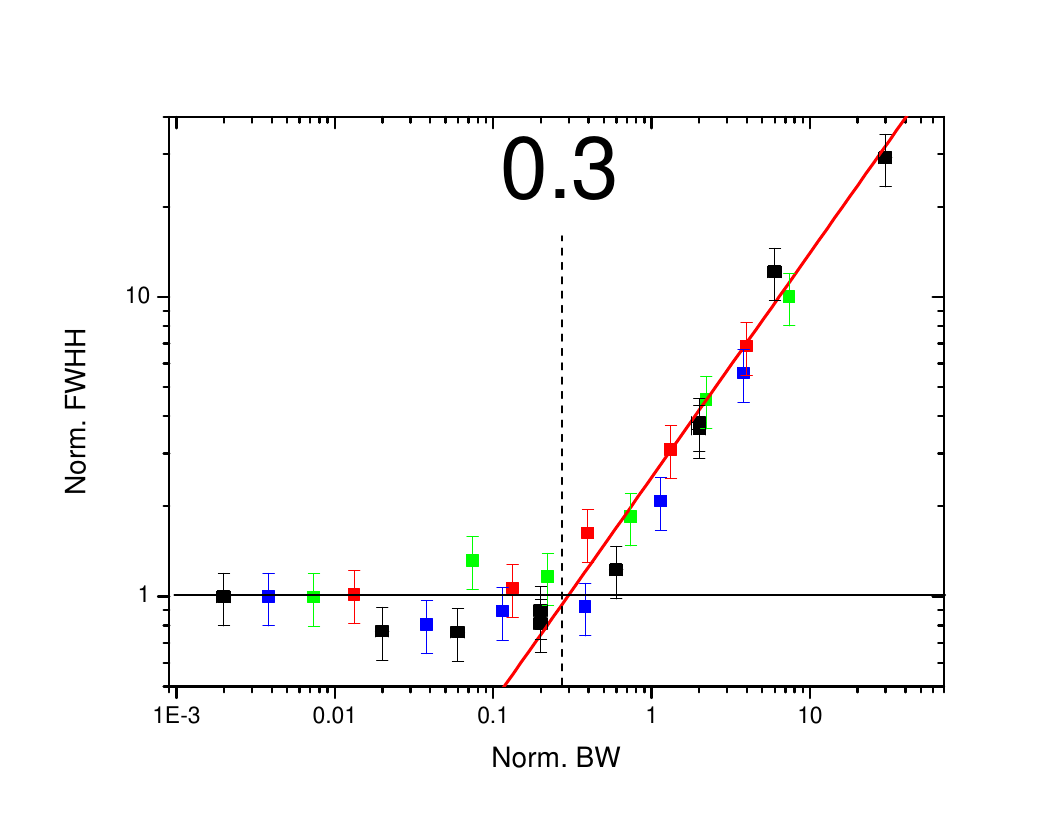}
			\caption{\small{\textbf{Measured peak width versus lock-in bandwidth.} 
			FWHH of the motion spectra normalized to its minimum value, with respect to the bandwidth of the lock-in detection normalized to the mechanical linewidth (same conditions as Fig. \ref{fig_amplBW}). Conventions are the same as in the previous graph. }}
			\label{fig_WidthBW}
		\end{figure}

The effect of the bandwidth $BW$ is twofold: on one hand (1) the larger it is, the more noise is collected by the lock-in and thus the larger the detected signal $C_X(0)$ is. However, (2) when the bandwidth is large compared to the width of the mechanical resonance, then the true mechanical signal is ``diluted'' within the background. As a result, the measured peak height decreases with respect to the background, and broadens; this is shown on both Figs. \ref{fig_amplBW} and \ref{fig_WidthBW}. At the same time, the background simply scales as the bandwith, as it should for a white noise. 
On these graphs, we represent the heights and FWHH of the resonance peak normalized to their actual value with respect to the $BW$ of the lock-in detection normalized to the mechanical resonance linewidth $\Delta \omega$. We see that the best trade-off is around $BW \approx 0.3~\Delta \omega_n$, where signal is rather large without a strong impact on the measured peak. \\

		\begin{figure}[h!]
		\centering
	\includegraphics[width=8.5cm]{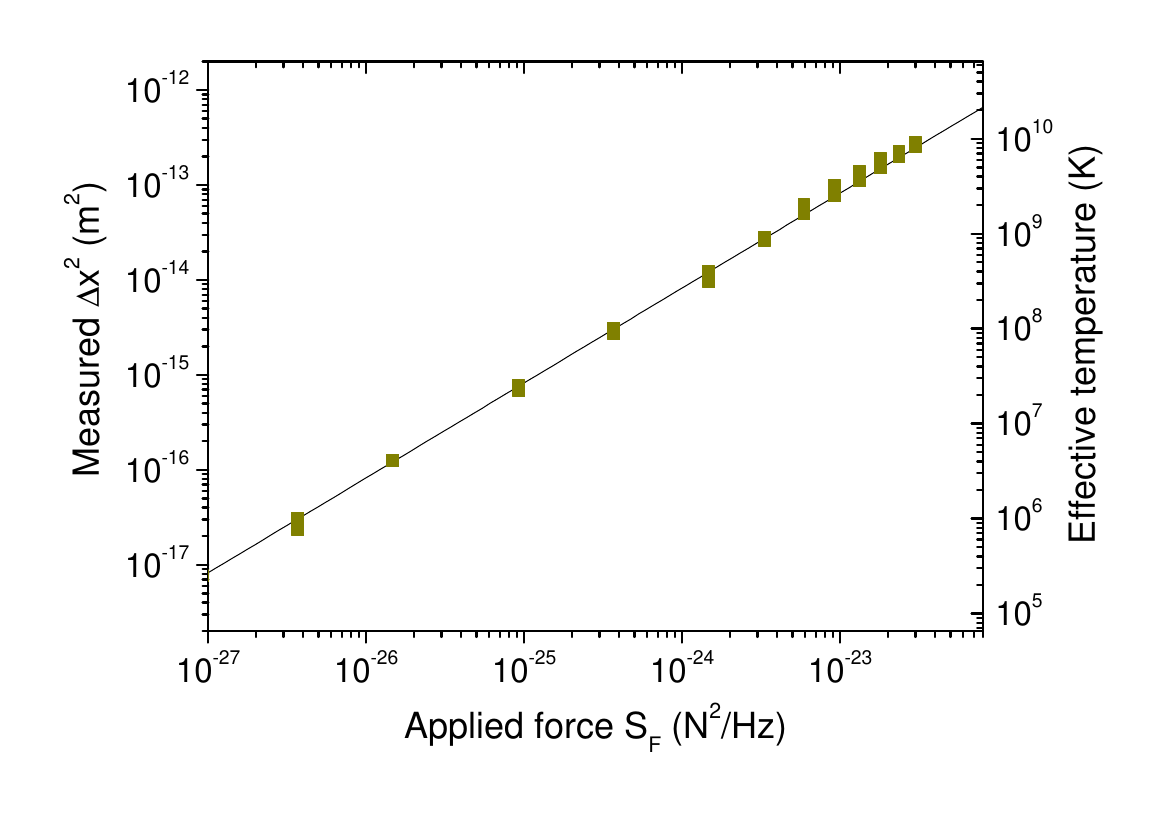}
			\caption{\small{\textbf{Measured Brownian motion versus applied stochastic force.} 
			Measured standard deviation $\Delta x_{1}^2$ for sample 300$~\mu$m-n$^\circ$2 at 0.84$~$T with respect to applied force noise (varying the noise current intensity). The line is the theoretical calculation (see text). On the right vertical axis, the motion is converted into effective temperature for mode $n=1$.}}
			\label{fig_sigma}
		\end{figure}
		
When the noise is not too large on mode $n$, the Duffing non-linearity is negligible and the mechanical spectrum is a Lorentzian peak. Its area is proportional to Height$~\times~$FWHH, and is directly related to the standard deviation $\Delta x_n^2$ of the mode position fluctuations. Furthermore,  $\Delta x_n^2$ is directly linked to the applied force noise through $\Delta x_n^2= S_F^n \, Q_n \omega_n/(2 k_n^2)$. This relation is shown in Fig. \ref{fig_sigma}, together with computed effective temperatures. \\

When the force noise becomes large, then the Duffing term is not negligible anymore: the spectrum becomes a ``Duffing spectrum'' (Fig. 2 of paper and Fig. \ref{fig_noise} below). However, for high-Q devices one can show solving the Fokker-Planck equation that the statistics remains Gaussian \cite{nonGaussFP}. Indeed, no anomalies are seen on the measured spectra (we still have $S_X(\omega)=S_Y(\omega)$ and $S_{XY}(\omega)=0$, see Section \ref{spectraSC}). Furthermore, since the measured peak is nothing but the convolution of the mechanical susceptibility with a ``distribution'' generated by the fluctuations, see Section \ref{adiabatic}, the area of this peak is preserved: thus the above proportionality relation between measured $\Delta x_n^2$ and force spectrum $S_F^n$ still applies for the distorted spectra. This can be seen experimentally at very large effective temperatures in Fig. \ref{fig_sigma}. Physically, this simply comes from the fact that the induced frequency noise by the non-linear term is purely dispersive, and no extra energy is transferred from the mode to the environment.
Note that even though the line distorts, the Height$~\times~$FWHH remains a rather good estimate of the area of the peak. For the sake of completeness, we should mention that for noise levels above about $\Delta x_1^2 \approx 4. \, 10^{-14}~$m$^2$, we do see a Joule heating of the structure due to the large currents injected in the aluminum layer. It is easy to correct for this effect, which produces here only a tiny {\it negative} frequency shift which is calibrated in the first place \cite{RSICollin}.  \\

Finally, let us comment the range of effective temperatures explored in this experiment. Even though the smallest stochastic drives already correspond to about $100\,000~$Kelvin,
they lie within the ``motional narrowing'' limit (see analytical expansions in Section \ref{adiabatic}, Figs. 3 \& \ref{fig_supplchar} and final discussion on actual thermal bath Section \ref{formulas}). On the other hand, the highest levels of noise used are {\it extremely high} and lie in the ``inhomogeneous broadening'' range: nonetheless, the basic properties of position fluctuations are not altered (Gaussianity, preserved stored energy in the resonance i.e. the expression of the equipartition theorem here). One needs quite peculiar conditions to destroy these properties, and e.g. create squeezed statistics. This is what can be achieved by adding a {\it large} sine-wave excitation force {\it on the same mode} where large fluctuations are present (see Section \ref{spectraSC}). 

\section{Non-linear mode coupling}
\label{sectionBernie}

The basic starting point is the Euler-Bernoulli equation in which the non-linear effect of tensioning has been incorporated \cite{CrossLifshitz}. It has been discussed extensively in the literature already \cite{RoukesNlin,KunalNlin}, and we just remind the maths here. The equation writes:
\begin{equation}
\!\!\!\!\! \rho_{beam} A \frac{\partial^2 u(z,t)}{\partial t^2} +\eta \frac{\partial {\cal L}  \left[u(z,t)\right]}{\partial t} + E_{beam}\, I_z \frac{\partial^4 u(z,t)}{\partial z^4} 
-\left[ T_0 + \frac{E_{beam}\, A}{2 L} \int^{L}_{0} \left( \frac{\partial u(z,t)}{\partial z}\right)^{\!\!2}  dz \right] \frac{\partial^2 u(z,t)}{\partial z^2} = \frac{\partial F(z,t)}{\partial z} \, \label{bernouille} ,
\end{equation}
with $\vec{z}$ the axis along the beam, $L$ the length, $A= w e$ its cross section ($w$ is the width and $e$ the thickness), $I_z$ its second moment of area. $T_0$ is the stored tension, $\rho_{beam}$ the mass density and $E_{beam}$ the Young's modulus. The parameters can be adapted to the case of a bilayer system (here Al on SiN, see e.g. Timoshenko's textbook). \\

In Eq. (\ref{bernouille}), $\partial F(z,t)/\partial z$ is the overall external force per unit length applied to the beam, and $u(z,t)$ its motion in the $\vec{x}$ direction. 
We do not mention any explicit model of mechanical dissipation, and just assume that the friction force is proportional to the time-derivative of some linear functional ${\cal L} $ of the local displacement $u(z,t)$.
We will limit the discussion to two modes, the extension to more being straitghforward, following the works of Refs. \cite{KunalNlin,RoukesNlin,VenstraNlin,CrossLifshitz}. 
Let $u(z,t) = x_n(t) \Psi_n(z)+ x_m(t) \Psi_m(z)$ with $\Psi_n(z), \Psi_m(z)$ the mode shapes of mode $n$ and $m$. In this modal decomposition, we take the $\Psi_n, \Psi_m$ functions to be normalized at 1 at their maximum. This defines our (time-dependent) amplitude parameters $x_n(t), x_m(t)$. \\

Replacing $u(z,t)$ into Eq. (\ref{bernouille}) and projecting on one of the modes (i.e. multiplying by $\Psi_n(z)$ or $\Psi_m(z)$  and integrating over the beam), we obtain:
\begin{eqnarray}
 & & \!\!\!\!\!\!\!\!\!\!\!\!\!\!\!\!\!\!\!\!\!\!\!\!\!\!\!\!\!\!\!\!\!\!\!\!\!\! \rho_{beam} A L J_{n,n} \, \ddot{x}_n(t) + \rho_{beam} A L J_{n,m} \, \ddot{x}_m(t) + \eta   K_{n,n} \, \dot{x}_n(t) + \eta   K_{n,m} \, \dot{x}_m(t) + \rho_{beam} A L \left[ \left(\omega_n^0 \right)^2 J_{n,n} \, \ddot{x}_n(t) + \left(\omega_m^0 \right)^2 J_{n,m} \, \ddot{x}_m(t) \right] \nonumber \\
 &\!\!\!\!\!\!\!\!\!\!\!\!\!\!\!\!\!\!\!\!\!\!\!\!\!\!\!\!\!\!\!\!\!\!\!\!\!\!\!\!\!\!\!\!\!\!\!\!\!\!\!\!\!\!\! + & \!\!\!\!\!\!\!\!\!\!\!\!\!\!\!\!\!\!\!\!\!\!\!\!\!\!\!\!\!\!\! \frac{E_{beam}\, A}{2 L^3} \!\! \left[ x_n(t)^3 I_{n,n}^2 + x_m(t)^3 I_{m,m} I_{n,m} + x_n(t)^2 x_m(t) \left( 3 I_{n,n} I_{n,m} \right)+ x_m(t)^2 x_n(t) \left( I_{n,n} I_{m,m} + 2 I_{n,m}^2 \right)  \right] \nonumber \\
& = & F_0 L\,\zeta_n \, \cos(\omega t)   \label{eqproj},
\end{eqnarray}
having assumed $\partial F(z,t)/\partial z=F_0 \cos(\omega t)$ independent of $z$, and used the modal relation $E_{beam} I_z \Psi_i^{''''} - T_0 \Psi_i^{''} = \rho_{beam} A \, \left(\omega_i^0\right)^2 \Psi_i$. $\omega_i^0$ is the natural resonance frequency of mode $i$. The parameters introduced above are $\zeta_n= \int^{L}_{0} \Psi_n(z) dz /L$, $J_{n,m} = \int^{L}_{0} \Psi_n(z) \Psi_m(z) dz /L $, $K_{n,m} = \int^{L}_{0} \Psi_n(z) {\cal L} \left[\Psi_m(z)\right] dz $ and $I_{n,m} = L \int^{L}_{0} \Psi_n(z)^{'}\Psi_m(z)^{'}  dz$. 
Eq. (\ref{eqproj}) applies to the projection on mode $n$; the projection on $m$ is obtained by inverting $n \leftrightarrow m$.
In the ideal case, the modes are orthogonal and $J_{n,m} = K_{n,m} =I_{n,m}=0$ when $n \neq m$. In the real case, orthogonality is not perfect; however, writing $x_i(t)=x_i^0 \cos(\omega t + \phi)$ and using a RWT one immediately sees that many terms in Eq. (\ref{eqproj}) are off-resonant with respect to mode $n$. It is thus perfectly enough to recast the above result into:
\begin{equation}
m_n \, \ddot{x}_n(t) + 2 \lambda_n \, \dot{x}_n(t) + m_n \left(\omega_n^0\right)^2 \, x_n(t) + k_{n,n} \, x_n(t)^3 + k_{n,m} \, x_n(t) x_m(t)^2 = F_n(t) \label{eqSaderArmour},
\end{equation}
for mode $n$. An equivalent expression holds for mode $m$. We have defined the mode parameters $m_n = \rho_{beam} A L J_{n,n} $, $\lambda_n = \eta   K_{n,n} /2$ together with $k_n = m_n \left(\omega_n^0\right)^2$. The projection of the force on the mode is $F_n(t)= F_n^0 \cos(\omega t)$ with $F_n^0 = F_0 L \zeta_n$. Finally, the non-linear coefficients write $k_{n,n} = \frac{E_{beam}\, A}{2 L^3} I_{n,n}^2$ and $k_{n,m} = \frac{E_{beam}\, A}{2 L^3} \left( I_{n,n} I_{m,m} +2 I_{n,m}^2 \right)$ for $n \neq m$.
Eq. (\ref{eqSaderArmour}) is noting but the driven Duffing equation (for mode $n$), with an extra coupling term to mode $m$. This is the formalism developed in Refs. \cite{KunalNlin,RoukesNlin,VenstraNlin,CrossLifshitz}. \\

We turn now to the situation where $F_n(t) = F_n^0 \cos(\omega t) + \delta F_n(t)$ and $F_m(t) = F_m^0 \cos(\omega t) + \delta F_m(t)$, introducing the stochastic variables $\delta F_n(t), \delta F_m(t)$. These are taken as being white, Gaussian, and uncorrelated. The obvious ansatz solving Eq. (\ref{eqSaderArmour}) consists in writing $x_i(t) = x_i^0(t) + \delta x_i(t)$ for $i=n,m$, with $x_i^0(t)$ a certain (and eventually sinusoidal) component and $\delta x_i(t)$ a random motion. Replacing in the above equation, we obtain:
\begin{eqnarray}
 & & \ddot{x}_n^0(t) + \Delta \omega_n  \, \dot{x}_n^0(t) + \left( \omega_n^0 \right)^2 \, x_n^0(t) + 2 \omega_n^0 \frac{4}{3} \beta_{n,n} \, x_n^0(t)^3 \nonumber \\
  & + & 2 \omega_n^0 \left[ + 2 \beta_{n,m} \, x_m^0 (t)^2 x_n^0(t) + 4 \beta_{n,n} \, \delta x_n(t)^2 x_n^0(t) +   2 \beta_{n,m} \, \delta x_m(t)^2 x_n^0(t)  \right]       \nonumber \\
 & + & \delta \ddot{x}_n(t) + \Delta \omega_n  \, \delta \dot{x}_n(t) + \left( \omega_n^0 \right)^2 \, \delta x_n(t) + 2 \omega_n^0 \frac{4}{3} \beta_{n,n} \, \delta x_n(t)^3 \nonumber \\
 & + & 2 \omega_n^0  \left[+2 \beta_{n,m} x_m^0 (t)^2 \delta x_n(t) + 4 \beta_{n,n} x_n^0(t)^2 \delta x_n(t) +  2 \beta_{n,m}  \delta x_m(t)^2 \delta x_n(t)  \right]   \nonumber \\
 &= & \frac{F_n^0}{m_n} \cos(\omega t) + \frac{\delta F_n(t)}{m_n} \label{finaleq},
\end{eqnarray}
with $\Delta \omega_n  = 2 \lambda_n/m_n$ the linewidth parameter associated to mode $n$. We introduced the Duffing parameters $\beta_{n,n} = \frac{3}{4} \frac{k_{n,n}/m_n}{2 \omega_n^0}$ and $\beta_{n,m} = \frac{1}{2} \frac{k_{n,m}/m_n}{2 \omega_n^0}$. Note that $k_{n,m}=k_{m,n}$ but $\beta_{n,m} \neq \beta_{m,n}$.
The first line and the third line of Eq. (\ref{finaleq}) correspond to the Duffing equation written for the certain component and the stochastic one respectively. The first bracket is the non-linear coupling onto the certain dynamics: the first element $\propto x_m^0 (t)^2 x_n^0(t)$ is nothing but the standard mode-coupling term between mode $m$ and $n$, while the two others correspond to the coupling of the fluctuations to the certain position variable. Note the factor of 2 between the coupling to mode $m$ fluctuations, and to the same mode $n$. 
The second bracket is the equivalent mode-coupling term onto the random motion of mode $n$: the two first elements depend on the certain components of $m$ and $n$, while the last one is due to the fluctuations of mode $m$. Note again the factor of 2 between ``inter-mode coupling'' and ``self-coupling'' (or ``intra-mode coupling'').
In the next section, we explain how to solve Eq. (\ref{finaleq}) for small sine-wave drives and give the explicit theoretical results relevant to the present work, taken from Ref. \cite{dykmanfluctu}.

\section{Adiabatic description of the noise coupling: from ``motional narrowing'' to ``inhomogeneous broadening''}
\label{adiabatic}

The theoretical analysis we use is the one developed in Ref. \cite{dykmanfluctu}.
Here, we quickly describe the maths leading to the analytic solution of the problem studied.
The first step is to write the dynamics equation Eq. (\ref{finaleq}) in the frame rotating at the speed of the mode's oscillation.
The transformed motion variable $\tilde{x}_n^0(t)$ writes $x_n^0 (t) = (\tilde{x}_n^0(t) \mathrm{Exp}[i \omega_n^0  t]+ \tilde{x}_n^{0\, \ast}(t) \mathrm{Exp}[-i \omega_n^0  t])/2$.
 Making a Rotating Wave Approximation (RWA), the component of the frequency fluctuations $\propto \delta x_n(t)^2$ at $2\omega_n$ is removed, keeping only its slow dynamics (and static average). Note that this high-frequency term corresponds somehow to a ``random parametric pumping'' for the certain dynamics $x_n^0$ of mode $n$; but one can show nonetheless that its impact is negligible, and can be safely neglected. \\

Thus, the dynamics of the certain component {\it adiabatically follows} the slow fluctuations, which can be interpreted as phase diffusion for the resonator. As such, the final result of the averaging of this phase diffusion can be written in time-domain as a multiplication of the mode's decay function by an expression which Fourier Transform we call a ``frequency distribution'' (in frequency-domain).
This is not strictly speaking a frequency distribution, since in the ``motional narrowing'' limit it is complex-valued. \\

Let us briefly remind the reader of the approach followed by Zhang and Dykman in Ref. \cite{dykmanfluctu}.
To simplify the discussion, we neglect for the time being the intra-mode non-linearity. Nonetheless, the following approach can be extended without loss of generality to the intra-mode case, provided that the driven motion is not too high. Under this assumption, the fluctuations of the driven mode and its certain component can be linearly separated. The linearized NEMS' dynamics equations for modes $n$ (driven, i.e. the probe mode) and $m$ (noisy, undriven) in the RWA, are then:
\begin{equation}
\label{dynamics_m}
\dot{\delta\tilde{x}}_m(t)+\frac{\Delta\omega_m}{2}\delta\tilde{x}_m(t)=\frac{\delta F_m(t)e^{-i\omega_m t}}{2m_m\omega_m} ,
\end{equation}
\begin{equation}
\label{dynamics_n}
\dot{\tilde{x}}_n^0(t)+\left(\frac{\Delta\omega_n}{2}+i[\omega-\omega_n-\delta\omega(t)]\right)\tilde{x}_n^0(t)=\frac{F_n^0e^{i\phi}}{4m_n\omega_n} ,
\end{equation}
with $\delta\omega_n(t)=\beta_{n,m}|\delta\tilde{x}_m^0(t)|^2=\beta_{n,m}\left[\delta X_m(t)^2+\delta Y_m(t)^2\right]$, where $\delta X_m,\delta Y_m$ are the quadratures of the fluctuating mode's motion. Integrating Eq.(\ref{dynamics_n}), we obtain the response of the slow variable in the time domain, which is stochastic:
\begin{equation}
\label{solution_n_time}
\tilde{x}_n^0(t)=\int_{-\infty}^t\frac{F_n^0e^{i\phi}}{4m_n\omega_n}\exp\left(-\left[\frac{\Delta\omega_n}{2}+i(\omega-\omega_n)\right](t-t')+i\int_{t'}^t\delta\omega(t'')\mathrm{d}t''\right)\mathrm{d}t' .
\end{equation}
The exponential term is nothing but a stochastic susceptibility in the time domain $\chi_{sl}(t,t')$:
\begin{equation}
\label{solution_n_time_susc}
\tilde{x}_n^0(t)=\int_{-\infty}^t\frac{F_n^0e^{i\phi}}{4m_n\omega_n}\chi_{sl}^*(t,t')\mathrm{d}t',
\end{equation}
such that the susceptibility in the frequency domain, i.e. the one that is measured, is:
\begin{equation}
\label{susc_freq}
\chi(\omega)=\frac{i}{2m_n\omega_n}\int_0^{+\infty}\langle\chi_{sl}(t,0)\rangle\mathrm{d}t  .
\end{equation}
The time-domain susceptibility is made stochastic only by the accumulated random phase created by the nonlinearly coupled Brownian motion of the noisy mode. We are then left with averaging over the slow susceptibility:
\begin{equation}
\label{avg_susc}
\langle\chi_{sl}(t,0)\rangle= 
\exp\left(-\left[\frac{\Delta\omega_n}{2}+i(\omega-\omega_n)\right](t-t')\right)\left\langle e^{\displaystyle i\int_0^t\delta\omega(t'')\mathrm{d}t''}\right\rangle .
\end{equation}
This averaging procedure is not trivial. The integral is made over a variable which is random at \textit{any time between $0$ and $t$}, with a \textit{finite correlation time}, meaning that this frequency noise is highly structured. Thus, to take into account the correlations, one has to average over \textit{every path accessible} for the accumulated phase in the quadrature space of the fluctuating mode between $0$ and $t$.
Assuming that the thermal bath is Markovian, the corresponding probability density functional is obtained through Eq. (\ref{dynamics_m}). 
The different phase paths interfere, with an efficiency set by the noisy mode dynamics that appears in the probability density. 
This so-called path integral approach is detailed in Ref. \cite{dykmanfluctu}. Using a discretization procedure for the time interval $[0,t]$, the averaging is then reduced to a cumbersome yet fully analytical calculation of Gaussian integrals, which is performed in Ref. \cite{dykmanfluctu}. A similar computation, also detailed in Ref. \cite{dykmanfluctu} leads to the averaging of the position spectrum. \\

 In frequency-space these calculations thus lead to convolutions, which are explicitly given from Ref.  \cite{dykmanfluctu} as:
\begin{eqnarray}
\!\!\!\!\!\! S_X^n (\omega ) & = & S_{Lorentz}^n (\omega) \ast FT \left[\frac{\exp(+2 \Gamma_n t)}{\left[\cosh( a _n t ) + \frac{\Gamma_n}{a_n} \left( 1+2 i \alpha_n \right) \sinh( a_n t ) \right]^2 }\right]\left(\omega \right) \,\,\,\,\, \mbox{Noise on $n$, measure noise on $n$} \nonumber \\
\!\!\!\!\!\! \tilde{x}_m^0 & = & X_{Lorentz}^m (\omega) \ast FT \left[\frac{\exp(+ \Gamma_m t)}{\cosh( a_m t ) + \frac{\Gamma_m}{a_m} \left( 1+2 i \alpha_m  \right) \sinh( a_m t )  }\right]\left(\omega \right) \,\,\,\,\, \mbox{Noise on $m$, measure sine on $m$} \nonumber \\
\!\!\!\!\!\! \tilde{x}_m^0 & = & X_{Lorentz}^m (\omega) \ast FT \left[\frac{\exp(+ \Gamma_n t)}{\cosh( a_{m,n} t ) + \frac{\Gamma_n}{a_{m,n}} \left( 1+2 i \alpha_{m,n}  \right) \sinh( a_{m,n} t )  }\right]\left(\omega \right) \,\,\,\,\, \mbox{Noise on $n$, measure sine on $m$} ,\nonumber
\end{eqnarray}
$FT$ meaning Fourier Transform, with:
\begin{eqnarray}
\Gamma_n & = & \Delta \omega_n / 2 \nonumber \\
\Gamma_m & = & \Delta \omega_m / 2 \nonumber \\
 a_n & = & \Gamma_n \sqrt{1+4 i \alpha_n} \nonumber \\
 a_m & = & \Gamma_m \sqrt{1+4 i \alpha_m} \nonumber \\
a_{m,n} & = & \Gamma_n \sqrt{1+4 i \alpha_{m,n}} \nonumber \\
\alpha_n & = & \frac{\beta_{n,n} \Delta x_n^2 }{\Gamma_n} \nonumber \\
\alpha_m & = & \frac{2 \beta_{m,m} \Delta x_m^2 }{\Gamma_m} \nonumber \\
\alpha_{m,n} & = & \frac{\beta_{m,n} \Delta x_n^2 }{\Gamma_n} ,\nonumber
\end{eqnarray}
with $S_X^n$ the Brownian motion spectrum of mode $n$, and the measured quadratures $X=Im[\tilde{x}_m^0]$ and $Y=-Re[\tilde{x}_m^0]$ of mode $m$ using our definitions (matching the phase of the magnetomotive scheme detection). We write $S_{Lorentz}^n (\omega)$ and $X_{Lorentz}^m (\omega)$ the standard Brownian motion spectrum and (complex-valued) response function when no noise is applied. In the RWA (high-$Q$ limit) these are simple Lorentzian peaks.
In these formula, the impact of the stochastic motion is given by the ``motional narrowing parameter'' $\alpha_\lambda$ (with $\lambda=n$, $m$ or $n,m$). 
This index describes the different situations encountered, respectively: the effect of the Brownian motion on the spectrum itself (``Duffing spectrum''), the coupling of the noise on mode $n$ to the sine-wave response on the same mode $n$ (``self-coupling'') and the coupling of the noise on one mode $n$ to the sine-wave response of another mode $m$ (``mode-coupling'').

		\begin{figure}[h!]
	\includegraphics[width=8cm]{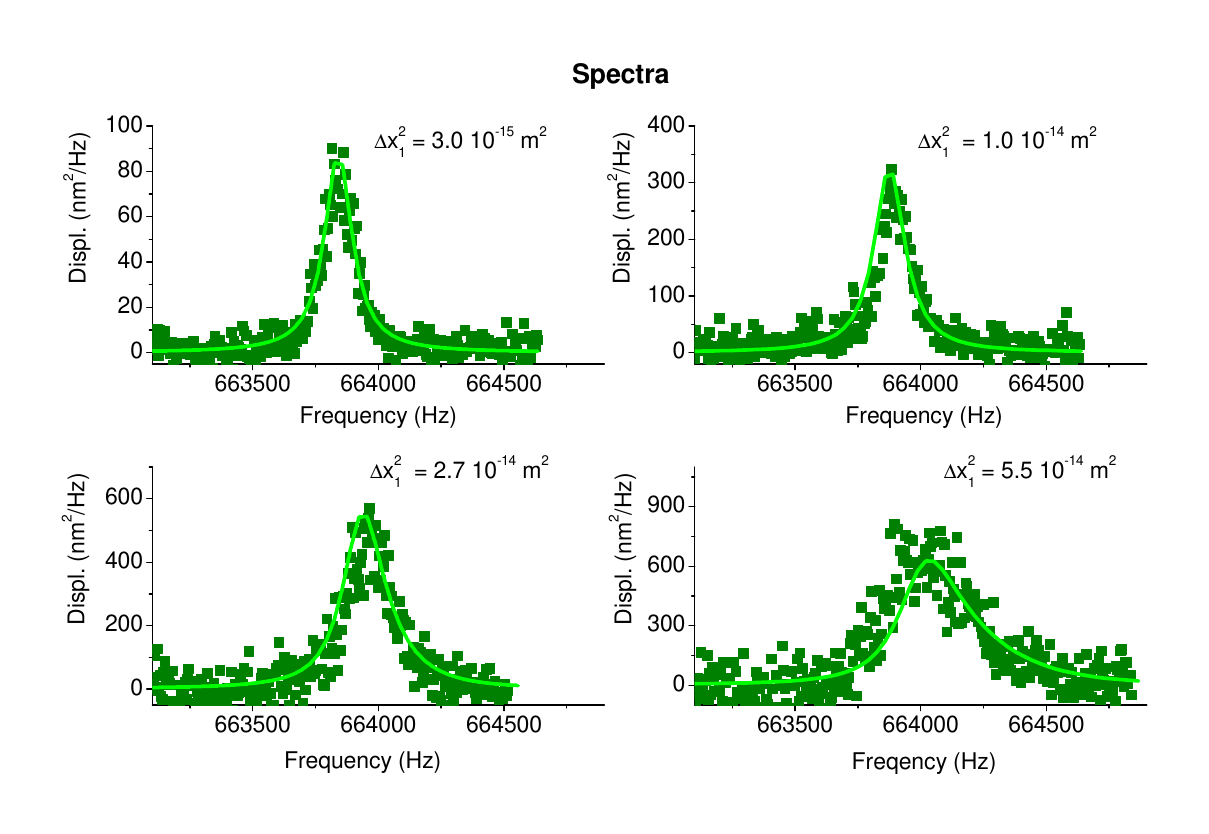} \includegraphics[width=8cm]{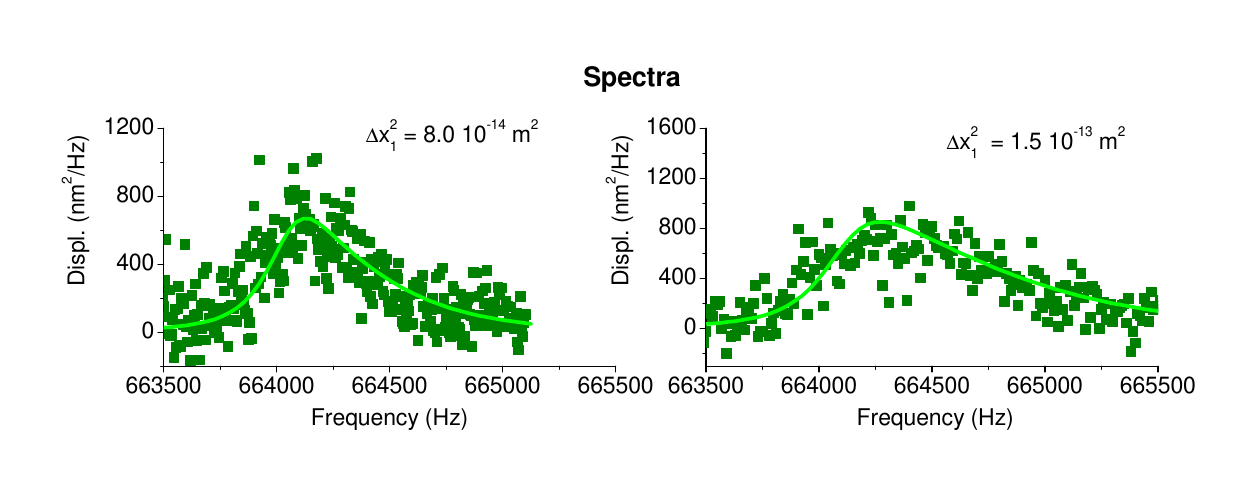}
			\caption{\small{\textbf{Brownian spectra in non-linear regime} 
			(Color online). Measured displacement spectra $S^n_X(\omega)$ for mode $n=1$ (first flexure, no sine-wave drive) on sample 300$~\mu$m-n$^\circ$2. The standard deviation $\Delta x_1^2$ (i.e Brownian motion level) is increased from top-left to bottom-right, and the lineshape distorts from a pure Lorentzian peak to a {\it ``Duffing spectrum''}. Lines are theoretical calculations (see text).}}
			\label{fig_noise}
		\end{figure}

These expressions are used in Fig. 2 of the main article to reproduce the data. We present in Fig. \ref{fig_noise} a broader set of measured spectral peaks, and in Fig. \ref{fig_reslines} a broader set of ``intra-mode'' coupling data (putting noise and measuring with a sine wave the same $n=1$ flexure). Below in Fig. \ref{fig_inter1} more resonance lines are presented in the case of ``inter-mode'' coupling, having the noise on mode $n=1$ and measuring with the sine-wave response mode $m=3$.

		\begin{figure}[h!]		 
			 \includegraphics[width=8cm]{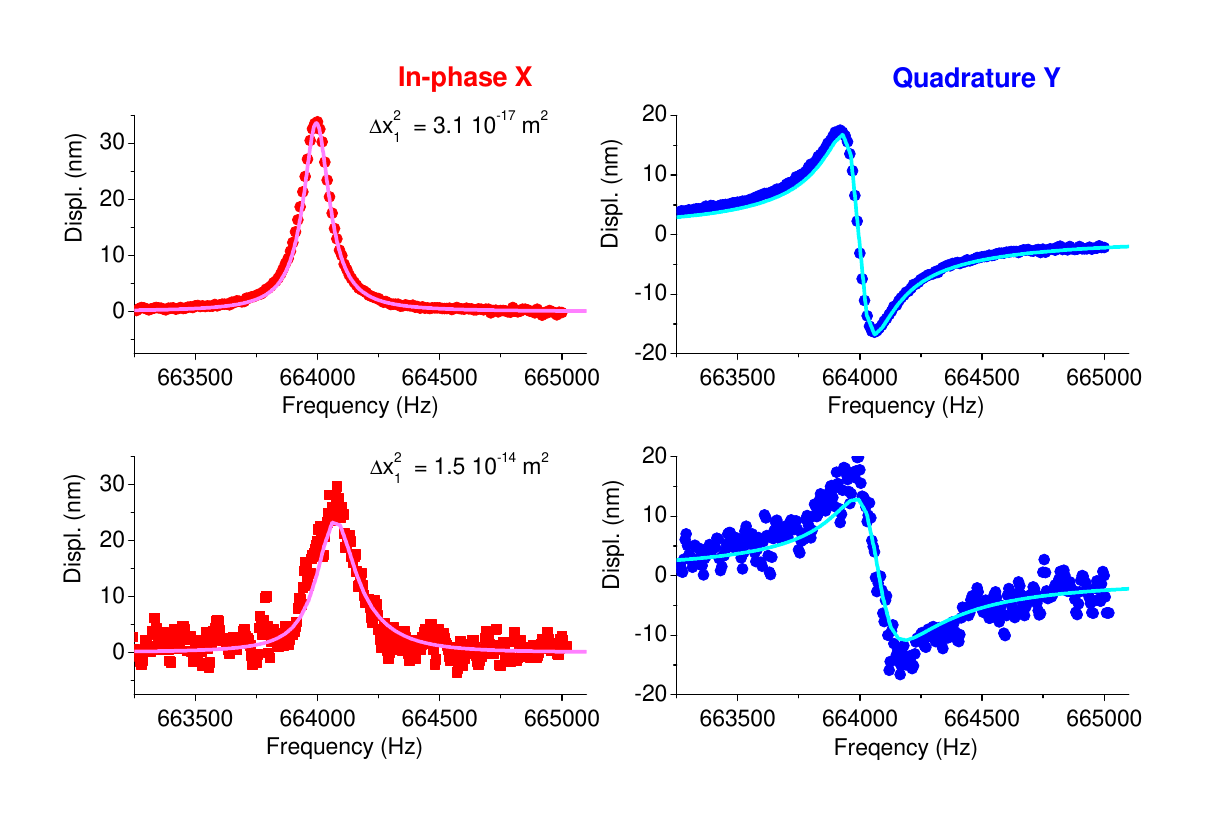}			 			 \includegraphics[width=8cm]{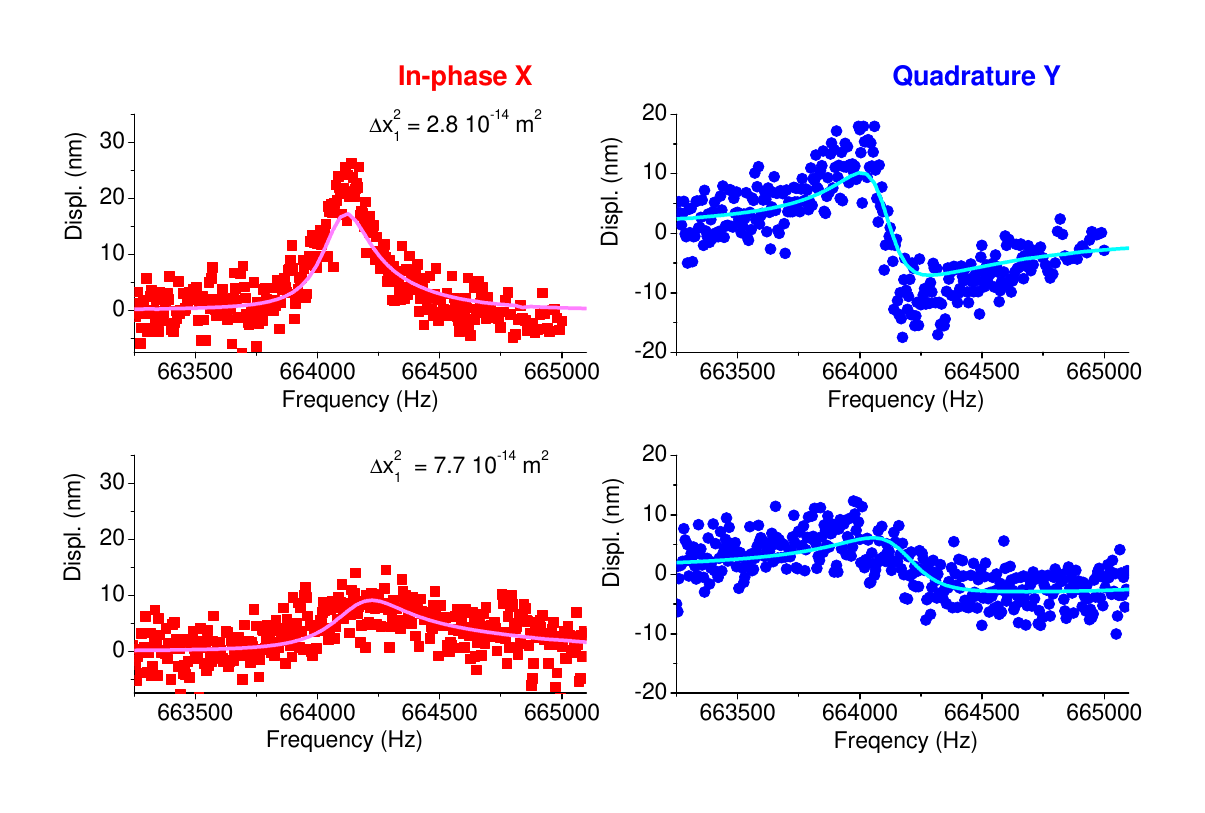}
			\caption{\small{\textbf{Intra-coupling fits}
			(Color online). In phase (X) and quadrature (Y) components measured for mode $n=1$ (first flexure) on sample 300$~\mu$m-n$^\circ$2. The sinusoidal force is kept at $F_1=91~$fN, while the standard deviation $\Delta x_1^2$ (i.e Brownian motion level) is increased (from top to bottom). Lines are theoretical calculations (see text).}}
			\label{fig_reslines}
		\end{figure}

		\begin{figure}[h!]
		\centering
	\includegraphics[width=8cm]{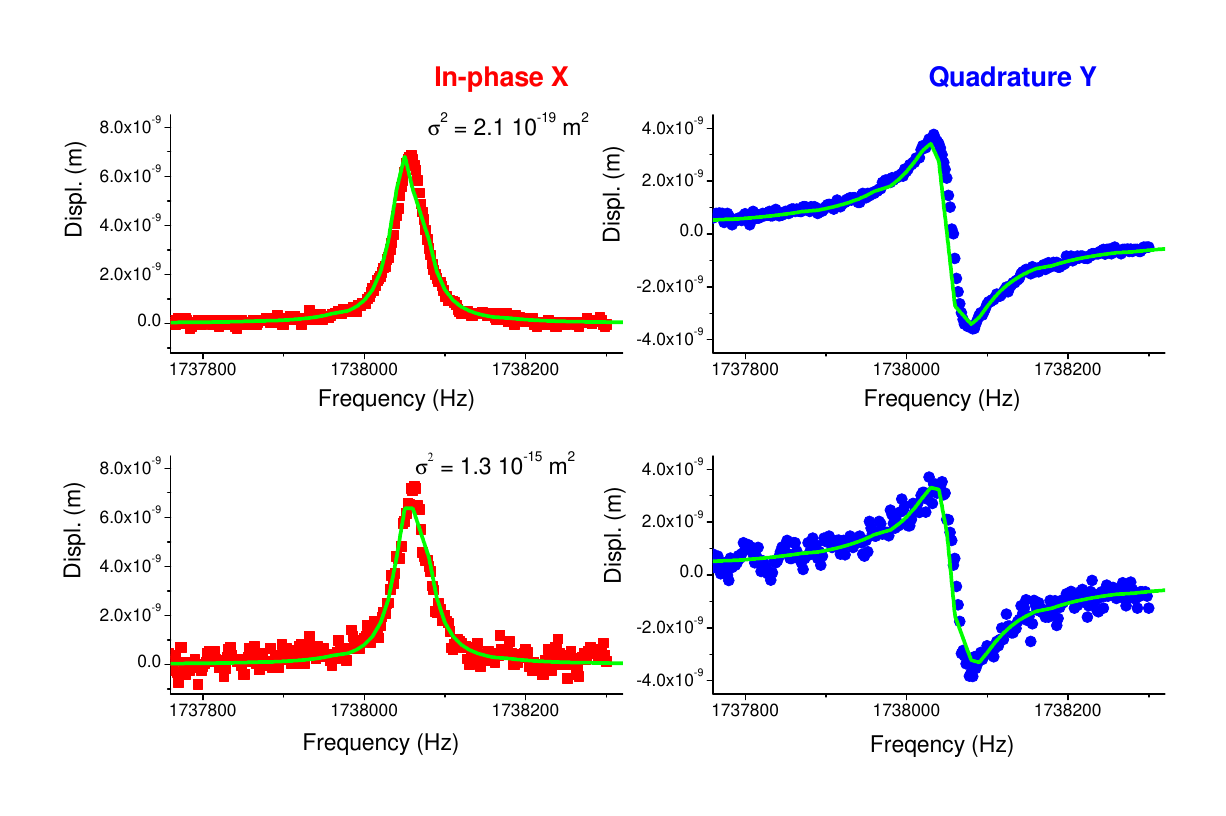} \includegraphics[width=8cm]{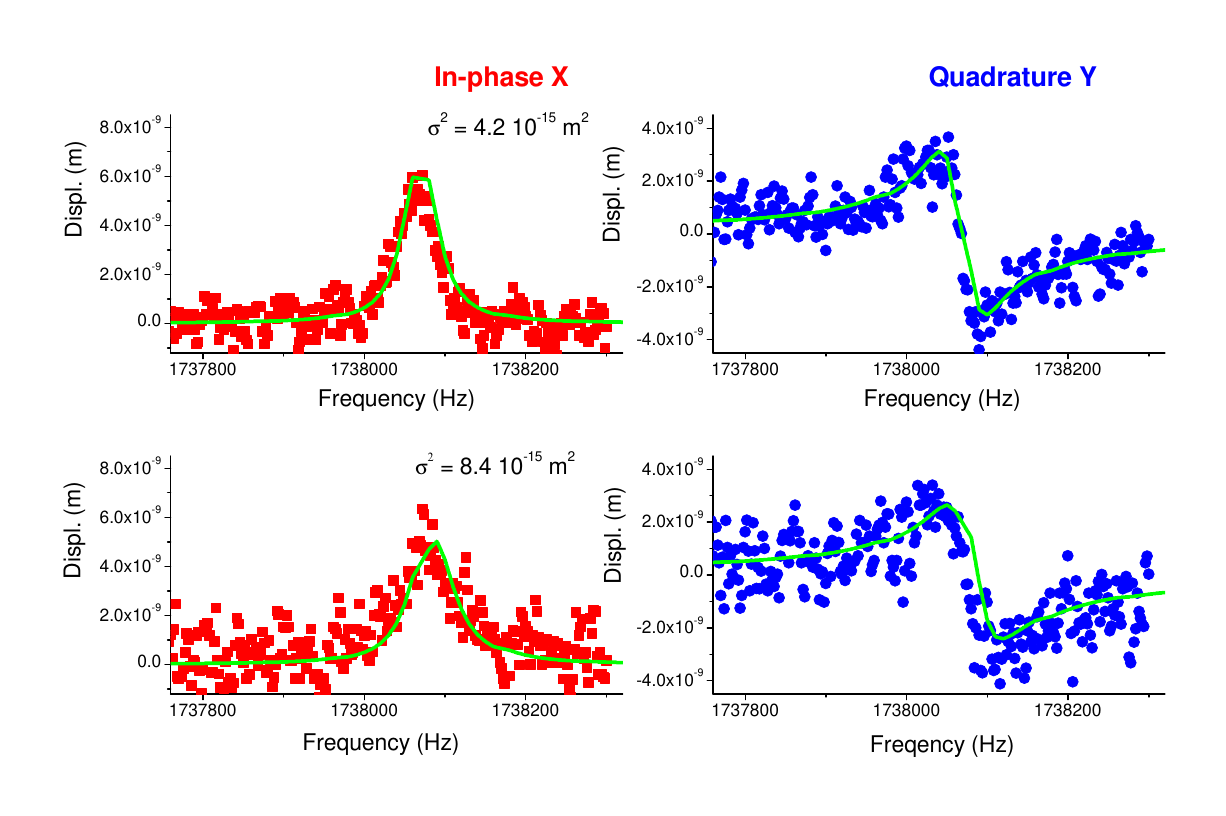}
			\caption{\small{\textbf{Inter-coupling fits.} 
			Data taken on sample 300$~\mu$m-n$^\circ$1, measuring mode $m=3$ while driving fluctuations on mode $n=1$. The applied field is 0.84$~$T for a force $F_3^0=0.6~$pN; the actual mode 1 position fluctuations are quoted in inset. Calculations (full green line) are explained in the text.}}
			\label{fig_inter1}
		\end{figure}
		
\begin{figure}[h!]
		\centering
	\includegraphics[width=10cm]{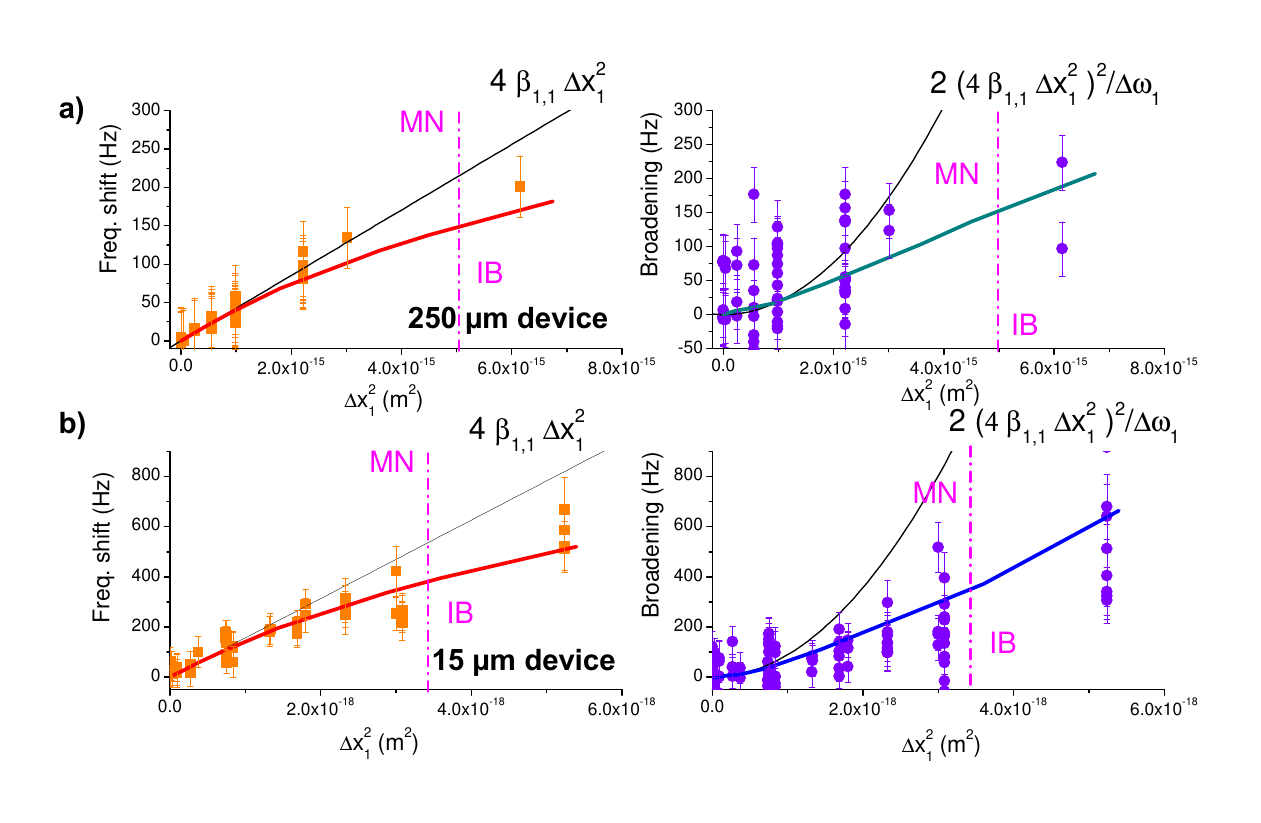} \includegraphics[width=10cm]{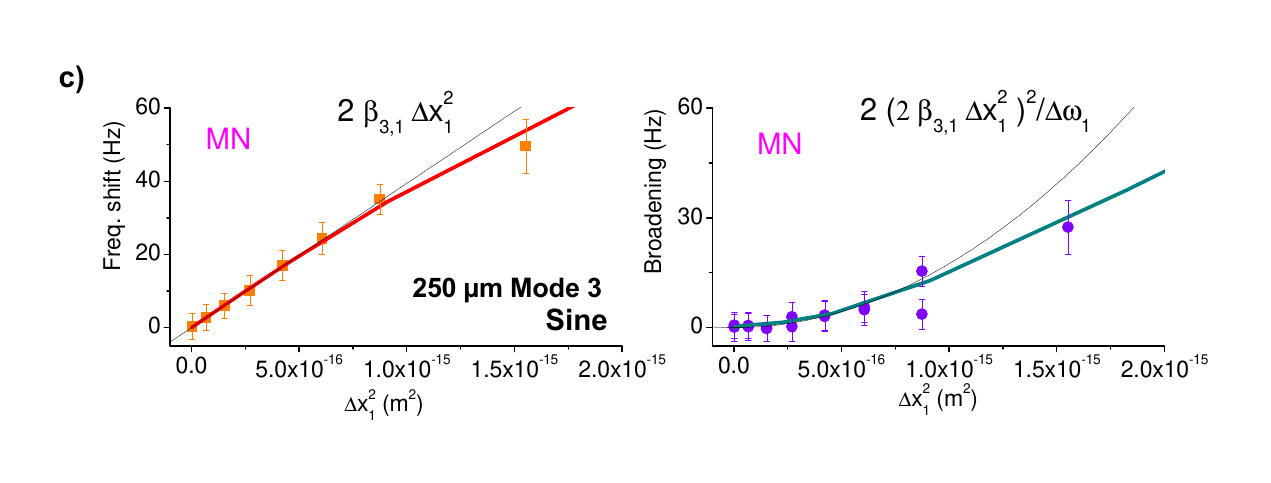}
			\caption{\small{\textbf{Characteristics for 15$~\mu$m and 250$~\mu$m.} 
			Same data as Fig. 3 of the paper, for 2 other devices (see tabular \ref{ValuesModes}). a) ``self-coupling'' on 250$~\mu$m device. b) ``self-coupling'' on 15$~\mu$m device. c) ``mode-coupling'' on 250$~\mu$m device. Note that home-made filter and setup characteristics were different, and had to be calibrated following the same procedure as for 300$~\mu$m devices. }}
			\label{fig_supplchar}
		\end{figure}

To prove the robustness of the effect, the same measurements have been performed on different devices: some rather similar (two high-stress 300$~\mu$m beams, and a 250$~\mu$m) but one quite different (a low-stress 15$~\mu$m beam). Indeed, the parameter $\alpha_\lambda$ depends strongly on the non-linear coefficients $\beta_{n,m}$ (and these depend strongly on the length, see Tab. \ref{ValuesModes}). The agreement between data and theory in Fig. \ref{fig_supplchar} using these extra devices is effectively as good as in Fig. 3. 
Note that the range explored in the ``mode-coupling'' case is smaller than for ``self-coupling'' for pure experimental reasons: the height of the detected peak is smaller, and does not allow to use as high noise intensities. 

All data show the same behavior, which is driven simply by the product of the frequency fluctuation amplitude ($\Sigma_{\delta \omega}$) with the fluctuation correlation time ($1/\Delta \omega_n$, the position fluctuation's correlation time being $2/\Delta \omega_n$): this is precisely the parameter $\alpha_\lambda$ called ``motional narrowing parameter'' in Ref. \cite{dykmanfluctu}.\\

If this parameter is small, we are in the ``motional narrowing'' limit: the fluctuations are too fast to resolve the full frequency distribution. As such, the function with which we have to convolve the linear response is not strictly speaking a distribution of frequency realization, since it is complex-valued. The result of the phase-diffusion process is actually to shift the resonance frequency (by the average of the frequency fluctuations, $\propto \beta_{m,n} \Delta x_n^2$) at first order. At {\it second order}, it slightly broadens the resonance line without changing its shape: the effect is thus $\propto (\beta_{m,n} \Delta x_n^2)^2$. The first term is simply the ``dressing'' of the mode by the interaction with all the others, while the second one is truly the ``decoherence'' effect; the first one is a {\it certain} quantity, while the second one quantifies by how much the frequency of the measured mode fluctuates. \\

If the motional narrowing parameter $\alpha_\lambda$ is large, then we are in the ``inhomogeneous broadening'' limit. This is the case of Ref. \cite{usdecoh} for instance. Note that with an artificial telegraph frequency noise generated by a gate electrode, in Ref. \cite{MotNarrowChan} Chan and co-workers could switch from one limit to the other not by tuning the amplitude of the noise, but its {\it correlation time} instead. 
In this limit, the full range of fluctuations is spanned by the oscillating mode, and one simply measures the averaged motion over all frequency realizations (assuming that the acquisition time is slow enough to capture all fluctuations) \cite{dyk_adiab}. In this case, the resonance expressions can also be simply described by a convolution with $\rho_\lambda(\delta \omega)$, $\lambda = n, m$ or $n,m$:
\begin{eqnarray}
S_X^n (\omega ) & = & S_{Lorentz}^n (\omega) \ast  \rho_n( \omega) \,\,\,\,\, \mbox{Noise on $n$, measure noise on $n$} \nonumber \\
 \tilde{x}_m^0 & = & X_{Lorentz}^m (\omega) \ast \rho_m(\omega) \,\,\,\,\, \mbox{Noise on $m$, measure sine on $m$} \nonumber \\
 \tilde{x}_m^0 & = & X_{Lorentz}^m (\omega) \ast \rho_{m,n}(\omega) \,\,\,\,\, \mbox{Noise on $n$, measure sine on $m$} ,\nonumber
\end{eqnarray}
with $\rho_\lambda(\delta \omega)$ the frequency distributions directly obtained from $\rho(\delta r)$ the distribution of position amplitudes. They simply write:
\begin{eqnarray}
\rho(\delta r) & = & \frac{1}{\Delta x_n^2} \delta r \exp \left[-\frac{1}{2 \, \Delta x_n^2} \delta r^2 \right] \, \Theta \left[\delta r\right] , \nonumber\\
\rho_n(\delta \omega) & = & \frac{1}{\Sigma_{\delta \omega}^n} \exp \left( - \frac{\delta \omega}{\Sigma_{\delta \omega}^n} \right) , \nonumber\\
\rho_m(\delta \omega) & = & \frac{1}{\Sigma_{\delta \omega}^m} \exp \left( - \frac{\delta \omega}{\Sigma_{\delta \omega}^m} \right) , \nonumber\\
\rho_{m,n}(\delta \omega) & = & \frac{1}{\Sigma_{\delta \omega}^{m,n}} \exp \left( - \frac{\delta \omega}{\Sigma_{\delta \omega}^{m,n}} \right)  \nonumber,
\end{eqnarray}
with the standard deviations of the frequency noises defined by:
\begin{eqnarray}
\Sigma_{\delta \omega}^n & = & 2 \beta_{n,n} \Delta x_n^2 , \nonumber \\
\Sigma_{\delta \omega}^m & = & 4 \beta_{m,m} \Delta x_n^2 , \nonumber\\
\Sigma_{\delta \omega}^{m,n} & = & 2 \beta_{m,n} \Delta x_n^2  \nonumber.
\end{eqnarray}
Above, $\Theta[x]$ is the Heaviside step function. The position distribution is a 2D-Gaussian, while the frequency distributions are exponentials; only the couplings are different in the above expressions, and in the paper we simply quote $\Sigma_{\delta \omega}$ without index. \\

\begin{figure}[h!]
		\centering
	\includegraphics[width=7.5cm]{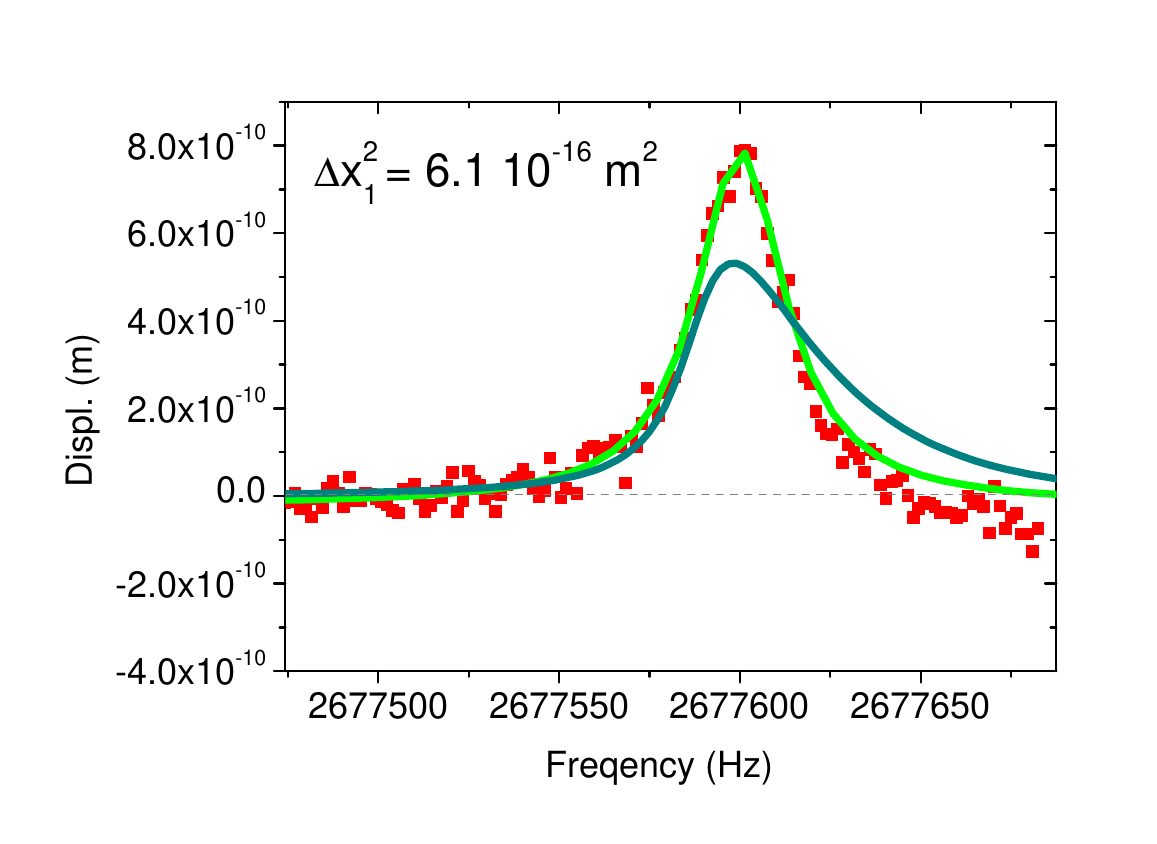} \includegraphics[width=7.5cm]{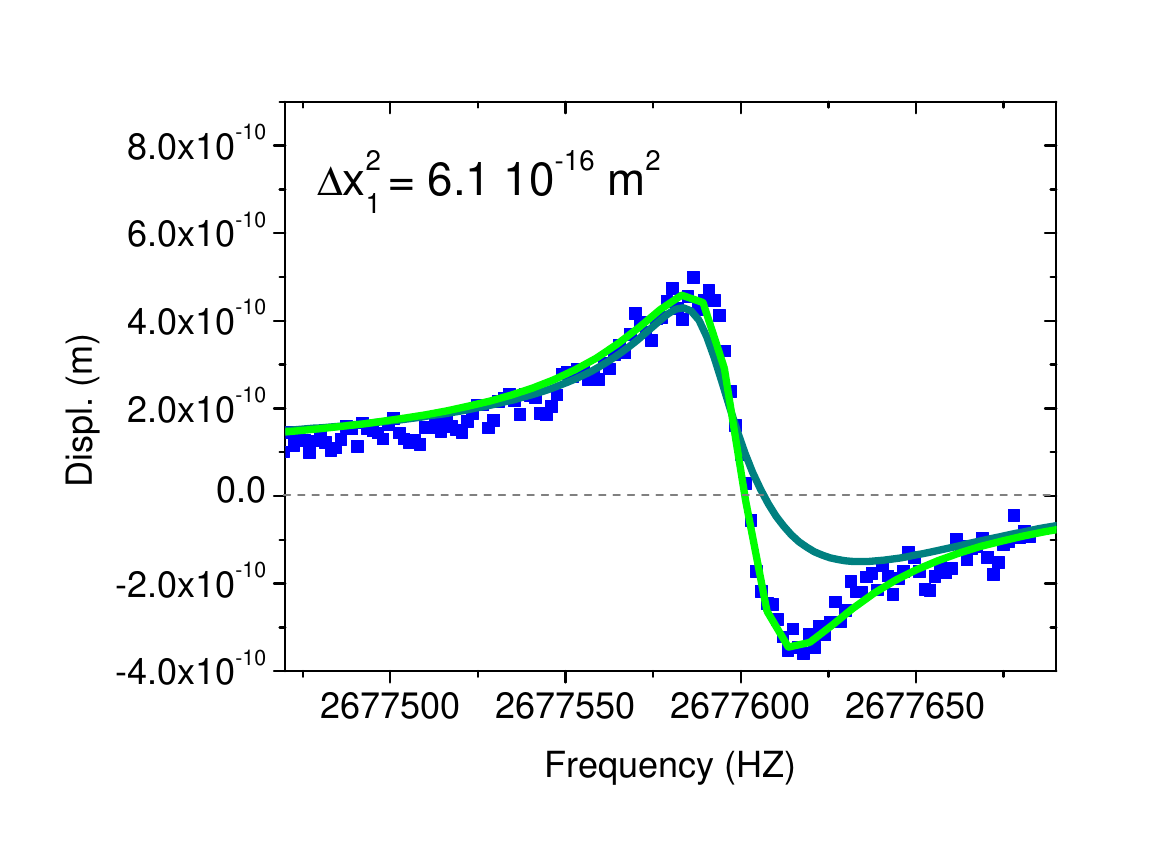} 
			\caption{\small{\textbf{Comparing fits.} 
			Data taken on sample 250$~\mu$m-n$^\circ$1, measuring mode $m=3$ while driving fluctuations on mode $n=1$ (X quadrature on the left, and Y on the right). The applied field is 1$~$T for a force $F_3^0=0.13~$pN; the actual mode $n=1$ position fluctuations is quoted in inset. The full green line is the exact theoretical fit, while the dark grey corresponds to the simple average (``inhomogeneous broadening'' theory, here peak shifted by about 15$~$Hz to match data, see text).}}
			\label{comp_MNIB}
		\end{figure}

The simple averaging procedure works well for large $\alpha_\lambda$ \cite{usdecoh}. However, even for moderate motional narrowing parameters it seems to reproduce not too badly the shapes measured (over estimating a bit the broadening); but it fails to capture the certain frequency shift, by construction. For very high $Q$ and small noise levels, the resolution of the measurement can nonetheless be good enough to demonstrate the difference between the two approaches: this is shown in Fig. \ref{comp_MNIB}. Again, we demonstrate very good agreement with the exact theoretical calculation \cite{dykmanfluctu}: the ``motional narrowing'' effect reduces the impact of the asymmetry of the actual frequency-distribution.

\section{Measured correlators and noise spectra}	
\label{spectraSC}	

The theory applied from Ref. \cite{dykmanfluctu} is valid for small sine wave excitations: one should remain in the linear response limit. In this limit, there should be no back-action of the sinusoidal drive onto the statistics of the fluctuations, even though their {\it spectra} can be altered. \\

\begin{figure*}[h!]
	\includegraphics[width=7.5cm]{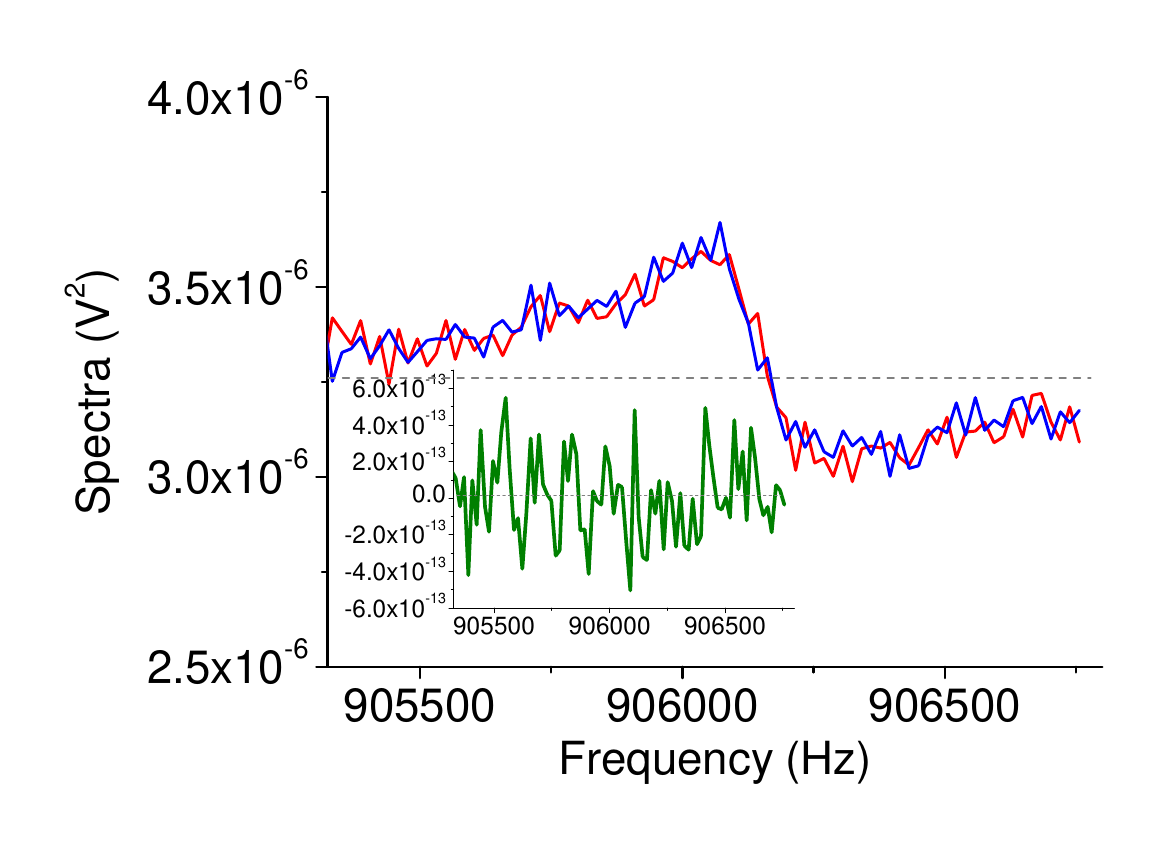}  \includegraphics[width=7.5cm]{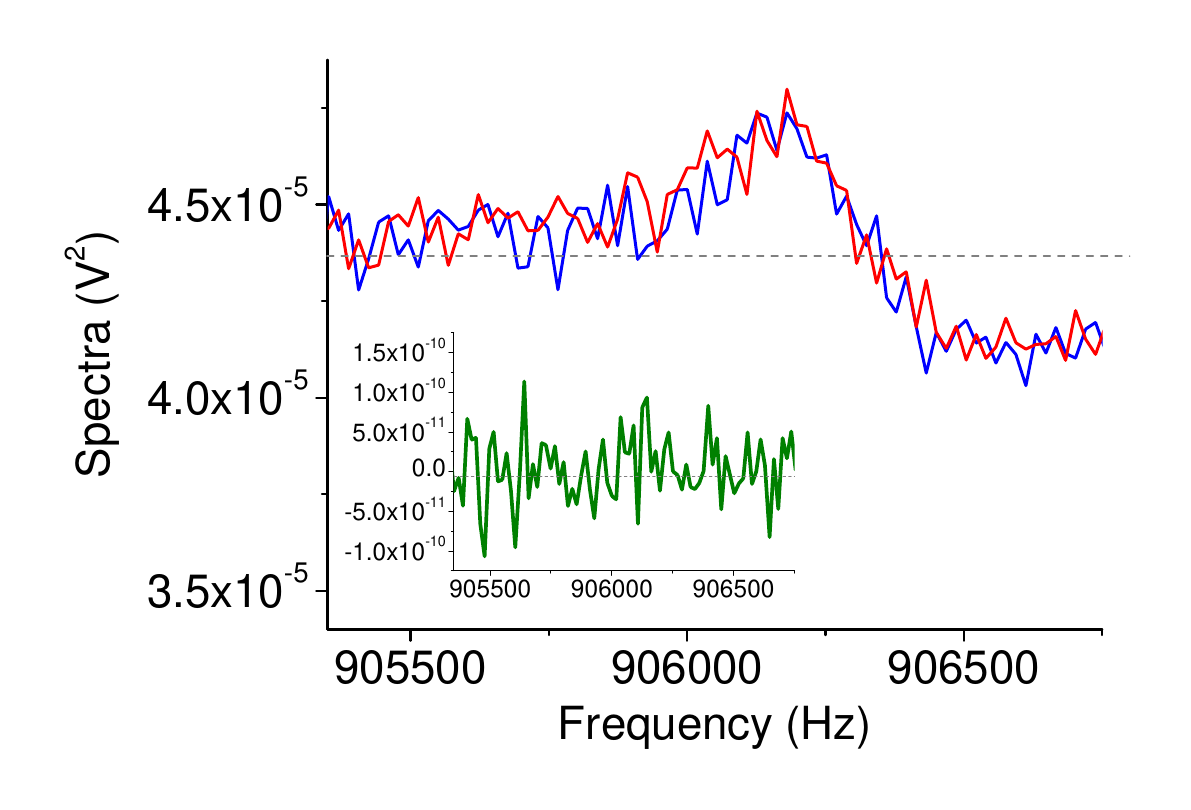}
			\caption{\small{\textbf{Raw spectra, no sine-wave drive.} 
			Raw spectra for mode n=1 measured for small (left, $\Delta x_1^2 =2.2\,10^{-17}~$m$^2$) and large (right, $\Delta x_1^2 =4.\,10^{-15}~$m$^2$) noise levels, with no sine-wave drive. Data from sample 250$~\mu$m-n$^\circ$1 taken at 1$~$T, with X quadrature in red, and Y in blue. The green inset is the cross-correlation spectrum. Acquisition bandwidth $78~$Hz. }}
			\label{fig_rawspectra}
		\end{figure*}

So, from the experimental point of view, it is important to check that this limit is satisfied. We thus first show in Fig. \ref{fig_rawspectra} the raw spectra obtained with no sine wave excitation applied for both small and large noise amplitudes. 
We see that even if the spectrum is {\it distorted } at large Brownian motion levels, we still confirm that X and Y quadratures are equivalent; no cross correlations are detected either. The physical situation is perfectly normal, as it should be for a high $Q$ device 
\cite{nonGaussFP}. \\
		
\begin{figure*}[h!]
	\includegraphics[width=7.5cm]{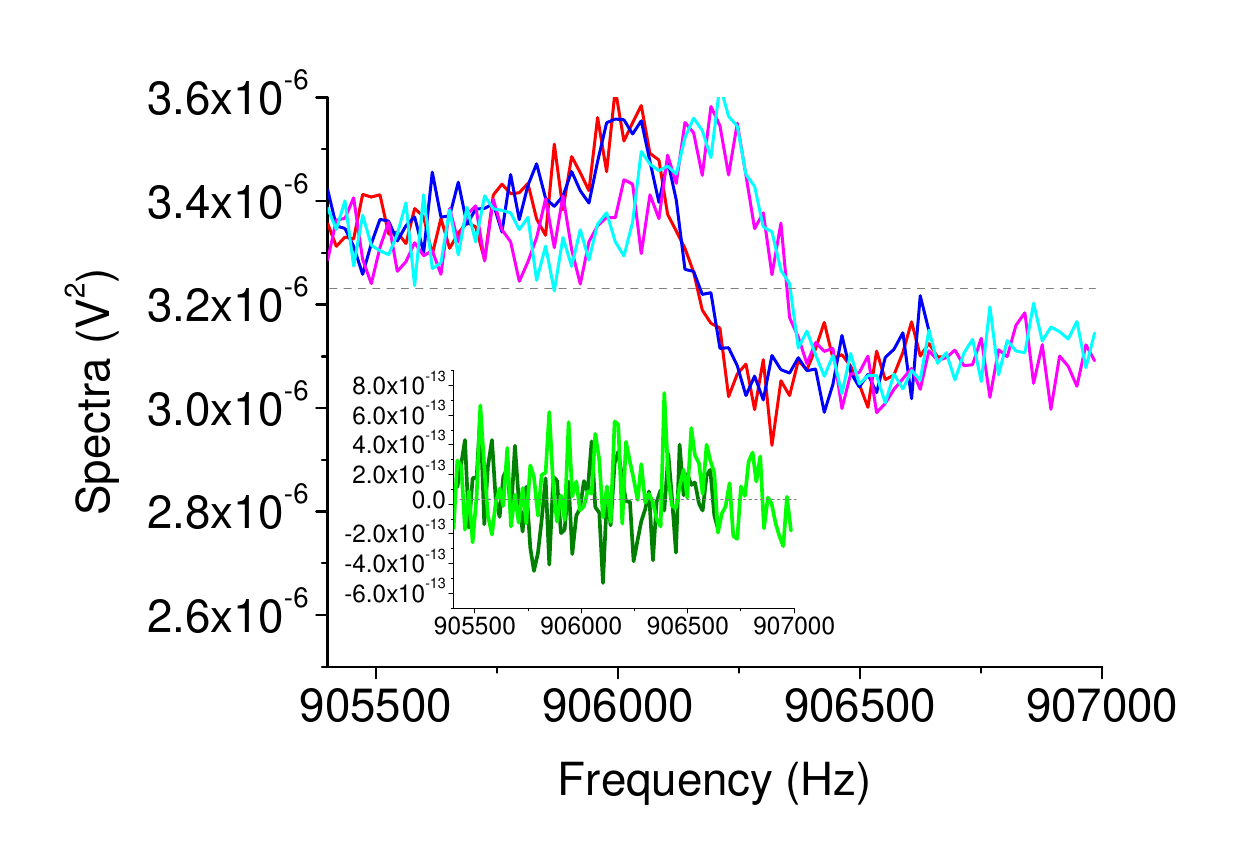}  \includegraphics[width=7.5cm]{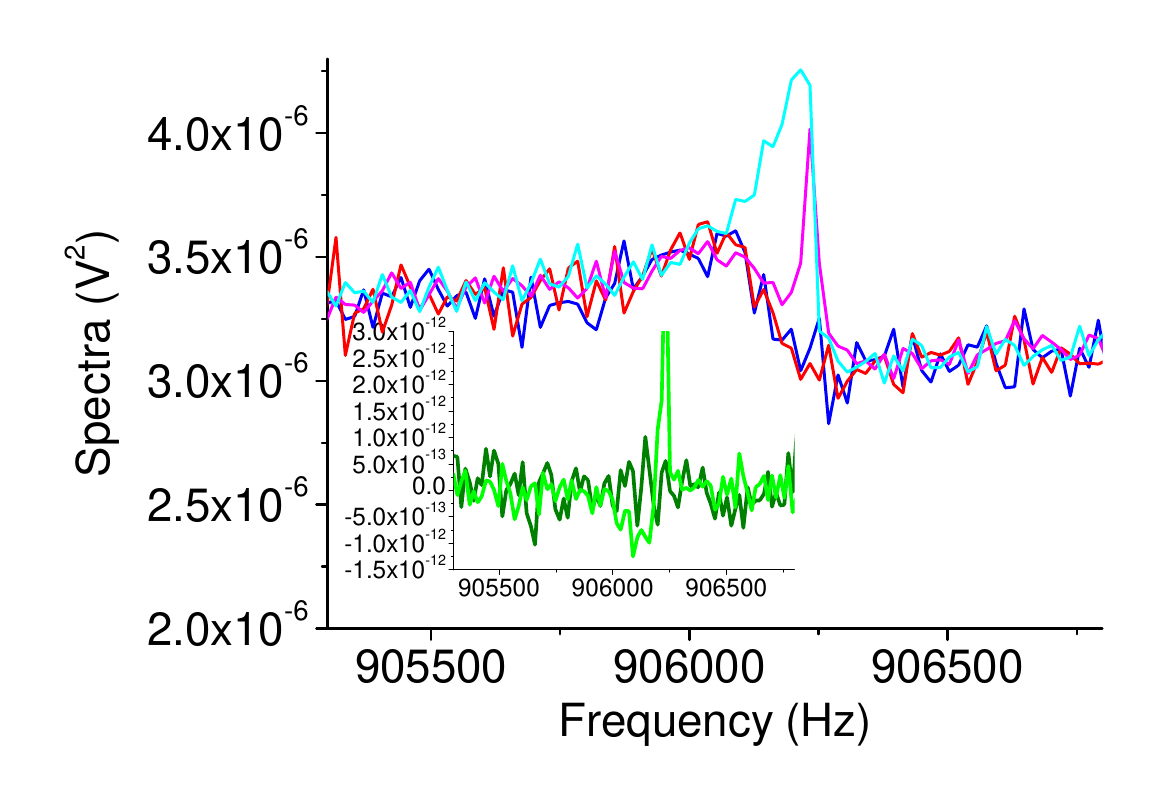}
			\caption{\small{\textbf{Raw spectra, with sine-wave drive.} 
			Raw spectra of Brownian motion of mode $n=1$ measured with a sinusoidal drive on the third mode $m=3$ (left), and on the first $m=1$ (right). Data from sample 250$~\mu$m-n$^\circ$1 taken at 1$~$T, with X quadrature in red, and Y in blue. The green inset is the cross-correlation spectrum. The dark color lines are the references with small (or none) sine-wave drive, while the light color corresponds to the strongly driven case. $F_3^0 = 0~$pN and $F_3^0 = 3.1~$pN (left), and $F_1^0 = 2.7~$pN, $F_1^0 = 11~$pN (right) for small and large settings respectively. For all graphs the motional noise level was $\Delta x_1^2 =2.2\,10^{-17}~$m$^2$. Acquisition bandwidth $78~$Hz. }}
			\label{fig_pump}
		\end{figure*}

But we expect new phenomena to show up if a strong sine-wave signal is applied, like e.g. noise squeezing \cite{fluctu_Buks}.
In Fig. \ref{fig_pump} we show raw spectra obtained with a rather strong sinusoidal force applied onto mode $m=1$ or $m=3$, compared to smaller ones (or none). 
Of course, the data presented in the main part of the paper are obtained with the smallest possible drive levels.
What we see is that when mode $m=3$ is excited, the noise spectrum measured on $n=1$ simply shifts with the amplitude of the sine wave motion. This is nothing but the usual ``mode-coupling'' effect \cite{KunalNlin}, but seen on a noise spectrum. From $\beta_{1,3}=4.\,10^{16}~$Hz/m$^2$, we compute a shift of the order of $300~$Hz consistent with Fig. \ref{fig_pump}. No correlations between X and Y are detected, and the two spectra are equivalent. \\
		
The situation becomes more interesting when one drives strongly with a sine-wave {\it the same mode} where the noise is. 
We see a peak appearing in the $XY$ correlation, and now spectra measured on X and Y are clearly different. We interpret these features as signatures of noise squeezing, as measured in Ref. \cite{fluctu_Buks}.
We illustrate empirically the effect of large sinusoidal excitation levels on the measured response in Fig. 4 of the paper. 
The amplitude of the detected mechanical peak {\it lies below} the calculation, as if the impact of frequency noise was stronger than expected. 
Both the usual Duffing non-linearity of the mode and the altered statistics are responsible for the apparent saturation of the peak height. A new and difficult theoretical work would be needed to further investigate this very interesting regime.

		
\section{Extension to thermally induced Brownian noise}
\label{formulas}

When the stochastic driving force is a real thermal bath, the system is always in the ``motional narrowing'' limit. 
In this case, the response is Lorentzian with a ``dressed'' frequency, and an additional ``thermal decoherence''. 
These can be written at lowest order with the simple expansions, respectively, reproducing the results of Ref. \cite{dykmanfluctu}:
\begin{eqnarray}
\omega_n &=& \omega_n^0 +  4 \beta_{n,n} \Delta x_n^2 + \sum_{m \neq n} 2 \beta_{n,m}  \Delta x_m^2 + \sum_{m'} 2 \bar{\beta}_{n,m'}  \Delta y_{m'}^2 \nonumber, \\
\Delta \omega_n & = & \Delta \omega_n^0 + 2 \frac{\left(4\beta_{n,n} \Delta x_n^2\right)^2}{\Delta \omega_n^0} + \sum_{m \neq n} 2\frac{\left(2\beta_{n,m} \Delta x_m^2\right)^2}{\Delta \omega_m^0} + \sum_{m'} 2\frac{\left(2\bar{\beta}_{n,m'} \Delta y_{m'}^2\right)^2}{\Delta \bar{\omega}_{m'}^0} .\nonumber
\end{eqnarray}
These are Eqs. (4) and (5) of the main paper, on which the thermal bath discussion is based.
The validity of these expansions has been experimentally verified in the present work for two modes only, Figs. 3 \& \ref{fig_supplchar}.
They can be extended in this simple way to many modes since the Brownian motion between $n\neq m$ is uncorrelated.
For the sake of completeness, we also added the sum over the {\it other family of transverse modes} (in $\vec{y}$ direction), which coefficients are designed with a bar, and the index with a prime (the position standard deviation simply writes $\Delta y_{m'}^2$). 
The nonlinear coupling between flexural modes of different family has been studied recently [D. Cadeddu et al. Nano Letters {\bf 16}, 926 (2016)]. The calculation of Section \ref{sectionBernie} is easily adapted with $u(z,t) = x_n(t) \Psi_n(z)+ y_{m'}(t) \bar{\Psi}_{m'}(z)$.
Note that the mode shapes (and mode numbers) can be a bit different along the $\vec{x}$, $\vec{y}$ axes because the beam may not be a perfect square; the second moment of area $I_z$ is a bit different for the two families. Hence the bar notation introduced above. \\

Using the integrals $I_{n,m}$ over mode shapes introduced in Section \ref{sectionBernie}, and the simple equipartition result $\Delta x_n^2 = k_B T/k_n$, $\Delta y_{m'}^2 = k_B T/\bar{k}_{m'}$ we can rewrite these expressions such that:
\begin{eqnarray}
\!\!\!\!\!\!\!\!\!\!\!\!\!\!\frac{\omega_n - \omega_n^0}{\omega_n^0} &\!\!\!\!\!\! =&  \!\!\!\!\!\! \left(\frac{E_{beam} A}{2 L^3}\right) \frac{(k_B T)}{(2 k_n^2)} \left[\sum_{m} \frac{(I_{n,n} I_{m,m} + 2 I_{n,m}^2)}{(k_m/k_n)} + \sum_{m'} \frac{(I_{n,n} \bar{I}_{m',m'} + 2 \tilde{I}_{n,m'}^2)}{(\bar{k}_{m'}/k_n)} \right] \!\!, \label{dressing} \\
\!\!\!\!\!\!\!\!\!\!\!\!\!\!\!\!\!\frac{\Delta \omega_n - \Delta \omega_n^0}{\Delta \omega_n^0}  &\!\!\!\!\!\!\!  = & \!\!\!\!\!\!\! \left(\frac{E_{beam} A}{2 L^3}\right)^2 \frac{(k_B T)^2}{(2 k_n^4)} Q_n^2 \left[ \sum_{m} \frac{Q_m}{Q_n} \frac{(I_{n,n} I_{m,m} + 2 I_{n,m}^2)^2}{(k_m/k_n)^2 (\omega_m^0/\omega_n^0)} + \sum_{m'} \frac{\bar{Q}_{m'}}{Q_n}\frac{(I_{n,n} \bar{I}_{m',m'} + 2 \tilde{I}_{n,m'}^2)^2}{(\bar{k}_{m'}/k_n)^2 (\bar{\omega}_{m'}^0/\omega_n^0)}  \right] \!\!  , \label{thermdecoh}
\end{eqnarray}
having introduced equivalent integrals with bar and tilde applying to the other mode family (and cross term):
\begin{eqnarray*}
I_{n,m} &=& L \int^{L}_{0} \Psi_n(z)^{'}\Psi_m(z)^{'}  dz, \\
\bar{I}_{n',m'} &=& L \int^{L}_{0} \bar{\Psi}_{n'}(z)^{'}\bar{\Psi}_{m'}(z)^{'}  dz, \\
\tilde{I}_{n,m'} &=& L \int^{L}_{0} \Psi_n(z)^{'}\bar{\Psi}_{m'}(z)^{'}  dz.
\end{eqnarray*}
The terms in brackets in Eqs. (\ref{dressing}-\ref{thermdecoh}) have no dimensions and can be calculated from mode parameters. This is what we describe below. The prefactor gives the strength of the effect from materials properties and geometry. \\

Up to date, there is no universal microscopic model explaining nano-mechanical damping, even though many theories exist. They are however phenomenological descriptions of experimental results on SiN beams, e.g. Quirin P. Unterreithmeier et al., PRL {\bf 105}, 027205 (2010)  and A. Suhel et al. APL {\bf 100}, 173111 (2012). For our devices, we have verified that these approaches apply rather well, see M. Defoort, PhD thesis {\it Universit\'e de Grenoble} (2014). 
To make our estimates of ``thermal decoherence'', we thus use the lowest order expression representing a damping proportional to the bending energy $Q_n \propto (E_{bending}+E_{tension})/E_{bending}$. The formulas are summarized in Tab. \ref{Qus}, computed from the mode shapes $\Psi_n(z)$ in the two limits of interest: low-stress and high-stress (see discussion below).
 While these expressions of $Q_n$ are clearly not enough to fit perfectly the experimental findings (it overestimates the $Q$s at high $n$), it is a good starting point since it does not involve any fit parameter in the computation of the bracket of Eq. (\ref{thermdecoh}), and leads to the proper tendency for the sums involved. This will at least produce a very reasonable upper bound for our estimates. \\

The resonance frequencies verify $(\omega_n^0)^2=k_n/m_n$, and both $k_n$ and $m_n$ can be readily computed from the mode shapes $\Psi_n(z)$. The mass writes $m_n= \rho_{beam} A \int^{L}_{0} \Psi_n(z)^{2} dz$, and the spring constant $k_n = E_{beam} I_z \int^{L}_{0} (\partial^2\Psi_n[z]/\partial z^2)^{2} dz+ T_0 \int^{L}_{0} (\partial^2\Psi_n[z]/\partial z^2)\Psi_n(z) dz$ (first term stands for bending energy, second for tensioning; the same integrals are used for the estimate of the $Q$s). We limit the discussion to perfectly clamped beams. The mode shapes $\Psi_n(z)$ (and $\bar{\Psi}_n[z]$) are then obtained by solving the (linear) Euler-Bernoulli equation. \\
	
				\begin{figure}[h!]
		\centering
	\includegraphics[width=8.5cm]{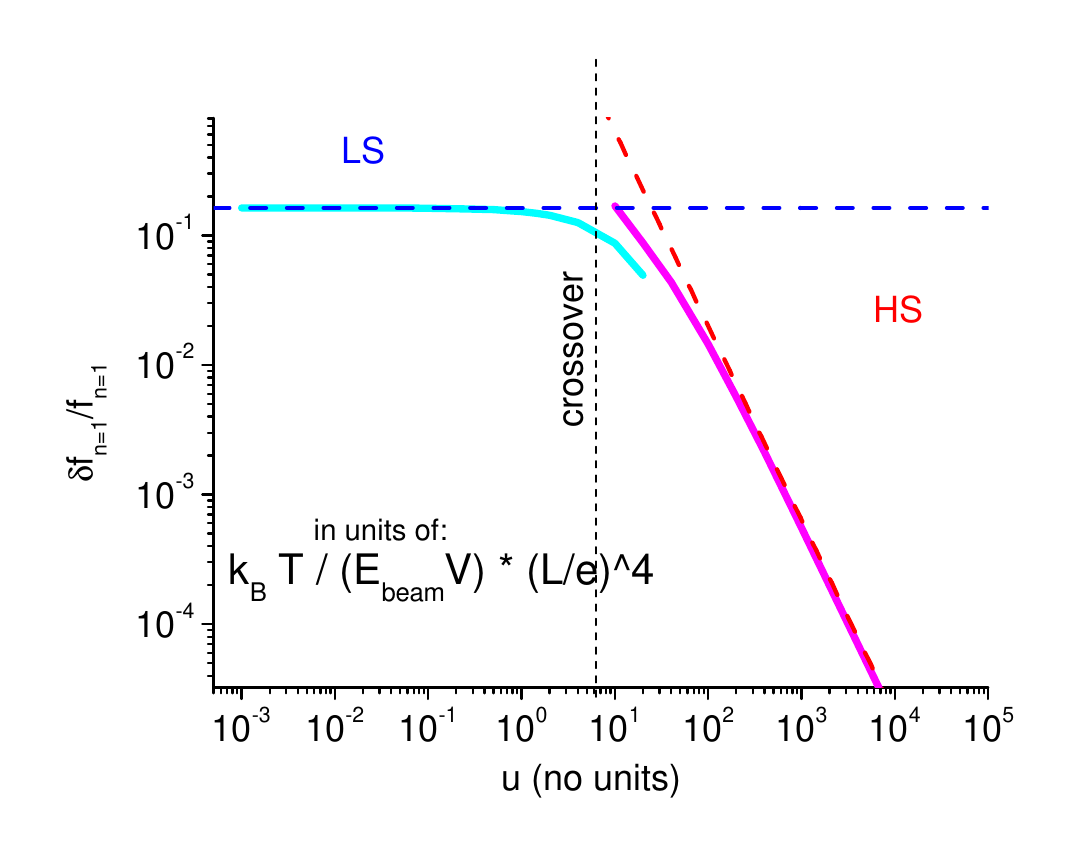}
			\caption{\small{\textbf{Thermal frequency ``dressing'' coefficient vs in-built stress.} 
			Calculated sum in the bracket of Eq. (\ref{dressing}) for a single family of modes. Here, numerics corresponding to mode $n=1$. The overall prefactor is given in inset (same definitions as for Fig. \ref{fig_dephas}). The two limits are depicted (low-stress LS and high-stress HS), with their asymptotic behaviors (dashed lines).}}
			\label{fig_dressed}
		\end{figure}
		
A full analytical solution of the Euler-Bernoulli equation does not exist. However, two limiting cases can be described: low-stress $u\ll 1$ and high-stress $\tilde{u} \ll 1$ where we define $u= T_0  L^2 / (E_{beam} I_z)$ and $\tilde{u}=E_{beam} I_z/(\left|T_0\right| L^2)$. Our definitions are $u<0$ for tensile, and $\tilde{u}>0$. In these regimes, analytic shapes can be found as Taylor expansions of $u$ (and $\sqrt{\tilde{u}}$), and integrated to obtain all the required parameters.
We summarize the results in Tabs. \ref{ls} and \ref{hs} respectively (and Tab. \ref{Qus} for $Q$ factors). In the low-stress case, we fit analytic functions that reproduce rather well the numerical results, and have the proper asymptotic dependence for large $n$. For high-stress, the expansion is exact at the lowest orders in $\tilde{u}$. This is the limit that applies to all our SiN beams. Note that if all modes would be perfectly orthogonal, the $I_{n,m}$ terms would be 0 for $n \neq m$  (and $\bar{I}_{n',m'}$, $\tilde{I}_{n,m'}$ as well). This is not strictly the case, as shown in Tab. \ref{nmls}. \\

\begin{table}[h!]
\begin{center}  \small  \hspace*{-1.5cm}
\begin{tabular}{|c|c|c|c|c|c|c|}    \hline
                     $n$ & ${\cal Q}_n^0$ & ${\cal Q}_n^1$ &   &${\cal Q}_n^0$& ${\cal Q}_n^1$ &${\cal Q}_n^2$      \\   \hline \hline
1 & $1$ & $-0.03027\cdots$ &    & $1$ & $2$  & $17.870\cdots$  \\    \hline
2 & $1$ & $-0.015 \cdots$ &     & $1$ & $2$  & $47.480\cdots$  \\    \hline
3 & $1$ & $-0.00760 \cdots$ &   & $\cdots$ &  $\cdots$ & $\cdots$  \\    \hline
4 & $1$ & $-0.0045 \cdots$ &     & $\cdots$ & $\cdots$ & $\cdots$  \\    \hline
5 & $1$ & $-0.00300 \cdots$ &     & $\cdots$ & $\cdots$ & $\cdots$  \\    \hline
$\cdots$ & $\cdots$ & $\cdots$ &  & $\cdots$ & $\cdots$ & $\cdots$     \\     \hline \hline
$n$         & $1$       & $-\frac{0.0161\cdots + 0.0322\cdots n + 0.101\cdots n^2}{n^3 (n+2.2)}$ &   & $1$ & 2  & $2\left[4+\frac{(n \pi)^2}{2}\right]$    \\    
     &        &  $\pm$1$~$\% $n\geq 4$   &       &      &    &       \\    \hline
\end{tabular}
\caption{\label{Qus} Numerical results for estimates of $Q$ factors of low-stress beams (left, $Q_n =Q_0 \times [ {\cal Q}_n^0 + u \, {\cal Q}_n^1 ] $) and high-stress beams (right, $Q_n = Q_0 \times [{\cal Q}_n^0 + \sqrt{\tilde{u}} \, {\cal Q}_n^1 + \tilde{u} \, {\cal Q}_n^2]/[ \sqrt{\tilde{u}} \, {\cal Q}_n^1 + \tilde{u} \, {\cal Q}_n^2 ]$), assuming losses proportional to bending energy (see text). Tensile means $u<0$, and $\tilde{u}>0$ by construction. $Q_0$ is a fit parameter corresponding to the bending-limited $Q$ factor. Expansions are exact for high-stress, and the fit error to numerics is specified for low-stress.}
\end{center}
\end{table}

\begin{figure}[h!]
		\centering
	\includegraphics[width=8.5cm]{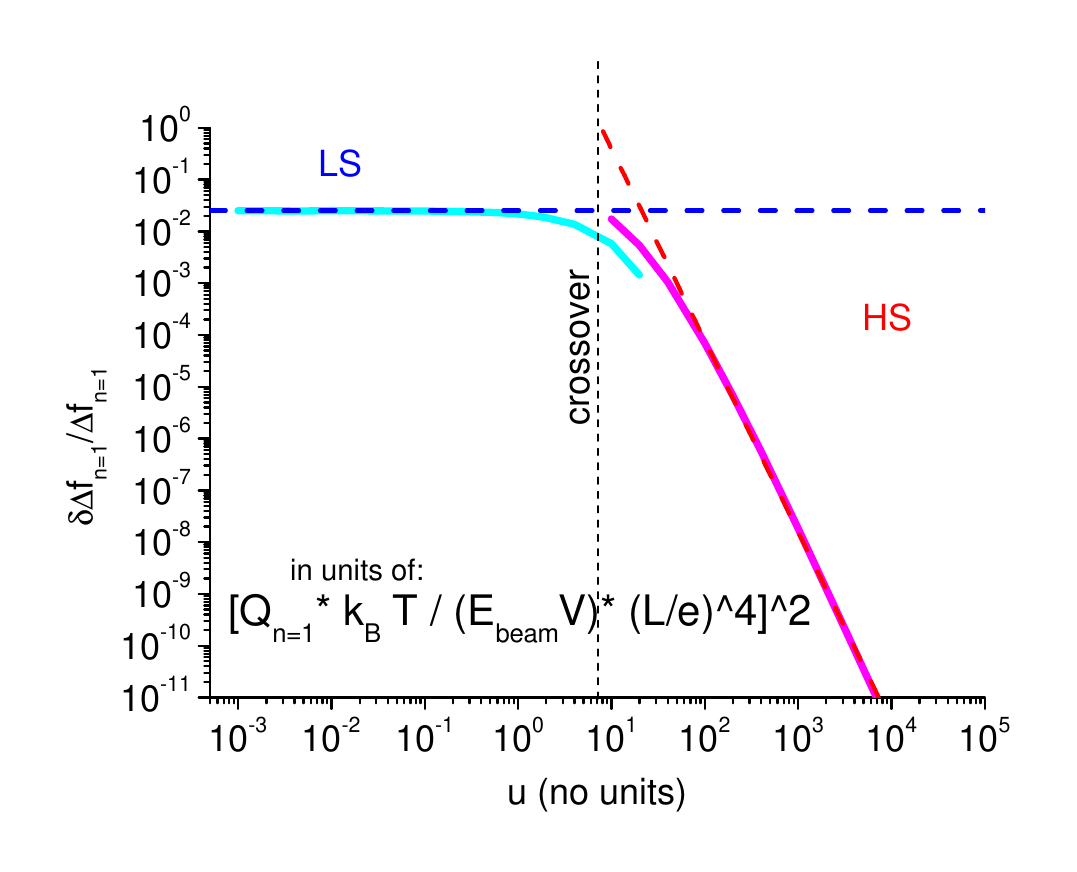}
			\caption{\small{\textbf{Thermal ``dephasing'' coefficient vs in-built stress.} 
			Calculated sum in the bracket of Eq. (\ref{thermdecoh}) for a single family of modes. Here, numerics corresponding to mode $n=1$. The overall prefactor is given in inset (see text). The two limits are depicted (low-stress LS and high-stress HS), with their asymptotic behaviors (dashed lines). }}
			\label{fig_dephas}
		\end{figure}

In Figs. \ref{fig_dressed} and \ref{fig_dephas} we show the $u$-dependence of the brackets in Eqs. (\ref{dressing}-\ref{thermdecoh}) calculated from the Tabulars for mode $n=1$ (first flexure). We compute only a {\it single sum}, the one on the same family of modes; for a perfectly squared (monolithic) beam, taking into account the second sum simply amounts to multiply the result by 2. In the more generic case of a bilayer and rectangular beam, the $\bar{I}_{n',m'}$, $\tilde{I}_{n,m'}$ should be computed correctly.  Thanks to the denominator in the sums, they converge reasonably quickly (much before the high-frequency cut-off that should delimit the validity range of the Euler-Bernoulli equation; namely the atomic size for the wavelength associated to high-frequency modes or the phonon correlation time, whichever comes first). \\

The exact numerical values obtained in Figs. \ref{fig_dressed} and \ref{fig_dephas} are of order $\approx 0.1$ for low-stress, but {\it fall very quickly} with increasing stress. Clearly, only for low-stress devices shall this effect be relevant. The ``dressing'' of the resonance frequency is a certain value if $T$ is fixed; any temperature instability will translate into a frequency noise which can be calculated from Fig. \ref{fig_dressed}. Furthermore, the strength of ``thermal decoherence'' for low-stress devices is essentially given by the prefactor $\left(\frac{E_{beam} A}{2 L^3}\right)^2 \frac{(k_B T)^2}{(2 k_n^4)} Q_n^2$. It goes quadratically with temperature, and since $k_n \propto E_{beam} (w e^3/L^3)$ it depends {\it very strongly on the aspect ratio} $e/L$ of the structure (see inset of Fig. \ref{fig_dephas}, $\propto (k_B T)^2 Q_n^2/(E_{beam} V)^2 \times (L/e)^8$ with $V=w e L$ the volume of the beam). It becomes obvious that low-stress bottom-up structures, especially with high-$Q$, shall be {\it very sensitive to this effect}. In practice, it means that an {\it unstressed} doubly-clamped nanotube can display a poor spectral $Q$ even if its intrinsic one is high, limiting thus sensing applications. A conclusion also reached by different means in Ref. \cite{nanotubeTheory}. \\

\begin{table}[h!]
\begin{center}  \small  \hspace*{-1.5cm}
\begin{tabular}{|c|c|c|c|c|c|c|c|c|}    \hline
                     $n$ & ${\cal L}_n^0$ & ${\cal L}_n^1$ & ${\cal K}_n^0$ &${\cal K}_n^1$& ${\cal M}_n^0$ &${\cal M}_n^1$& ${\cal I}_n^0$ &${\cal I}_n^1$    \\   \hline \hline
1 & $4.73004\cdots$ & $-0.02906\cdots$ & $198.46\cdots$    & $-5.09494\cdots$ & $0.396478\cdots$ & $-0.0004340\cdots$ & $4.8777\cdots$ & $0.0006361\cdots$ \\    \hline
2 & $7.85320\cdots$ & $-0.02377\cdots$ & $1\,669.9\cdots$  & $-20.2173\cdots$ & $0.439028\cdots$ & $-0.0001160\cdots$ & $20.217\cdots$ & $0.003449\cdots$   \\    \hline
3 & $10.9956\cdots$ & $-0.01860\cdots$ & $7\,394.5\cdots$  & $-42.4997\cdots$ & $0.505860\cdots$ & $0.00051528\cdots$ & $50.032\cdots$ & $0.05501\cdots$    \\    \hline
4 & $14.1372\cdots$ &$-0.015182\cdots$ & $20\,082.8\cdots$ & $-77.6713\cdots$ & $0.502777\cdots$ & $0.00021525\cdots$ & $86.270\cdots$ & $0.03945\cdots$    \\    \hline
5 & $17.2788\cdots$ &$-0.012794\cdots$ & $44\,545.4\cdots$ & $-123.614\cdots$ & $0.499750\cdots$ & $0.00009326\cdots$ & $131.93\cdots$ & $0.02726\cdots$    \\    \hline
6 & $20.4203\cdots$ & $-0.01104\cdots$ & $86\,919.8\cdots$ & $-178.034\cdots$ & $0.499880\cdots$ & $0.00005748\cdots$ & $189.12\cdots$ & $0.02282\cdots$    \\    \hline
7 & $23.5619\cdots$ & $-0.00971\cdots$ & $154\,107.5\cdots$& $-242.210\cdots$ & $0.500011\cdots$ & $0.00003834\cdots$ & $254.02\cdots$ & $0.02032\cdots$    \\     \hline 
$\cdots$ & $\cdots$ & $\cdots$ & $\cdots$ & $\cdots$ & $\cdots$ & $\cdots$ & $\cdots$ & $\cdots$    \\     \hline \hline
$n$         & $n \pi \left[ 1+ \frac{1}{2 n}\right]$       & $-\frac{1}{4\pi n}\left[ 1- \frac{1}{n}\right]$ & $\frac{(n \pi)^4}{2}\left[ 1+ \frac{2.2\cdots}{n}\right]$     & $-\frac{(n\pi)^2}{2}$ & $\frac{1}{2}$   & $0$  & $\frac{(n+\frac{1}{2})^2\pi^2-2(n+\frac{1}{2})\pi}{2}$   &  $\frac{1}{2}\frac{(n+\frac{1}{2})\pi-1}{(n+\frac{1}{2})^2\pi^2}$    \\    
     &   $\pm$0.5$~$\% $n\geq 1$    &  $\pm$1$~$\% $n\geq 5$   &  $\pm$1.6$~$\% $n\geq 5$      &  $\pm$1$~$\% $n\geq 1$   &  $\pm$0.6$~$\% $n\geq 4$   &  Negligible  &   $\pm$1$~$\% $n\geq 3$  &   $\pm$2$~$\% $n\geq 5$   \\    \hline
\end{tabular}
\caption{\label{ls} Numerical results for low-stress beams (see text). The coefficients listed stand for $\omega_n^0=({\cal L}_n^0+u \, {\cal L}_n^1)^2 \sqrt{E_{beam} I_z/L^3/(\rho_{beam} A L)}$, $k_n=(E_{beam} I_z/L^3) ({\cal K}_n^0+u \,{\cal K}_n^1)$, $m_n=(\rho_{beam} A L) ({\cal M}_n^0+u \, {\cal M}_n^1)$, and $I_{n,n}={\cal I}_n^0+u\,{\cal I}_n^1$. We remind that our definition of motion amplitude $x_n$ is the maximum value along the beam; and tensile means $u<0$. We give fits to these coefficients as a function of $n$, with the stated accuracy in the last line.}
\end{center}
\end{table}
\begin{table}[h!]
\begin{center}  \small   \hspace*{-1.5cm}
\begin{tabular}{|c|c|c|c|c|c|c|c|c|c|c|c|c|}    \hline
                     $n$ & ${\cal L}_n^0$ & ${\cal L}_n^1$& ${\cal L}_n^2$ & ${\cal K}_n^0$ &${\cal K}_n^1$ &${\cal K}_n^2$ & ${\cal M}_n^0$ &${\cal M}_n^1$ &${\cal M}_n^2$ & ${\cal I}_n^0$ & ${\cal I}_n^1$ &${\cal I}_n^2$  \\   \hline \hline
1 & $1.7724\cdots$ & $1.7724\cdots$ &  $7.0320\cdots$ & $4.9348\cdots$  & $9.8696\cdots$   & $68.4437\cdots$  & $0.5$  & $-1$  & $0$   & $4.9348\cdots$ & $0$  &  $-19.739\cdots$   \\    \hline
$\cdots$          & $\cdots$  & $\cdots$      & $\cdots$ & $\cdots$ & $\cdots$  & $\cdots$  & $\cdots$  & $\cdots$  & $\cdots$ & $\cdots$ & $\cdots$  &  $\cdots$ \\    \hline \hline
$n$         & $\sqrt{n\pi}$ & $\sqrt{n\pi}$ & $\sqrt{n\pi} \left[\frac{6+ (n \pi)^2}{4} \right]$ & $\frac{(n \pi)^2}{2}$ & $(n \pi)^2$  & $2 (n \pi)^2 + \frac{(n \pi)^4}{2}$   & $\frac{1}{2}$   & $-1$  &  $0$  & $\frac{(n \pi)^2}{2}$ & 0  &  $-2 (n \pi)^2$ \\    \hline
\end{tabular}
\caption{\label{hs} Numerical results for high-stress beams (see text). The coefficients listed stand for $\omega_n^0=({\cal L}_n^0+\sqrt{\tilde{u}} \, {\cal L}_n^1+ \tilde{u}  \, {\cal L}_n^2)^2 \sqrt{\left|T_0\right|/L/(\rho_{beam} A L)}$, $k_n=(\left|T_0\right|/L) ({\cal K}_n^0+\sqrt{\tilde{u}} \,{\cal K}_n^1+\tilde{u} \,{\cal K}_n^2)$, $m_n=(\rho_{beam} A L) ({\cal M}_n^0+\sqrt{\tilde{u}} \, {\cal M}_n^1+\tilde{u} \, {\cal M}_n^2)$, and $I_{n,n}={\cal I}_n^0+\sqrt{\tilde{u}}\,{\cal I}_n^1+\tilde{u}\,{\cal I}_n^2$. We remind that our definition of motion amplitude $x_n$ is the maximum value along the beam; and $\tilde{u}>0$ by construction. These expansions are exact. }
\end{center}
\end{table}

\newpage

\begin{table}[h!]
\begin{center}  \small  \hspace*{-1.9cm}
\begin{tabular}{|c|c|c|c|c|c|}    \hline
                     $n,m$ & ${\cal I}_{n,m}^0$  & ${\cal I}_{n,m}^1$  &  & ${\cal I}_{n,m}^1$ &${\cal I}_{n,m}^2$  \\   \hline \hline
1,2           & $0$        & $0$ &  & $0$  & $0$      \\    \hline
1,3           & $4.35786\cdots$ & $0.046108\cdots$ &  & $29.6088\cdots$  & $118.435\cdots$     \\    \hline
1,4           & $0$        & $0$ &  & $\cdots$  & $\cdots$    \\    \hline
1,5           & $-3.38985\cdots$& $-0.017825\cdots$ &  & $\cdots$  & $\cdots$     \\    \hline 
1,6           & $0$& $0$   &  & $\cdots$  & $\cdots$     \\   \hline
1,7           & $2.71958\cdots$& $0.0079437\cdots$ &  & $\cdots$  & $\cdots$     \\    \hline 
1,8           & $0$& $0$   &  & $\cdots$  & $\cdots$     \\   \hline
1,9           & $-2.2558\cdots$& $-0.00374\cdots$ &  & $\cdots$  & $\cdots$     \\    \hline 
$\cdots$      & $\cdots$   & $\cdots$ &  & $\cdots$  & $\cdots$     \\           \hline \hline
2,3           & $0$        & $0$ &  & $0$  & $0$     \\    \hline
2,4           & $8.04755\cdots$       & $0.02438\cdots$ &   & $78.957\cdots$  & $315.83\cdots$     \\    \hline
2,5           & $0$        & $0$ &   & $\cdots$  & $\cdots$      \\    \hline
2,6           & $-7.11816\cdots$       & $-0.005967\cdots$ &   & $\cdots$  & $\cdots$     \\    \hline 
$\cdots$      & $\cdots$       & $\cdots$ &   & $\cdots$  & $\cdots$     \\    \hline \hline
3,4           & $0$        & $0$ &   & $0$  & $0$     \\    \hline 
3,5           & $12.243\cdots$ & $0.0322\cdots$ &   & $\cdots$  & $\cdots$     \\    \hline 
3,6           & $0$        & $0$ &   & $\cdots$  & $\cdots$     \\    \hline 
3,7           & $-11.5594\cdots$ & $-0.01651\cdots$ &   & $\cdots$  & $\cdots$     \\    \hline 
$\cdots$      & $\cdots$       & $\cdots$ &   & $\cdots$  & $\cdots$     \\    \hline \hline
4,5           & $0$        & $0$ &   & $0$  & $0$     \\    \hline 
4,6           & $15.6797\cdots$ & Negligible &   & $\cdots$  & $\cdots$     \\    \hline 
4,7           & $0$        & $0$ &   & $\cdots$  & $\cdots$     \\    \hline 
4,8           & $-15.3318\cdots$ & Negligible &   & $\cdots$  & $\cdots$     \\    \hline 
$\cdots$      & $\cdots$       & $\cdots$ &   & $\cdots$  & $\cdots$     \\    \hline \hline
$n,m$         & $-\frac{8 \pi (n+\frac{1}{2})^2 (m+\frac{1}{2})^2}{(n+m+1)[(n+m+1)^2+(n-m)^2]} \times $       & $0$ &   & $-n m \pi^2 \times  $  & $-4 n m \pi^2 \times  $      \\ 
          & $ \frac{(1+(-1)^{n+m})}{2} (-1)^{C[\frac{(n-1)}{2}]} (-1)^{C[\frac{(m-1)}{2}]} $       &  &   & $ \frac{(1+(-1)^{n+m})}{2} (-1)^{C[\frac{(n-1)}{2}]} (-1)^{C[\frac{(m-1)}{2}]}  $  & $\frac{(1+(-1)^{n+m})}{2} (-1)^{C[\frac{(n-1)}{2}]} (-1)^{C[\frac{(m-1)}{2}]}  $      \\   
            &   $\pm$11$~$\% $n=1, m\geq 9$    &  Negligible   &   &     &    \\   
						&   $\pm$6$~$\% $n \geq 2, m > 2$    &               &   &     &    \\   \hline
\end{tabular}
\caption{\label{nmls} Numerical results for cross term $I_{n,m}$ ($n \neq m$) for low-stress beams (${\cal I}_{n,m}^0+u\,{\cal I}_{n,m}^1$, left) and high-stress beams ($\sqrt{\tilde{u}}\,{\cal I}_{n,m}^1+\tilde{u}\,{\cal I}_{n,m}^2$, right). Note that $I_{n,m}=I_{m,n}$. 
Expansions are exact for high-stress, and the fit error to numerics is specified for low-stress. $C[x]$ stands for the Integer part function (Ceil). Tensile means $u<0$, and $\tilde{u}>0$ by construction. }
\end{center}
\end{table}

\end{document}